\documentclass[copyright,creativecommons]{eptcs} 
\providecommand{\event}{ICE 26} % Name of the event you are submitting to
\usepackage{underscore}           % Only needed if you use pdflatex.
 
\usepackage{breakurl}

\usepackage{wrapfig}

\usepackage{hyperref}
\usepackage{amsfonts}
\usepackage{amsmath}
\usepackage{amssymb}
\usepackage{amsthm}
\usepackage{amsbsy}
\usepackage{enumerate}
\usepackage{stackrel}
\usepackage{bm}
\usepackage{stmaryrd}
\usepackage{wasysym}
\usepackage{color}
\usepackage[utf8]{inputenc}
\usepackage{dashbox}
\usepackage[capitalise]{cleveref}
\usepackage{accents}
\usepackage{lineno}
\crefformat{enumi}{condition~#2#1#3}
\crefname{fact}{Fact}{Facts}
\Crefname{fact}{Fact}{Facts}
\crefformat{fact}{Fact~#2#1#3}

\usepackage{mathtools}
\def\colorPtp{\color{blue}}
\def\colorNode{\color{cyan}}
\def\colorOp{\color{OliveGreen}}
\def\colorMsg{\color{BrickRed}}

\usepackage{amsmath}
\usepackage{amssymb}
\usepackage{mfirstuc}
\newcommand{\aG}{\mathsf{G}}
\newcommand{\gatedistancein}{3pt}
\newcommand{\gatedistanceinand}{2pt}
\newcommand{\gname}[1][i]{{\colorNode{\scriptstyle\textsf{#1}}}}
\usepackage{xifthen}        % for conditional commands
\usepackage{xstring}
\usepackage{xargs}
\usepackage[usenames,dvipsnames,svgnames,table]{xcolor} % must be loaded before tikz
\usepackage{tikz}

\usetikzlibrary{
  arrows,snakes,shapes,automata,backgrounds,petri,positioning,fit,calc,
  decorations.markings,decorations.text,shadows,fadings,patterns,
  decorations.pathreplacing,chains,spy
}

\tikzset{
  src/.style={draw,circle,fill=white,
    minimum size=2mm,
    inner sep=0pt
  },
  sink/.style={draw,circle,double,fill=white,
    minimum size=1.5mm,
    inner sep=0pt
  },
  node/.style={draw,circle,fill=black,
    minimum size=2mm,
    inner sep=0pt
  },
  source/.style={draw,circle,fill=white,
    minimum size=3mm,
    inner sep=0pt
  },
  sink/.style={draw,circle,double,fill=white,
    minimum size=3mm,
    inner sep=0pt
  },
  block/.style = {rectangle, draw=gray, align=center, fill=orange!25, rounded corners=0.1cm,
    minimum size=5mm, inner sep=2pt},
  prenode/.style = {minimum size=9pt,inner sep=2pt, font=\Large},
  bblock/.style = {rectangle, draw=blue!50, opacity=.5, line width=1pt, align=center, fill=white, rounded corners=0.1cm,
    minimum size=7mm, inner sep=2pt},
  prenode/.style = {minimum size=9pt,inner sep=2pt, font=\Large},
  agate/.style={draw, rectangle,
    minimum size=3mm,
    inner sep=0pt,
    fill=orange!25,
    postaction={path picture={% 
        \draw[red]
        ([yshift=\gatedistanceinand]path picture bounding box.south) --
        ([yshift=-\gatedistanceinand]path picture bounding box.north) ;}}
  },
  ogate/.style = {
    diamond, draw, fill=orange!25,
    minimum size=4mm,
    inner sep=0pt,
    postaction={path picture={% 
        \draw[red]
        ([yshift=\gatedistancein]path picture bounding box.south) -- ([yshift=-\gatedistancein]path picture bounding box.north)
        ([xshift=-\gatedistancein]path picture bounding box.east) -- ([xshift=\gatedistancein]path picture bounding box.west)
        ;}}},
  altogate/.style = {
    diamond, draw,
    minimum size=4mm,
    inner sep=0pt,
    postaction={path picture={% 
        \draw
        ([yshift=\gatedistancein]path picture bounding box.south) -- ([yshift=-\gatedistancein]path picture bounding box.north)
        ([xshift=-\gatedistancein]path picture bounding box.east) -- ([xshift=\gatedistancein]path picture bounding box.west)
        ;}}},
  altgate/.style={draw, rectangle,
    minimum size=3mm,
    inner sep=0pt,
    postaction={path picture={% 
        \draw
        ([yshift=\gatedistanceinand]path picture bounding box.south) --
        ([yshift=-\gatedistanceinand]path picture bounding box.north) ;}}},
  anygate/.style = {circle, draw, fill=white,
    minimum size=4mm,
    inner sep=0pt,
    postaction={path picture={% 
        \draw[black]
        ([xshift=-\gatedistancein,yshift=\gatedistancein]path picture bounding box.south east) --
        ([xshift=\gatedistancein,yshift=-\gatedistancein]path picture bounding box.north west)
        ([xshift=-\gatedistancein,yshift=-\gatedistancein]path picture bounding box.north east) --
        ([xshift=\gatedistancein,yshift=\gatedistancein]path picture bounding box.south west)
        ;}}
  },
  smallglobal/.style={
        node distance=1cm and 0.8cm, semithick, scale=0.8, every node/.style={transform shape}
  },
  elli/.style = {draw,densely dotted,-},
  line/.style = {draw,->, rounded corners=0.07cm,>=latex},
  nline/.style = {draw,semithick, ->},
  pline/.style = {draw,->,>=latex},
  node distance=1cm and 0.7cm,
  baseline=(current  bounding  box.center),
  local/.style={rectangle, draw, fill=\fillcolor, drop shadow,
    text centered, rounded corners, minimum height=5em
  },
  bigar/.style={
    draw,very thick, ->
  },
  process/.style={rectangle, draw=gray, fill=\fillcolor, drop shadow,
    text centered, minimum height=5em,text=gray
  },
  choreo/.style={rectangle, draw, fill=\fillcolor, drop shadow,
    text centered, rounded corners, minimum height=5em
  },
  mycfsm/.style={
        font=\footnotesize,
        initial where=above,
        ->,>=stealth,auto, node distance=1cm and 1cm,
        scale=1, every node/.style={transform shape},
        every state/.style=inner sep=2pt,
        baseline=(current  bounding  box.center)
  },
  machinecloud/.style={
    cloud, cloud puffs=10, cloud ignores aspect, minimum height=.1cm, minimum width=2cm, draw
  },
  fitting node/.style={
    inner sep=0pt,
    fill=none,
    draw=none,
    reset transform,
    fit={(\pgf@pathminx,\pgf@pathminy) (\pgf@pathmaxx,\pgf@pathmaxy)}
  },
  mypetri/.style={
    font=\footnotesize,
    baseline=(current  bounding  box.center)
  },
  silentrans/.style = {rectangle, draw=black, align=center, fill=black,
    minimum height=1pt,
    minimum width=15pt,
    inner sep=1.5pt
  },
  reset transform/.code={\pgftransformreset},
  tmtape/.style={draw,minimum size=1.2cm}
}

\newcommand{\p}{\ptp}
\newcommand{\q}{{\ptp[b]}}
\newcommand{\msg}[1][m]{\mathsf{\colorMsg{#1}}}
\newcommand{\ifempty}[3]{%
  \ifthenelse{\isempty{#1}}{#2}{#3}%
}
\newcommandx{\nmerge}[2][1={i},2={},usedefault=@]{
  \ifempty{#2}{
    \ifempty{#1}{\mu}{-\gname[{#1}]}
  }{-{#2}}
}
\newcommandx{\gint}[4][1=i,2=\ptp,3=\msg,4=\q,usedefault=@]{
  \ptp[{#2}] {\colorOp \xrightarrow{\scriptscriptstyle\gname[#1]}} \ptp[{#4}] \colon {\msg[{#3}]}
}
\newcommandx{\gout}[4][1=\gname,2=\ptp,3=\msg,4={\ptp[C]},usedefault=@]{
  \achan[{#2}][{#4}] {\colorOp {\colorOp{!}}} {\msg[{#3}]}
}
\newcommandx{\gin}[4][1=\gname,2=\ptp,3=\msg,4={\ptp[C]},usedefault=@]{
  \achan[{#2}][{#4}] {\colorOp {\colorOp{?}}} {\msg[{#3}]}
}
\newcommandx{\gseq}[3][1=i,2={\aG},3={\aG'},usedefault=@]{
  \gnode[{#1}][{#2} \gseqop {#3}]
}
\newcommandx{\gpar}[3][1=i,2={\aG},3={\aG'},usedefault=@]{
  \gnode[{#1}][\ifempty{#1}{{#2} \gparop {#3}}{({#2} \gparop {#3})}]
}
\newcommandx{\gcho}[3][1=i,2={\aG},3={\aG'},usedefault=@]{
  \gnode[{#1}][\ifempty{#1}{{#2} \gchoop {#3}}{\big({#2} \gchoop {#3}\big)}]
}
\newcommandx{\gchov}[3][1=i,2={\aG},3={\aG'},usedefault=@]{
  \gnode[{#1}][\left(
  \begin{array}l
    \ifempty{#1}{{#2} \\ \gchoop \\ {#3}}{\!\!{#2} \\ \gchoop \\ {#3}}
  \end{array}\right)
  ]
}
\newcommandx{\grec}[3][1=i,2={\aG},3={\p},usedefault=@]{
  \gnode[{#1}][\ifempty{#1}{\grecop {#2} \grecopp {#3}}{\big(\grecop {#2} \grecopp {#3}\big)}]
}
\newcommand{\ptp}[1][A]{
  \ensuremath{\mathtt{\colorPtp{%\capitalisewords
  {#1}}}}
}

\newcommandx{\mkint}[6][3=i,4=\p,5=\msg,6=\q,usedefault=@]{
  \node[bblock,{#1}] (#2) {$\gint[#3][#4][#5][#6]$};
}

\newcommandx{\mkgraph}[3][1=.5cm]{
  \node[source,above = #1 of {#2}] (src#2) {};
  \node[sink,below  = #1 of {#3}] (sink#3) {};
  \path[line] (src#2) -- (#2);
  \path[line] (#3) -- (sink#3);
}

\newcommandx{\mkloop}[4][1=.5,2=1.5]{ %it does insert an agate below and one above
  \node[ogate,above = #1 of {#3}] (entry#3) {};
  \pgfgetlastxy \xentry \yentry;
  \pgfmathtruncatemacro{\xentryrounded}{\xentry};
  \node[ogate,below  = #1 of {#4}] (exit#4) {};
  \pgfgetlastxy \xexit \yexit;
  \pgfmathtruncatemacro{\xexitrounded}{\xexit};
  \path[line] (entry#3) -- (#3);
  \path[line] (#4) -- (exit#4);
  \pgfmathsetmacro\tmpdiff{abs(\xentryrounded - \xexitrounded)}
  \path[line] (exit#4) -|  ($(exit#4)+(\tmpdiff,0)+(#2,0)$) |- (entry#3);
}

\newcommandx{\mklooptwo}[4][1=.5,2=1.5]{ %does not insert the a gate below
  \node[ogate,above = #1 of {#3}] (entry#3) {};
  \pgfgetlastxy \xentry \yentry;
  \pgfmathtruncatemacro{\xentryrounded}{\xentry};
  \path (#4);
   \pgfgetlastxy \xexit \yexit;
  \pgfmathtruncatemacro{\xexitrounded}{\xexit};
  \path[line] (entry#3) -- (#3);
  \pgfmathsetmacro\tmpdiff{abs(\xentryrounded - \xexitrounded)}
  \path[line] (#4) -|  ($(#4)+(\tmpdiff,0)+(#2,0)$) |- (entry#3);
}

\newcommandx{\mklooptwobelow}[4][1=.5,2=1.5]{ %does not insert the a gate below, add extra line down
  \node[ogate,above = #1 of {#3}] (entry#3) {};
  \pgfgetlastxy \xentry \yentry;
  \pgfmathtruncatemacro{\xentryrounded}{\xentry};
  \path (#4);
   \pgfgetlastxy \xexit \yexit;
  \pgfmathtruncatemacro{\xexitrounded}{\xexit};
  \path[line] (entry#3) -- (#3);
  \pgfmathsetmacro\tmpdiff{abs(\xentryrounded - \xexitrounded)}
  \path[line] (#4)  |- ($(#4)+(0,-0.5)$) -|  ($(#4)+(\tmpdiff,0)+(#2,0)$) |- (entry#3);
}

\newcommandx{\mklooponetwo}[4][1=.5,2=1.5]{ %does not insert the a gate below
  \path (#3);
  \pgfgetlastxy \xentry \yentry;
  \pgfmathtruncatemacro{\xentryrounded}{\xentry};
  \path (#4);
   \pgfgetlastxy \xexit \yexit;
  \pgfmathtruncatemacro{\xexitrounded}{\xexit};
  \path[line] (entry#3) -- (#3);
  \pgfmathsetmacro\tmpdiff{abs(\xentryrounded - \xexitrounded)}
  \path[line] (#4) -|  ($(#4)+(\tmpdiff,0)+(#2,0)$) |- (entry#3);
}

\newcommandx{\mkfork}[4][2=gatenode,3=i,4=.6,usedefault=@]{
  \mkgatebegin{#1}[{\gname[#3]}][agate][#4]{#2}
}

\newcommandx{\mkbranch}[4][2=gatenode,3=i,4=.6,usedefault=@]{
  \mkgatebegin{#1}[{\gname[#3]}][ogate][#4]{#2}
}

\newcommandx{\mkgatebegin}[5][2={},3=ogate,4=.5]{
  \coordinate (gatecord) at (0,0);
  \foreach \n [count=\i] in {#1}{
    \pgfgetlastxy \xc \yc;
    \path (\n);
    \pgfgetlastxy \xn \yn;
    \coordinate (gatecord) at ($(gatecord) + (\xn,0)$);
    \coordinate (gatecord) at ($1/\i*(gatecord)$);
    \ifdim \yn < \yc
    \node (max) at (0,\yc) {};
    \else
    \node (max) at (0,\yn) {};
    \fi
  }
  \coordinate (gatecord) at ($(gatecord) + (0,#4) + (max)$);
  \node[#3,label={below:$#2$}] (#5) at (gatecord) {};
  \pgfgetlastxy{\xgate}{\ygate};
  \pgfmathtruncatemacro{\xgateround}{\xgate};
  \StrCount{#1,}{,}[\l] % from package xxstring
  \ifnum \l < 2 {\errmessage{#1 argument should be a comma-separated list of lenght >= 2}}
  \else{
    \foreach \n in {#1}{
      \path (\n);
      \pgfgetlastxy{\xnode}{\ynode};
      \pgfmathtruncatemacro{\xnround}{\xnode};
      \pgfmathsetmacro\tmpdiff{abs(\xnround - \xgateround)}
      \ifdim \tmpdiff pt > 1 pt \path[line] (#5) -| (\n);
      \else
        \path[line] (#5) -- (\n);
      \fi
    }
  }
  \fi
}

\newcommandx{\mkmerge}[4][2=gatenode,3=i,4=0,usedefault=@]{\mkgateend{#1}[{\ifempty{#3}{}{\nmerge[#3]}}][ogate][#4]{#2}}

\newcommandx{\mkjoin}[4][2=gatenode,3=i,4=0,usedefault=@]{\mkgateend{#1}[{\ifempty{#3}{}{\nmerge[#3]}}][agate][#4]{#2}}

\newcommandx{\mkgateend}[5][2={},3=ogate,4=.5]{
  \coordinate (gatecord) at (0,0);
  \foreach \n [count=\i] in {#1}{
    \pgfgetlastxy \xc \yc;
    \path (\n);
    \pgfgetlastxy \xn \yn;
    \coordinate (gatecord) at ($(gatecord) + (\xn,0)$);
    \coordinate (gatecord) at ($1/\i*(gatecord)$);
    \ifdim \yn > \yc
    \node (min) at (0,\yc) {};
    \else
    \node (min) at (0,\yn) {};
    \fi
  }
  \coordinate (gatecord) at ($(gatecord) - (0,#4) + (min)$);
  \node[#3,label={above:$#2$}] (#5) at (gatecord) {};
  \pgfgetlastxy{\xgate}{\ygate};
  \pgfmathtruncatemacro{\xgateround}{\xgate};
  \StrCount{#1,}{,}[\l] % from package xxstring
  \ifnum \l < 2 {\errmessage{#1 argument should be a comma-separated list of lenght >= 2}}
  \else{
    \foreach \n in {#1}{
      \path (\n);
      \pgfgetlastxy{\xnode}{\ynode};
      \pgfmathtruncatemacro{\xnround}{\xnode};
      \pgfmathsetmacro\tmpdiff{abs(\xnround - \xgateround)}
      \ifdim \tmpdiff pt > 1 pt \path[line] (\n) |- (#5);
      \else
        \path[line] (\n) -- (#5);
      \fi
    }
  }
  \fi
}

\newtheorem{definition}{Definition}[section]

\newtheorem{proposition}[definition]{Proposition}
\newtheorem{theorem}[definition]{Theorem}

\newtheorem{remark}[definition]{Remark}
\newtheorem{example}[definition]{Example}

\DeclareMathAlphabet{\mathpzc}{OT1}{pzc}{m}{it}

\newcommand{\brunningex}{\smallskip\noindent$\blacktriangleright$\hspace{4pt}}
\newcommand{\erunningex}{\hfill $\blacktriangleleft$\smallskip}

\newcommand{\Comment}[1]{ }

\def\Pred[#1]{~[\,#1\,]}

\newcommand{\RS}{\mathsf{RC}}

\newcommand{\IDS}{\mathsf{IDS}\!}

\newcommand{\Commented}[1]{}

\newcommand{\gts}{\downdownarrows}

\newcommand{\proj}[2]{#1\!\!\downharpoonright\!_{#2}}

\newcommand{\emb}[5]{#4\ \,{^{\mbox{\tiny $?$}}_{\mbox{\tiny $!$}}\hspace{-8pt}\hookrightarrow_{\!(#1,#2,#3)\,\,}}#5}
\newcommand{\embd}{\,\,{^{\mbox{\tiny $?$}}_{\mbox{\tiny $!$}}\hspace{-8pt}\hookrightarrow}\,}
\newcommand{\intf}{\iota}
\newcommand{\nintf}{\not\intf}
\newcommand{\subj}[1]{\mathsf{sbj}(#1)}

  \def\finex{{\unskip\nobreak\hfil\penalty50\hskip1em\null\nobreak\hfil{\Large $\diamond$}\parfillskip=0pt\finalhyphendemerits=0\endgraf}}

\newcommand{\cs}{\mathbb{C}}

\newcommand{\HH}{{\ptp[h]}}
\newcommand{\hh}{\HH}

\newcommand{\KK}{{\ptp[k]}}

\newcommand{\roles}{\mathbf{P}}

\newcommand{\ttp}{{\ptp[p]}}
\newcommand{\ttq}{{\ptp[q]}}

\newcommand{\ttr}{{\ptp[r]}}
\newcommand{\pr}{{\ttr}}
\newcommand{\tts}{{\ptp[s]}}

\newcommand{\ttu}{{\ptp[u]}}
\newcommand{\ttx}{{\ptp[x]}}

\newcommand{\ttv}{{\ptp[v]}}
\newcommand{\ttw}{{\ptp[w]}}

\newcommand{\elle}{\mathit{l}}

\newcommand{\PC}{\mathcal{P}\hspace{-2pt}\mathcal{C}\!}
\newcommand{\fusion}{{\mathcal{F}}}
\newcommand{\fusioncomp}{{\mathcal{F}\!\mathcal{C}}}

\newcommand{\noeps}{\not\hspace{-1pt}\bm{\varepsilon}}
\newcommand{\Set}[1]{\{#1\}}

\renewcommand{\implies}{~\Rightarrow~ }

\newcommand{\trans}[2][{}]{\,\xrightarrow{#2}_{#1}\,}
\newcommand{\lts}[1]{\trans{#1}}
\newcommand{\LTS}[1]{\xRightarrow{#1}}

\newcommand{\notlts}[1]{\stackrel{#1}{\;\;\not\!\!\longrightarrow}}

\newcommandx{\achan}[2][1=A,2=B,usedefault=@]{{\ptp[#1]\,\ptp[#2]}}

\newcommandx{\outop}[2][1=\gname,2={}]{{\colorOp{!}}^{{#1}{#2}}}
\newcommandx{\inop}[2][1=\gname,2={}]{{\colorOp{?}}^{{#1}{#2}}}

\newcommandx{\aout}[5][1={\p},2={\q},3={},4=m,5={},usedefault=@]{
  \achan[#1][#2] \outop[{#3}] {\msg[#4]}{#5}
}
\newcommandx{\ain}[5][1={\p},2={\q},3={},4=m,5={},usedefault=@]{
  \achan[#1][#2] \inop[{#3}] {\msg[#4]}{#5}
}

\tikzset{
  cnode/.style={
    shape=circle,
    minimum size = 0mm,
    inner sep = 1pt,
    font=\tiny,
    draw
  },
  carrow/.style={
    ->,
    shorten >=1pt,
    >=stealth',
    auto,
    draw,
    sloped
  }
}

\tikzset{
  src/.style={draw,circle,fill=white,
    minimum size=2mm,
    inner sep=0pt
  },
  sink/.style={draw,circle,double,fill=white,
    minimum size=1.5mm,
    inner sep=0pt
  },
  node/.style={draw,circle,fill=black,
    minimum size=2mm,
    inner sep=0pt
  },
  source/.style={draw,circle,fill=white,
    minimum size=3mm,
    inner sep=0pt
  },
  sink/.style={draw,circle,double,fill=white,
    minimum size=3mm,
    inner sep=0pt
  },
  block/.style = {rectangle, draw=gray, align=center, fill=orange!25, rounded corners=0.1cm,
    minimum size=5mm, inner sep=2pt},
  prenode/.style = {minimum size=9pt,inner sep=2pt, font=\Large},
  bblock/.style = {rectangle, draw=blue!50, opacity=.7, line width=.5pt, align=center, fill=white, rounded corners=0.1cm,
    minimum size=4mm, inner sep=1pt},
  prenode/.style = {minimum size=9pt,inner sep=2pt, font=\Large},
  agate/.style={draw, rectangle,
    minimum size=3mm,
    inner sep=0pt,
    fill=orange!25,
    label={[red]center:$\mid$}
  },
  ogate/.style = {
    diamond, draw, fill=orange!25,
    minimum size=4mm,
    inner sep=0pt,
    label={[red]center:$+$}
  },
  lgate/.style = {
    diamond, draw, fill=orange!25,
    minimum size=4mm,
    inner sep=0pt,
    label={[red]center:$\circlearrowleft$}
    },
  altogate/.style = {
    diamond, draw,
    minimum size=4mm,
    inner sep=0pt,
    postaction={path picture={% 
        \draw
        ([yshift=\gatedistancein]path picture bounding box.south) -- ([yshift=-\gatedistancein]path picture bounding box.north)
        ([xshift=-\gatedistancein]path picture bounding box.east) -- ([xshift=\gatedistancein]path picture bounding box.west)
        ;}}},
  altgate/.style={draw, rectangle,
    minimum size=3mm,
    inner sep=0pt,
    postaction={path picture={% 
        \draw
        ([yshift=\gatedistanceinand]path picture bounding box.south) --
        ([yshift=-\gatedistanceinand]path picture bounding box.north) ;}}},
  anygate/.style = {circle, draw, fill=white,
    minimum size=4mm,
    inner sep=0pt,
    postaction={path picture={% 
        \draw[black]
        ([xshift=-\gatedistancein,yshift=\gatedistancein]path picture bounding box.south east) --
        ([xshift=\gatedistancein,yshift=-\gatedistancein]path picture bounding box.north west)
        ([xshift=-\gatedistancein,yshift=-\gatedistancein]path picture bounding box.north east) --
        ([xshift=\gatedistancein,yshift=\gatedistancein]path picture bounding box.south west)
        ;}}
  },
  smallglobal/.style={
        node distance=1cm and 0.8cm, semithick, scale=0.8, every node/.style={transform shape}
  },
  elli/.style = {draw,densely dotted,-},
  line/.style = {draw,->, rounded corners=0.07cm,>=latex},
  nline/.style = {draw,semithick, ->},
  pline/.style = {draw,->,>=latex},
  node distance=1cm and 0.7cm,
  baseline=(current  bounding  box.center),
  local/.style={rectangle, draw, fill=\fillcolor, drop shadow,
    text centered, rounded corners, minimum height=5em
  },
  bigar/.style={
    draw,very thick, ->
  },
  process/.style={rectangle, draw=gray, fill=\fillcolor, drop shadow,
    text centered, minimum height=5em,text=gray
  },
  choreo/.style={rectangle, draw, fill=\fillcolor, drop shadow,
    text centered, rounded corners, minimum height=5em
  },
  mycfsm/.style={
        font=\footnotesize,
        initial where=above,
        ->,>=stealth,auto,
		  node distance=1.9cm,
        scale=.85,
		  every node/.style={transform shape},
        every state/.style={cnode, inner sep=1pt, transform shape},
		  every edge/.style={carrow},
        baseline=(current  bounding  box.center),
        initial text={}
  },
  machinecloud/.style={
    cloud, cloud puffs=10, cloud ignores aspect, minimum height=.1cm, minimum width=2cm, draw
  },
  fitting node/.style={
    inner sep=0pt,
    fill=none,
    draw=none,
    reset transform,
    fit={(\pgf@pathminx,\pgf@pathminy) (\pgf@pathmaxx,\pgf@pathmaxy)}
  },
  mypetri/.style={
    font=\footnotesize,
    baseline=(current  bounding  box.center)
  },
  silentrans/.style = {rectangle, draw=black, align=center, fill=black,
    minimum height=1pt,
    minimum width=15pt,
    inner sep=1.5pt
  },
  reset transform/.code={\pgftransformreset},
  tmtape/.style={draw,minimum size=1.2cm}
}

\def \bmr {\begin{color}{red}} 
\def \emr {\end{color}}
\def \bfr {\begin{color}{Fuchsia}} 
\def \efr {\end{color}}
\def \bmc {\begin{color}{magenta}Mariangiola: } 
\def \emc {\end{color}}
\def \bfc {\begin{color}{brown}Franco: } 
\def \efc {\end{color}}
\def \brc {\begin{color}{blue}Rolf: } 
\def \erc {\end{color}}
\def \brr {\begin{color}{olive}} 
\def \err {\end{color}}

\author{
Franco Barbanera\thanks{ Partially supported by 
Project “National Center for HPC, Big Data e Quantum Computing”,  Programma M4C2, Investimento 1.3.; and by Project
PIA.CE.RI (PIAno di inCEntivi per la RIcerca di Ateneo) UniCT 2024/2026.}
\institute{Dipartimento di Matematica e Informatica\\
University of Catania (Italy)}
\email{franco.barbanera@unict.it}
}

\begin{document}

%\title{Multicomposition
% of Systems \\ of Communicating Finite State Machines
%}

\title{Safe Composition of CFSM Systems via Partial Gateways
%PaI System Composition via Partial Gateways
}

\def\titlerunning{Composition via Partial Gateways}
\def\authorrunning{
F.\,Barbanera % \& R.\,Hennicker
}

\maketitle

\begin{abstract}
The {\em Participants-as-Interfaces} (PaI) methodology for system composition proposes that system participants can be regarded as interfaces. For each system in a given collection, one participant is designated to serve as its interface. When the systems are composed, these interface participants are replaced with gateways that communicate with one another by forwarding messages.
We generalise the approach to 
\emph{partial gateways}, where 
gateways can forward only a chosen
set of messages. 
As for the standard PaI approach, we exploit such extended version for
systems of communicating finite state machines (CFSMs).
This extension is fully detailed for the binary case, and can be scaled up to the multicomposition case.
We prove that many relevant 
communication properties (deadlock-freeness, reception-error-freeness, etc.) are preserved by {\em PaI composition via partial gateways\/} in case also the
connection policy (i.e. the system representing the way we wish gateways interact with each other) enjoys the same properties.
Such a proof turns out to be just a corollary of a property preservation result for a restricted
and partial composition method, dubbed \emph{fusion-composition}.
Fusion-composition hence turns out to be at the heart of the PaI composition approach.
\end{abstract}

\section{Introduction}
\label{sec:Intro}

PaI is an approach to the composition of concurrent/distributed systems introduced in \cite{BdLH18,BdLH19} and further investigated in papers among which \cite{BH24,BH26,BDLT21,BDGY23,BBD25,BLT23}. 
It is specifically designed for systems that communicate via message passing, in which the communication behaviours of participants can also be interpreted as interfaces.
Our intended meaning of ``interface'' (a term actually used in the literature with  several different connotations) is, informally, a description of the behaviour of an outer system. 
In the drawing (\ref{eq:twointerfaces}) below we sketch two systems  $S_1$ and $S_2$.
For the sake of simplicity and to focus only on the most relevant issues, we abstract the participants' behaviours from everything but the possibility of sending/receiving messages in such a drawing and throughout the present introduction. 
In particular, we abstract away from dynamic issues such as the logical order of the exchanged messages, the representation of which depends on the chosen formalism.
System $S_1$ has a participant, $\hh_1$, which can receive and send messages,
 $\msg[sbs]$ and $\msg[inf]$, respectively, from and to other participants of $S_1$, 
say $\ttr$ and $\ttr'$. 
Meanwhile, in $S_2$, $\hh_2$ can send and receive messages, $\msg[sbs]$ and $\msg[inf]$, respectively, to and from another participant of $S_2$, say $\tts$. 
For simplicity, only participants $\hh_1$ and $\hh_2$ are shown inside the dashed boxes representing systems  $S_1$ and $S_2$, respectively.
% \begin{figure}[h]
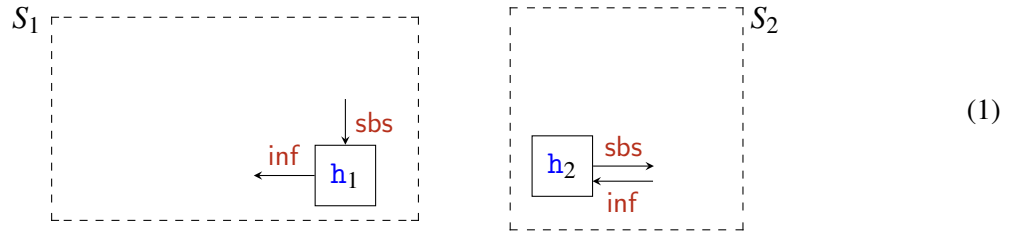
\begin{equation}
\label{eq:twointerfaces}
\raisebox{12mm}{\text{\large $S_1$}\,\,}
    \dbox{
\hspace{14mm} \begin{tikzpicture}[node distance=1.5cm,scale=1]
        \node (square-h) [draw,minimum width=0.8cm,minimum height=0.8cm] {\large $\hh_1$};
        \node [state] (h-a) [above of = square-h, draw=none] {};
        \node [state] (h-c) [left of = square-h, draw=none, xshift=-2mm] {};
        \draw [-stealth] (h-a) --  node[right] {$\msg[sbs]$} (square-h);
        \draw [stealth-] (h-c) --  node[above] {$\msg[inf]$} (square-h);
 \end{tikzpicture}
            }
\hspace{12mm}
     \dbox{
 \begin{tikzpicture}[node distance=1.5cm,scale=1]
        \node (square-k) [draw,minimum width=0.8cm,minimum height=0.8cm] {\large $\hh_2$};
        \node [state] (k-b) [above of = square-k, draw=none] {};
        \node [state] (k-a) [right of = square-k, draw=none, xshift=2mm] {};
        %\draw [stealth-] (k-b) --  node[right] {$\msg[inf]$} (square-k);
        \draw [-stealth] (square-k) --  node[above] {$\msg[sbs]$} (k-a);
        \draw  [stealth-] (0.4,-0.2)   --  node [below] {$\msg[inf]$} (1.2,-0.2);
 \end{tikzpicture}
             }
 \raisebox{12mm}{\text{\large $\,\,S_2$}}
\end{equation}
% \vspace{-2mm}
% \caption{\label{fig:twointerfaces} Two interfaces participants belonging to, respectively, systems $S_1$ and $S_2$.}
% }
%\end{figure}

\noindent
$S_1$ may represent a domotic system in which component $\hh_1$ controls the diffusion of a given substance upon receiving the ``set-by-sensor'' message $\msg[sbs]$ from either $\ttr$ or  $\ttr'$ (the sensors). 
Through the message $\msg[inf]$, $\hh_1$ can also provide the other participants with some information about the effects of the diffusion.
$S_2$, instead, can be seen as an aroma diffusion system where, following a request from a sensor 
$\hh_2$ participant $\tts$ coordinates the release of an aromatic substance. Participant $\hh_2$ may likewise receive information about the resulting effects.

Now, rather than interpreting $\hh_1$ as the behaviour of a participant of $S_1$,
 we could interpret it as the behaviour of an outer system interacting with $S_1$ by sending the message $\msg[sbs]$ and receiving $\msg[inf]$. 
  In other words, it can be viewed as an {\em interface}, in the previously hinted at sense.
%By treating $\hh_1$ as an interface, it can be interpreted as representing an external  
Analogously, $\hh_2$ can be viewed as an external system interacting with $S_2$ by receiving 
$\msg[sbs]$ messages and sending $\msg[inf]$.
The idea underlying {\em PaI composition} is that, once two participants are identified as interfaces according to our current needs, they are replaced by forwarders, referred to as ``gateways''. 
In our simple example, the resulting system would look like (\ref{fig:bincomp1}) below, where dotted lines indicate message forwarding.
The gateway that now replaces $\hh_1$ (while retaining the same name) forwards
any $\msg[sbs]$ messages received from the other participants of $S_1$,
to the gateway that has replaced $\hh_2$ (which also retains the same name). Upon receiving such a message, the gateway $\hh_2$ forwards it to $\tts$.
The process is symmetrical for the $\msg[inf]$ messages received by $\hh_2$. 
%\begin{figure}[h]
%    \centering{\small
%    $
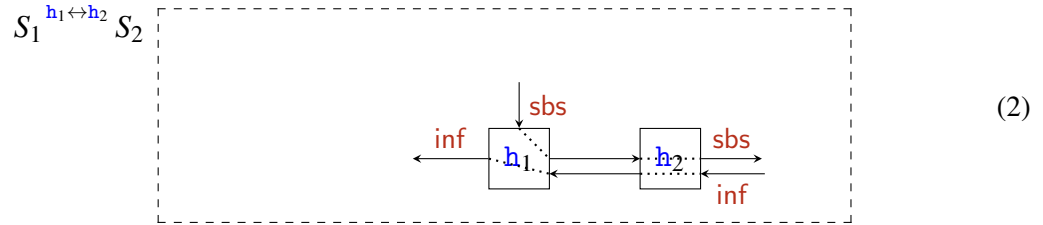
\begin{equation}
\label{fig:bincomp1}
\begin{array}{l}
\text{\large $S_1$}\!\stackrel{\hh_1{\leftrightarrow}\hh_2}{ \phantom{\mathtt{comp}}}\text{\large $S_2$}\\[22mm]
\end{array}
 \dbox{ \hspace{28mm}
 \begin{tikzpicture}[node distance=1.5cm,scale=1]
        \node (square-h) [draw,minimum width=0.8cm,minimum height=0.8cm] {\large $\hh_1$};
        \draw [-stealth] (h-a) --  node [right] {$\msg[sbs]$} (square-h);
        \node (square-k) [draw,minimum width=0.8cm,minimum height=0.8cm, right of = square-h, xshift=5mm] {\large $\hh_2$};
        \node [state] (k-a) [right of = square-k, draw=none,xshift=2mm] {};
         \node[draw=none,fill=none] (phantom) [above = 12mm  of square-h]{};
         \node[draw=none,fill=none] (phantom2) [left = 10mm  of square-h]{};
        \draw [-stealth] (square-k) --  node [above]{$\msg[sbs]$} (k-a);
        \draw [stealth-] (phantom2) --  node [above]{$\msg[inf]$} (square-h);
        \draw (0,0.4)[dotted,thick]  --  (0.4,0); % carrying sbs inside h 
        \draw (-0.4,0)[dotted,thick]  --  (0.4,-0.2); % carrying par inside h 
        \draw [-stealth] (0.4,0)  --  (1.6,0); % carrying sbs from h1 to h2
        \draw [stealth-] (0.4,-0.2)  --  (1.6,-0.2); % carrying inf from h1 to h2
        \draw (1.6,-0.2) [dotted,thick]  --  (2.4,-0.2); % carrying inf inside h2
        \draw  [stealth-] (2.4,-0.2)   --  node [below] {$\msg[inf]$} (3.25,-0.2); % carrying inf outside h2
        \draw (1.6,0)[dotted,thick]  --  (2.4,0); % carrying sbs inside h2
 \end{tikzpicture}
       }
\end{equation}
% \hspace{4mm}
% $
% \caption{\label{fig:bincomp} The PaI idea for binary composition via gateways}
% }
%\end{figure}

% \begin{figure}[h]
%    \centering{\small
%    $
Note that any participant can be treated as an {\em interface}.
This makes the approach suitable for the composition of closed systems.
However, some conditions are required for the chosen interfaces in order to ensure that PaI composition is safe, i.e. property preserving. 
These conditions clearly depend on the formalism used to describe participant behaviours.
Several results have been established using the automata-based formalism of Communicating Finite State Machines (CFSM)~\cite{BZ83}.
\Commented{It has been shown in \cite{BdLH19} that a number of communication properties (e.g. deadlock freedom, orphan-message freedom) are preserved once interfaces are {\em compatible}. This roughly means that, when we abstract away from the local identities of senders and receivers, their message traces are dual, in the sense that each is obtained from the other by swapping input and output symbols. 
It is also possible to focus on the preservation of individual properties.} %END Commented
It has been shown in \cite{BH24,BH26} that a given communication property $P$ (e.g. deadlock freedom and orphan-message freedom among others) is preserved by composition if the connection policy used for the composition also enjoys $P$. 
A {\em connection policy} is intended as a description of of how messages should be forwarded by 
the gateways in a composition and it can be formalised in terms of another system.
In our present simple example a natural connection policy would consist of a two-CFSM system,
which is abstractly represented as follows.
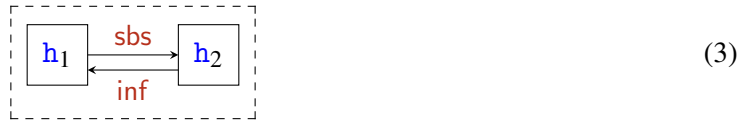
\begin{equation}
\label{eq:introcm}
 \dbox{
 \begin{tikzpicture}[node distance=1.5cm,scale=1]
        \node (square-h) [draw,minimum width=0.8cm,minimum height=0.8cm] {\large $\hh_1$};
        \node (square-k) [draw,minimum width=0.8cm,minimum height=0.8cm, right of = square-h, xshift=5mm] {\large $\hh_2$};
        \draw [-stealth] (0.4,0)  -- node [above]{$\msg[sbs]$} (1.6,0); % carrying sbs from h1 to h2
        \draw [stealth-] (0.4,-0.2)  -- node [below]{$\msg[inf]$} (1.6,-0.2); % carrying inf from h2 to h12
 \end{tikzpicture}
       }
\end{equation}
In PaI binary composition, there is always just one connection policy since the participant to which a message can be forwarded is uniquely determined.
Therefore, the gateways that we substitute for interfaces are also uniquely determined.
The results in \cite{BH24,BH26} consider the composition of more than two systems, where the forwarding policies among several interface participants can be chosen with greater freedom.
It is worth noting that the notion of a connection policy and the aforementioned property preservation results do not depend on any ``duality'' relationship between the chosen interfaces for the composition. This means that we could choose interfaces and connection policies that result in gateways, some of which cannot actually exchange messages, although they are forwarded or expected by one of the gateways. 
Although this unseemly conduct does not necessarily hinder property preservation results, it can be prevented by imposing connection policies that comply with {\em connection models}, as defined in references \cite{BH24,BH26}. (We do not address this notion in the present paper). Alternatively, one could argue that certain messages, although present in an interface, should not be forwarded by the gateways at all. 
% so making the property preservation results of \cite {BH24,BH26} depend solely
%on the properties of the systems we compose and of the connection policy.
%In our boxes-and-arrows abstract representation of a connection policy, an  $\msg[a]$ labelled arrow from, say, $\hh$ to $\kk$  simply recalls that $(a)$ in case the gateway substituted for $\hh$ receives $\msg[a]$, this message is immediately be forwarded to the gateway for $\kk$,
%regardless of whether $\kk$ is able to receive the message;
%$(b)$  the gateway substituted for $\kk$ can send a message $\msg[a]$ only after it has been
%first forwarded from the gateway for $\hh$, regardless of whether the latter gateway is actually able to send the message. In our simple example

%according to are not uniquely determined in the presence of several systems (and therefore several interfaces).  

\medskip

In the present paper, partly inspired by \cite{BDL22}, we generalise the PaI approach by considering participants as interfaces only in part.
Let us consider the following example. (From now on, unless necessary, we will avoid representing systems by dashed boxes and will only represent the participants identified as interfaces.
Their belonging to different systems is indicated by dashed vertical lines.)\\[-14mm]
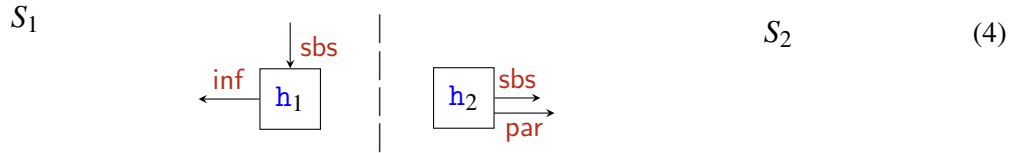
\begin{equation}
\label{eq:sbspar}
\raisebox{2mm}{\text{\large $S_1$}\,\,}
%    \dbox{
\hspace{10mm} \begin{tikzpicture}[node distance=1.5cm,scale=1]
        \node (square-h) [draw,minimum width=0.8cm,minimum height=0.8cm] {\large $\hh_1$};
        \node [state] (h-a) [above of = square-h, draw=none] {};
        \node [state] (h-c) [left of = square-h, draw=none, xshift=-2mm] {};
        \draw [-stealth] (h-a) --  node[right] {$\msg[sbs]$} (square-h);
        \draw [stealth-] (h-c) --  node[above] {$\msg[inf]$} (square-h);
 \end{tikzpicture}
%            }
\hspace{2mm}
 \begin{array}{c}
 \\[8mm]
| \\
| \\
|\\
|\\
\end{array}
\hspace{4mm}
%     \dbox{
 \raisebox{-1mm}{\begin{tikzpicture}[node distance=1.5cm,scale=1]
        \node (square-k) [draw,minimum width=0.8cm,minimum height=0.8cm] {\large $\hh_2$};
        \node [state] (k-b) [above of = square-k, draw=none] {};
        \node [state] (k-a) [right of = square-k, draw=none] {};
        %\draw [stealth-] (k-b) --  node[right] {$\msg[inf]$} (square-k);
        \draw [-stealth] (square-k) --  node[above] {$\msg[sbs]$} (k-a);
        \draw  [-stealth] (0.4,-0.2)   --  node [below] {$\msg[par]$} (1.2,-0.2);
 \end{tikzpicture} } \hspace{18mm}
 %            }
 \raisebox{0mm}{\text{\large $\,\,S_2$}}
 \end{equation}
 \\[-4mm]
 System $S_1$ remains as before, whereas in system $S_2$ the participant $\hh_2$ can, 
 in addition to sending the message $\msg[sbs]$, also provide parameters for properly adjusting the diffusion of the aromatic substance via the message $\msg[par]$.
% In such a case, the results of \cite{BdLH19} do not apply.
%  In fact, we can turn both interfaces into gateways, but they are definitely not compatible.
%The results in \cite{BH24} still apply, since they do not rely on compatibility. 
 In this example, 
% orphan-message freedom property is likely to be preserved, but not most of the others.
substituting $\hh_1$ and $\hh_2$ with uniformly built gateways, as required by the PaI approach, would result rather unreasonable.
 The gateway for $\hh_1$ would never be able to emit $\msg[inf]$ since, as a forwarder,
 it would never be able to receive it from the gateway for $\hh_2$. This might disrupt most of the communication properties that the two systems satisfy.
 
A more reasonable approach would be to consider only the actions concerning the message 
$\msg[sbs]$ as 
 ``interface actions'', leaving the interpretation of the others as pertaining to their respective 
 participants. 
Following this approach, we would replace the interface participants with {\em partial gateways} that only forward messages relating to interface actions, as informally shown below.\\[-4mm]
\vspace{-8mm}
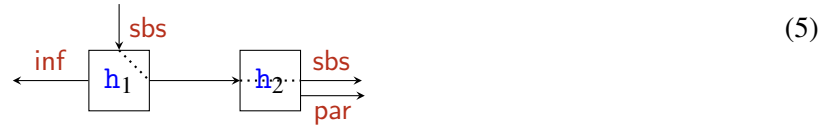
\begin{equation}
\label{eq:pgcompsbspar}
\begin{array}{l}
\end{array}
 \begin{tikzpicture}[node distance=1.5cm,scale=1]
        \node (square-h) [draw,minimum width=0.8cm,minimum height=0.8cm] {\large $\hh_1$};
        \draw [-stealth] (h-a) --  node [right] {$\msg[sbs]$} (square-h);
        \node (square-k) [draw,minimum width=0.8cm,minimum height=0.8cm, right of = square-h, xshift=5mm] {\large $\hh_2$};
        \node [state] (k-a) [right of = square-k, draw=none,xshift=2mm] {};
         \node[draw=none,fill=none] (phantom) [above = 12mm  of square-h]{};
         \node[draw=none,fill=none] (phantom2) [left = 10mm  of square-h]{};
        \draw [-stealth] (square-k) --  node [above]{$\msg[sbs]$} (k-a);
        \draw [stealth-] (phantom2) --  node [above]{$\msg[inf]$} (square-h);
        \draw (0,0.4)[dotted,thick]  --  (0.4,0); % carrying sbs inside h 
        \draw [-stealth] (0.4,0)  --  (1.6,0); % carrying sbs from h1 to h2
        \draw  [-stealth] (2.4,-0.2)   --  node [below] {$\msg[par]$} (3.25,-0.2); % carrying par outside h2
        \draw (1.6,0)[dotted,thick]  --  (2.4,0); % carrying sbs inside h2
 \end{tikzpicture}
 \end{equation}
 \\[-4mm]
In this paper, we extend the PaI approach to include the composition of systems via partial gateways and identify the conditions under which this is applicable. We achieve this in the context of the CFSM formalism. The safety of the approach is proven by resorting to a restricted and simplified form of composition that we refer to as {\em partial-fusion composition}. \\

\noindent
{\em Overview.} In \cref{sec:infdesfus} we present an informal description of PaI composition via partial gateways and show how this can be derived from the restricted form of composition
we refer to as partial-fusion composition.
The CFSM formalism is described in \cref{sect:cfsm}, where some relevant communication properties are also defined. 
A complete formalisation of the PaI composition via partial gateways in the
CFSM setting is provided in \cref{sec:cpg}.  
Before their formalisation, the underlying notions will be introduced and discussed
by means of a running example. 
The property preservation result for  PaI composition via partial gateways is stated at the end of that section.
\cref{sec:pfc} is  devoted to the formalisation of partial-fusion composition and how 
to obtain  PaI composition from it. 
The latter's property preservation result can therefore be obtained
by applying a similar result for the former. A concluding section briefly summarises the paper and provides some guidelines for future work.

 \section{Informal description of PaI composition with partial gateways.}
 \label{sec:infdesfus}
While maintaining an abstract level of description, we now elaborate on the idea of partial gateways mentioned in the introduction, using a further example.
By considering the possibility of forwarding only some messages in a composition,
partial gateways are generally not uniquely determined by the interface participants chosen for a composition, even in the binary case.
Consider, for example, the following interface participants for systems $S_1$ and $S_2$.
\\[-2mm]
\vspace{-8mm}
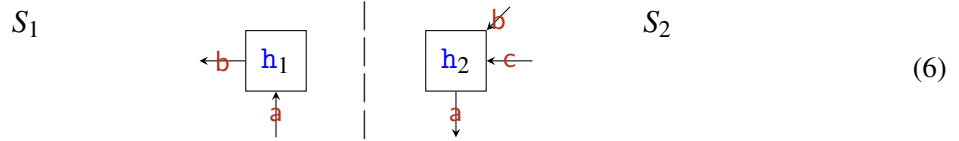
\begin{equation}
\label{eq:twointf}
\raisebox{6mm}{\text{\large $S_1$}\,\,}
%    \dbox{
\hspace{10mm}  
\begin{tikzpicture}[node distance=1.5cm,scale=1]
        \node (square-v) [draw,minimum width=0.8cm,minimum height=0.8cm] {\large $\hh_1$};
        \node [state] (v-b) [left of = square-v, draw=none] {};
        \node [state] (v-a) [below of = square-v, draw=none] {};
        \node [state] (w-b) [above right of = square-v, draw=none] {};
        \draw [-stealth] (v-a) --  node {$\msg[a]$} (square-v);
        \draw [-stealth] (square-v) --  node {$\msg[b]$} (v-b);
 \end{tikzpicture}
%            }
\hspace{-6mm}
 \begin{array}{c}
| \\
| \\
|\\
|\\
\end{array}
\hspace{5mm}
%     \dbox{
  \begin{tikzpicture}[node distance=1.5cm,scale=1]
        \node (square-w) [draw,minimum width=0.8cm,minimum height=0.8cm] {\large $\hh_2$};
        \node [state] (w-c) [right of = square-w, draw=none] {};
        \node [state] (w-a) [below of = square-w, draw=none] {};
        \node [state] (w-b) [above right of = square-w, draw=none] {};
        \draw [-stealth] (w-c) --  node {$\msg[c]$} (square-w);
        \draw [-stealth] (square-w) --  node {$\msg[a]$} (w-a);
        \draw [stealth-] (square-w) --  node {$\msg[b]$} (w-b);
 \end{tikzpicture} 
\hspace{4mm}            
%}
 \raisebox{6mm}{\text{\large $\,\,S_2$}}
 \end{equation}
 \\[-12mm]
 Message $\msg[c]$ cannot reasonably be considered one to be forwarded (although the usual binary
 PaI composition would make it). 
We have, however, definitely a choice concerning messages $\msg[a]$ and $\msg[b]$. 
According to our intentions, we can decide whether they have to be exchanged between the partial gateways (i.e. whether they belong to the interface parts of $\hh_1$ and $\hh_2$) or not (i.e. whether the partial gateways will deal with them as $\hh_1$ and $\hh_2$ did prior to the composition). 
This type of decision can be represented graphically by using thicker lines for the actions intended for the interface parts of $\hh_1$ and $\hh_2$. The following reasonable choices are possible.\\[-8mm]
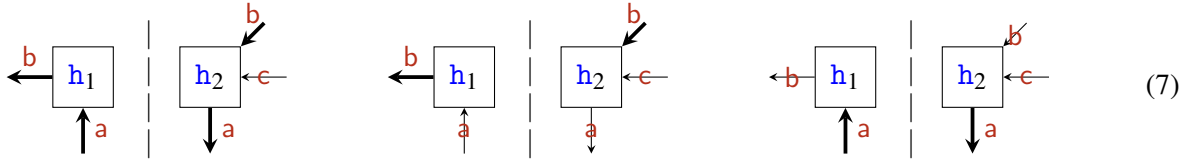
\begin{equation}
\label{fig:somecms}
\hspace{-8mm}
\begin{tikzpicture}[node distance=1.5cm,scale=1]
        \node (square-v) [draw,minimum width=0.8cm,minimum height=0.8cm] {\large $\hh_1$};
        \node [state] (v-b) [left of = square-v, draw=none] {};
        \node [state] (w-b) [above right of = square-w, draw=none] {};
        \node [state] (v-a) [below of = square-v, draw=none] {};
        \draw [-stealth,line width=0.5mm] (v-a) --  node [right] {$\msg[a]$} (square-v); 
        \draw [-stealth,line width=0.5mm] (square-v) --  node [above] {$\msg[b]$} (v-b);
 \end{tikzpicture}
\hspace{-9mm}
 \begin{array}{c}
 \\[-4mm]
| \\
| \\
|\\
|\\
\end{array}
\hspace{1mm}
\begin{tikzpicture}[node distance=1.5cm,scale=1]
        \node (square-w) [draw,minimum width=0.8cm,minimum height=0.8cm] {\large $\hh_2$};
        \node [state] (w-c) [right of = square-w, draw=none] {};
        \node [state] (w-a) [below of = square-w, draw=none] {};
        \node [state] (w-b) [above right of = square-w, draw=none] {};
        \draw [-stealth] (w-c) --  node {$\msg[c]$} (square-w);
        \draw [-stealth,line width=0.5mm] (square-w) --  node [right] {$\msg[a]$} (w-a);
        \draw [stealth-,line width=0.5mm] (square-w) --  node [above]{$\msg[b]$} (w-b);
\end{tikzpicture}
\hspace{-6mm}
\begin{tikzpicture}[node distance=1.5cm,scale=1]
        \node (square-v) [draw,minimum width=0.8cm,minimum height=0.8cm] {\large $\hh_1$};
        \node [state] (v-b) [left of = square-v, draw=none] {};
        \node [state] (w-b) [above right of = square-w, draw=none] {};
        \node [state] (v-a) [below of = square-v, draw=none] {};
        \draw [-stealth] (v-a) --  node  {$\msg[a]$} (square-v); 
        \draw [-stealth,line width=0.5mm] (square-v) --  node [above]{$\msg[b]$} (v-b);
 \end{tikzpicture}
\hspace{-9mm}
 \begin{array}{c}
 \\[-4mm]
| \\
| \\
|\\
|\\
\end{array}
\hspace{1mm}
\begin{tikzpicture}[node distance=1.5cm,scale=1]
        \node (square-w) [draw,minimum width=0.8cm,minimum height=0.8cm] {\large $\hh_2$};
        \node [state] (w-c) [right of = square-w, draw=none] {};
        \node [state] (w-a) [below of = square-w, draw=none] {};
        \node [state] (w-b) [above right of = square-w, draw=none] {};
        \draw [-stealth] (w-c) --  node {$\msg[c]$} (square-w);
        \draw [-stealth] (square-w) --  node {$\msg[a]$} (w-a);
        \draw [stealth-,line width=0.5mm] (square-w) --  node [above]{$\msg[b]$} (w-b);
\end{tikzpicture}
\hspace{-6mm}
\begin{tikzpicture}[node distance=1.5cm,scale=1]
        \node (square-v) [draw,minimum width=0.8cm,minimum height=0.8cm] {\large $\hh_1$};
        \node [state] (v-b) [left of = square-v, draw=none] {};
        \node [state] (w-b) [above right of = square-w, draw=none] {};
        \node [state] (v-a) [below of = square-v, draw=none] {};
        \draw [-stealth,line width=0.5mm] (v-a) --  node [right] {$\msg[a]$} (square-v); 
        \draw [-stealth] (square-v) --  node {$\msg[b]$} (v-b);
 \end{tikzpicture}
\hspace{-9mm}
 \begin{array}{c}
 \\[-4mm]
| \\
| \\
|\\
|\\
\end{array}
\hspace{1mm}
\begin{tikzpicture}[node distance=1.5cm,scale=1]
        \node (square-w) [draw,minimum width=0.8cm,minimum height=0.8cm] {\large $\hh_2$};
        \node [state] (w-c) [right of = square-w, draw=none] {};
        \node [state] (w-a) [below of = square-w, draw=none] {};
        \node [state] (w-b) [above right of = square-w, draw=none] {};
        \draw [-stealth] (w-c) --  node  {$\msg[c]$} (square-w);
        \draw [-stealth,line width=0.5mm] (square-w) --  node [right] {$\msg[a]$} (w-a);
        \draw [stealth-] (square-w) --  node {$\msg[b]$} (w-b);
\end{tikzpicture}
 \end{equation}
 \\[-10mm]
It is also possible to make choices that would hardly result in safe compositions.\\[-8mm]
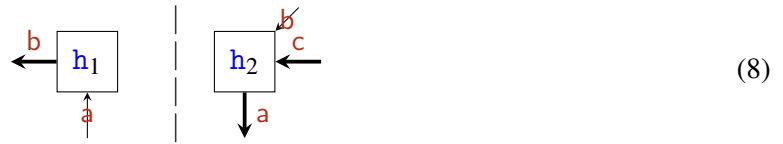
\begin{equation}
\begin{tikzpicture}[node distance=1.5cm,scale=1]
        \node (square-v) [draw,minimum width=0.8cm,minimum height=0.8cm] {\large $\hh_1$};
        \node [state] (v-b) [left of = square-v, draw=none] {};
        \node [state] (w-b) [above right of = square-w, draw=none] {};
        \node [state] (v-a) [below of = square-v, draw=none] {};
        \draw [-stealth] (v-a) --  node  {$\msg[a]$} (square-v); 
        \draw [-stealth,line width=0.5mm] (square-v) --  node [above] {$\msg[b]$} (v-b);
 \end{tikzpicture}
\hspace{-6mm}
 \begin{array}{c}
 \\[-4mm]
| \\
| \\
|\\
|\\
\end{array}
\hspace{2mm}
\begin{tikzpicture}[node distance=1.5cm,scale=1]
        \node (square-w) [draw,minimum width=0.8cm,minimum height=0.8cm] {\large $\hh_2$};
        \node [state] (w-c) [right of = square-w, draw=none] {};
        \node [state] (w-a) [below of = square-w, draw=none] {};
        \node [state] (w-b) [above right of = square-w, draw=none] {};
        \draw [-stealth,line width=0.5mm] (w-c) --  node [above] {$\msg[c]$} (square-w);
        \draw [-stealth,line width=0.5mm] (square-w) --  node [right] {$\msg[a]$} (w-a);
        \draw [stealth-] (square-w) --  node {$\msg[b]$} (w-b);
\end{tikzpicture}
\vspace{-8mm}
 \end{equation}
 \\[-6mm]
The notion of ``participant with interface actions'', hinted at above, will be formalised in terms of CFSMs, as {\em CFSM with interface transitions} (see \cref{def:cfsmintftrans}). 

Our goal is to generalise the results in \cite{BH24, BH26}, where the preservation of a property by composition is implied by the property also being satisfied by the connection policy used. 
In our setting, a connection policy determines how messages labelling interface actions are exchanged among partial gateways in the intended composition.
The interface actions represented in (\ref{fig:somecms}) correspond to the three connection
policies abstractly described as follows.\\[-2mm]
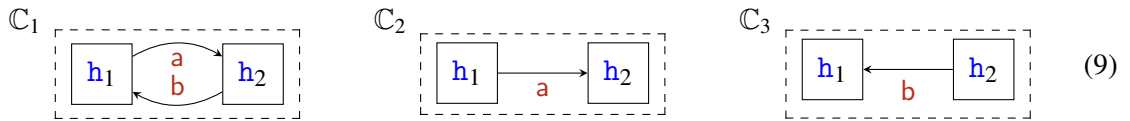
\begin{equation}
\label{fig:threecms}
\begin{array}{l}
\text{$\cs_1$}\\[12mm]
\end{array}
 \dbox{
 \begin{tikzpicture}[node distance=1.5cm,scale=1]
        \node (square-h) [draw,minimum width=0.8cm,minimum height=0.8cm] {\large $\hh_1$};
        \node (square-k) [draw,minimum width=0.8cm,minimum height=0.8cm, right of = square-h, xshift=5mm] {\large $\hh_2$};
      \path
      (square-h) edge[-stealth,bend left] node[below] {$\msg[a]$} (square-k)
      (square-k) edge[-stealth,bend left] node[above] {$\msg[b]$} (square-h)
      ;
 \end{tikzpicture}
 }
 \hspace{8mm}
\begin{array}{l}
\text{$\cs_2$}\\[12mm]
\end{array}
 \dbox{
 \begin{tikzpicture}[node distance=1.5cm,scale=1]
        \node (square-h) [draw,minimum width=0.8cm,minimum height=0.8cm] {\large $\hh_1$};
        \node (square-k) [draw,minimum width=0.8cm,minimum height=0.8cm, right of = square-h, xshift=5mm] {\large $\hh_2$};
      \path
      (square-h) edge[-stealth] node[below] {$\msg[a]$} (square-k)
      ;
 \end{tikzpicture} 
        }
 \hspace{8mm}
\begin{array}{l}
\text{$\cs_3$}\\[12mm]
\end{array}
 \dbox{
 \begin{tikzpicture}[node distance=1.5cm,scale=1]
        \node (square-h) [draw,minimum width=0.8cm,minimum height=0.8cm] {\large $\hh_1$};
        \node (square-k) [draw,minimum width=0.8cm,minimum height=0.8cm, right of = square-h, xshift=5mm] {\large $\hh_2$};
      \path
      (square-k) edge[-stealth] node[below] {$\msg[b]$} (square-h)
      ;
 \end{tikzpicture} 
        }
 \end{equation}
%We remark that, for the sake of simplicity and lack of space, we are dealing here only with binary PaI composition.
%Our results concerns however the multicomposition case, where even the diagrammatic
%representation of a connection policy is not uniquely detemined by the representation of
%the actions and interface actions of the participants considered for the composition.
From the connection policy represented in (\ref{fig:threecms}) we get different compositions
by means of the following partial gateways.
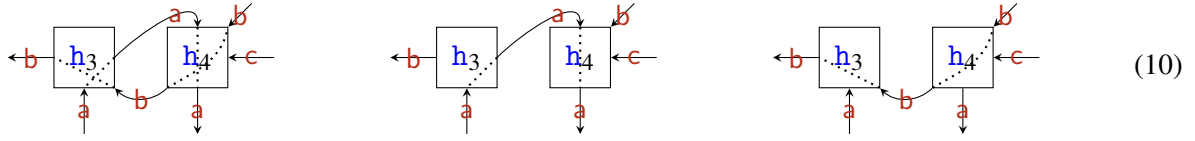
\vspace{-10mm}
\begin{equation}
\hspace{-8mm}
\begin{tikzpicture}[node distance=1.5cm,scale=1]
        \node (square-v)  [draw,minimum width=0.8cm,minimum height=0.8cm] {\large $\hh_3$};
        \node [state] (v-b) [left of = square-v, draw=none] {};
        \node [state] (v-a) [below of = square-v, draw=none] {};
        \draw [-stealth] (v-a) --  node {$\msg[a]$} (square-v);
        \draw [-stealth] (square-v) --  node {$\msg[b]$} (v-b);
        \node (square-w)  [draw,minimum width=0.8cm,minimum height=0.8cm, right of = square-v] {\large $\hh_4$};
        \node [state] (w-c) [right of = square-w, draw=none] {};
        \node [state] (w-a) [below of = square-w, draw=none] {};
        \node [state] (w-b) [above right of = square-w, draw=none] {};
        \draw [-stealth] (w-c) --  node {$\msg[c]$} (square-w);
        \draw [-stealth] (square-w) --  node {$\msg[a]$} (w-a);
        \draw [stealth-] (square-w) --  node {$\msg[b]$} (w-b);
        \draw [-stealth] (square-w) to[out=-135,in=-45]  node {$\msg[b]$} (square-v);
        \draw (0.4,0)[dotted,thick]  --  (0,-0.4); % carrying a inside v
        \draw (0.4,-0.4)[dotted,thick]  --  (-0.4,0); % carrying b inside v
        \draw (1.5,0.4)[dotted,thick]  --  (1.5,-0.4); % carrying a inside w
        \draw (1.1,-0.4)[dotted,thick]  to[out=30,in=260]  (1.9,0.4); % carrying b inside w
        \draw [-stealth]  (0.4,0) to[out=45,in=90]  node [pos=0.6] {$\msg[a]$} (1.5,0.4 ); % carryng a from v to w  
 \end{tikzpicture}
 \hspace{-4mm}
 \begin{tikzpicture}[node distance=1.5cm,scale=1]
        \node (square-v)  [draw,minimum width=0.8cm,minimum height=0.8cm] {\large $\hh_3$};
        \node [state] (v-b) [left of = square-v, draw=none] {};
        \node [state] (v-a) [below of = square-v, draw=none] {};
        \draw [-stealth] (v-a) --  node {$\msg[a]$} (square-v);
        \draw [-stealth] (square-v) --  node {$\msg[b]$} (v-b);
        \node (square-w)  [draw,minimum width=0.8cm,minimum height=0.8cm, right of = square-v] {\large $\hh_4$};
        \node [state] (w-c) [right of = square-w, draw=none] {};
        \node [state] (w-a) [below of = square-w, draw=none] {};
        \node [state] (w-b) [above right of = square-w, draw=none] {};
        \draw [-stealth] (w-c) --  node {$\msg[c]$} (square-w);
        \draw [-stealth] (square-w) --  node {$\msg[a]$} (w-a);
        \draw [stealth-] (square-w) --  node {$\msg[b]$} (w-b);
        %
        %\draw [-stealth] (square-w) to[out=-135,in=-45]  node {$\msg[b]$} (square-v);
        %
        \draw (0.4,0)[dotted,thick]  --  (0,-0.4); % carrying a inside v
        %\draw (0.4,-0.4)[dotted,thick]  --  (-0.4,0); % carrying b inside v
        %
        \draw (1.5,0.4)[dotted,thick]  --  (1.5,-0.4); % carrying a inside w
        \draw [-stealth]  (0.4,0) to[out=45,in=90]  node [pos=0.6] {$\msg[a]$} (1.5,0.4 ); % carryng a from v to w  
 \end{tikzpicture}
  \hspace{-4mm}
 \begin{tikzpicture}[node distance=1.5cm,scale=1]
        \node (square-v)  [draw,minimum width=0.8cm,minimum height=0.8cm] {\large $\hh_3$};
        \node [state] (v-b) [left of = square-v, draw=none] {};
        \node [state] (v-a) [below of = square-v, draw=none] {};
        \draw [-stealth] (v-a) --  node {$\msg[a]$} (square-v);
        \draw [-stealth] (square-v) --  node {$\msg[b]$} (v-b);
        \node (square-w)  [draw,minimum width=0.8cm,minimum height=0.8cm, right of = square-v] {\large $\hh_4$};
        \node [state] (w-c) [right of = square-w, draw=none] {};
        \node [state] (w-a) [below of = square-w, draw=none] {};
        \node [state] (w-b) [above right of = square-w, draw=none] {};
        \draw [-stealth] (w-c) --  node {$\msg[c]$} (square-w);
        \draw [-stealth] (square-w) --  node {$\msg[a]$} (w-a);
        \draw [stealth-] (square-w) --  node {$\msg[b]$} (w-b);
        \draw [-stealth] (square-w) to[out=-135,in=-45]  node {$\msg[b]$} (square-v);
        %
        %\draw (0.4,0)[dotted,thick]  --  (0,-0.4); % carrying a inside v
        \draw (0.4,-0.4)[dotted,thick]  --  (-0.4,0); % carrying b inside v
        %
        %\draw (1.5,0.4)[dotted,thick]  --  (1.5,-0.4); % carrying a inside w
        \draw (1.1,-0.4)[dotted,thick]  to[out=30,in=260]  (1.9,0.4); % carrying b inside w
        %
        %
       % \draw [-stealth]  (0.4,0) to[out=45,in=90]  node [pos=0.6] {$\msg[a]$} (1.5,0.4 ); % carryng a from v to w  
 \end{tikzpicture}
\end{equation}
\\[-12mm]
In general, not all choices of interface actions make sense. 
In some cases, it is clearly unreasonable to select an action as an interface action. For example, consider the message $\msg[c]$ in (\ref{eq:twointf}).
However, there are other cases where the selection of an interface action is not sound for subtler reasons.
Let us consider the participant $\hh_2$ in (\ref{fig:abnoint}) below.
Assume also that the CFSM specifying the dynamics of the behaviour of $\hh_2$
contains an actual branching between sending either $\msg[a]$ or  $\msg[b]$.
In this case, the choice of interface actions described in (\ref{fig:abint}) would not make sense.\\[-8mm]
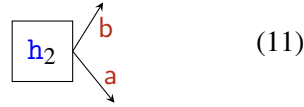
\begin{minipage}{0.45\textwidth}
\begin{equation}
\label{fig:abnoint}
 \begin{tikzpicture}[node distance=1.5cm,scale=1]
        \node (square-k) [draw,minimum width=0.8cm,minimum height=0.8cm] {\large $\hh_2$};
        \node [state] (k-b) [above right of = square-k, draw=none] {};
        \node [state] (k-a) [below right of = square-k, draw=none, xshift=2mm] {};
        %\draw [stealth-] (k-b) --  node[right] {$\msg[inf]$} (square-k);
        \draw [-stealth] (0.4,0) --  node[right] {$\msg[a]$} (k-a);
        \draw  [-stealth] (0.4,0)   --  node [right] {$\msg[b]$} (k-b);
 \end{tikzpicture}
 \hspace{-16mm}
\end{equation} 
\end{minipage}
\qquad
\begin{minipage}{0.45\textwidth}
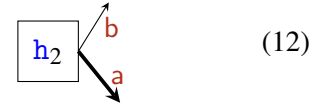
\begin{equation}
\label{fig:abint}
 \begin{tikzpicture}[node distance=1.5cm,scale=1]
        \node (square-k) [draw,minimum width=0.8cm,minimum height=0.8cm] {\large $\hh_2$};
        \node [state] (k-b) [above right of = square-k, draw=none] {};
        \node [state] (k-a) [below right of = square-k, draw=none, xshift=2mm] {};
        %\draw [stealth-] (k-b) --  node[right] {$\msg[inf]$} (square-k);
        \draw [-stealth,line width=0.5mm] (0.4,0) --  node[right] {$\msg[a]$} (k-a);
        \draw  [-stealth] (0.4,0)   --  node [right] {$\msg[b]$} (k-b);
 \end{tikzpicture}
  \hspace{-16mm}
\end{equation}
\end{minipage}
\\[-6mm]
In fact, selecting the transmission of message $\msg[a]$ as interface action implies that when the interface participant $\hh_2$ is substituted with a partial gateway in a composition, the latter should decide whether to wait for a message $\msg[a]$ to be forwarded or to autonomously send the
message $\msg[b]$.
Therefore, it is possible that the partial gateway sends message $\msg[b]$, even though message $\msg[a]$ is about to arrive.
Since CFSM is a formalism involving asynchronous communication, the message $\msg[a]$ might remain in the communication channel indefinitely once received, thereby disrupting some communication property of the composition.
In particular, orphan-message freedom.
It could be argued that the aforementioned problem actually involves the indirect presence of
{\em mixed states} (states containing both input and output outgoing transitions). 
The presence of mixed states is often problematic in CFSM systems. 
However, we shall discuss other cases of problematic interface action selection that do not involve mixed states, neither directly in the interfaces nor indirectly in the partial gateways.
In order to prevent issues like the one described above from occurring, we shall provide a syntactic condition that must be satisfied by the selection of interface actions.
\smallskip

The above sketched PaI approach to binary composition via partial gateways naturally scales up to PaI multicomposition and PaI  orchestrated multicomposition \cite{BH24,BH26}.
However, for the sake of clarity and due to space limitations, we consider only the binary case.
\smallskip
\Commented{ 
Let us consider an example from~\cite{BDGY23} having four systems $S_1$,  $S_2$, $S_3$ and $S_4$.
As shown in (\ref{eq:four-ips}) below, we have selected for each system one participant
as an interface, named respectively $\hh_1$, $\hh_2$, $\hh_3$ and $\hh_4$.
 
%\begin{wrapfigure}{r}{0.45\textwidth}
%\begin{figure}[h]
%    %\vspace{-8mm}
%     \centering{\small
%    $
\begin{equation}
\label{eq:four-ips}
    \begin{array}{@{\hspace{0mm}}c@{\hspace{-2mm}}}
    \begin{array}{c@{\hspace{-2mm}}c}
    \text{\large $S_1$}
    &
 \begin{tikzpicture}[node distance=1.5cm,scale=1]
        \node (square-h) [draw,minimum width=0.8cm,minimum height=0.8cm] {\large $\hh_1$};
        \node [state] (h-a) [above of = square-h, draw=none] {};
        \node [state] (h-c) [left of = square-h, draw=none] {};
        \draw [-stealth] (h-a) --  node {$\msg[a]$} (square-h);
        \draw [-stealth] (square-h) --  node {$\msg[c]$} (h-c);
 \end{tikzpicture}
 \end{array}
 \hspace{4mm}
\begin{array}{c}
 \\
 \\
| \\
| \\
|\\
|\\
\end{array}
 \hspace{4mm}
 \begin{array}{c@{\hspace{-2mm}}c}
\begin{tikzpicture}[node distance=1.5cm,scale=1]
        \node (square-k) [draw,minimum width=0.8cm,minimum height=0.8cm] {\large $\hh_2$};
        \node [state] (k-b) [above of = square-k, draw=none] {};
        \node [state] (k-a) [right of = square-k, draw=none] {};
        \draw [-stealth] (k-b) --  node {$\msg[b]$} (square-k);
        \draw [-stealth] (square-k) --  node {$\msg[a]$} (k-a);
 \end{tikzpicture}
 &
 \text{\large $S_2$} 
 \end{array}
 \\[12mm]
\hspace{3mm}- - - -    \hspace{8mm}- - - - -  \\[-5mm]
\begin{array}{@{\hspace{0mm}}c@{\hspace{0mm}}c}
\\[4mm]
\text{\large $S_3\hspace{-2mm}$} 
&
 \begin{tikzpicture}[node distance=1.5cm,scale=1]
        \node (square-v) [draw,minimum width=0.8cm,minimum height=0.8cm] {\large $\hh_3$};
        \node [state] (v-b) [left of = square-v, draw=none] {};
        \node [state] (v-a) [below of = square-v, draw=none] {};
        \draw [-stealth] (v-a) --  node {$\msg[a]$} (square-v);
        \draw [-stealth] (square-v) --  node {$\msg[b]$} (v-b);
 \end{tikzpicture}
 \end{array}
 \hspace{4mm}
\begin{array}{c}
 \\[-8mm]
| \\
| \\
| \\
|
\end{array}
 \hspace{4mm}
 \begin{array}{c@{\hspace{-2mm}}}
 \\[-6mm]
 \begin{tikzpicture}[node distance=1.5cm,scale=1]
        \node (square-w) [draw,minimum width=0.8cm,minimum height=0.8cm] {\large $\hh_4$};
        \node [state] (w-c) [right of = square-w, draw=none] {};
        \node [state] (w-a) [below of = square-w, draw=none] {};
        \node [state] (w-b) [above right of = square-w, draw=none] {};
        \draw [-stealth] (w-c) --  node {$\msg[c]$} (square-w);
        \draw [-stealth] (square-w) --  node {$\msg[a]$} (w-a);
        \draw [stealth-] (square-w) --  node {$\msg[b]$} (w-b);
 \end{tikzpicture}
 \hspace{-6mm}
 \begin{array}{l}
 \\[6mm]
 \text{\large\ \ $S_4$}
 \end{array} 
 \end{array}
 \\[-4mm]
 \end{array}
 \end{equation}
% $
% }
%\caption{\label{fig:four-ips}
%Four systems with their respective interface participants}
% \end{figure}
% \vspace{-5mm}
% \end{wrapfigure}

We could take into account the following interface actions.

\begin{equation}
\label{fig:four-ips}
    \begin{array}{@{\hspace{0mm}}c@{\hspace{-2mm}}}
    \begin{array}{c@{\hspace{-2mm}}c}
    \text{\large $S_1$}
    &
 \begin{tikzpicture}[node distance=1.5cm,scale=1]
        \node (square-h) [draw,minimum width=0.8cm,minimum height=0.8cm] {\large $\hh_1$};
        \node [state] (h-a) [above of = square-h, draw=none] {};
        \node [state] (h-c) [left of = square-h, draw=none] {};
        \draw [-stealth,line width=0.5mm] (h-a) --  node [right] {$\msg[a]$} (square-h);
        \draw [-stealth] (square-h) --  node {$\msg[c]$} (h-c);
 \end{tikzpicture}
 \end{array}
 \hspace{4mm}
\begin{array}{c}
 \\
 \\
| \\
| \\
|\\
|\\
\end{array}
 \hspace{4mm}
 \begin{array}{c@{\hspace{-2mm}}c}
\begin{tikzpicture}[node distance=1.5cm,scale=1]
        \node (square-k) [draw,minimum width=0.8cm,minimum height=0.8cm] {\large $\hh_2$};
        \node [state] (k-b) [above of = square-k, draw=none] {};
        \node [state] (k-a) [right of = square-k, draw=none] {};
        \draw [-stealth, line width=0.5mm] (k-b) --  node [right] {$\msg[b]$} (square-k);
        \draw [-stealth, line width=0.5mm] (square-k) --  node [below] {$\msg[a]$} (k-a);
 \end{tikzpicture}
 &
 \text{\large $S_2$} 
 \end{array}
 \\[12mm]
\hspace{3mm}- - - -    \hspace{8mm}- - - - -  \\[-5mm]
\begin{array}{@{\hspace{0mm}}c@{\hspace{0mm}}c}
\\[4mm]
\text{\large $S_3\hspace{-2mm}$} 
&
 \begin{tikzpicture}[node distance=1.5cm,scale=1]
        \node (square-v) [draw,minimum width=0.8cm,minimum height=0.8cm] {\large $\hh_3$};
        \node [state] (v-b) [left of = square-v, draw=none] {};
        \node [state] (v-a) [below of = square-v, draw=none] {};
        \draw [-stealth,line width=0.5mm] (v-a) --  node [right]   {$\msg[a]$} (square-v);
        \draw [-stealth,line width=0.5mm] (square-v) --  node [below] {$\msg[b]$} (v-b);
 \end{tikzpicture}
 \end{array}
 \hspace{4mm}
\begin{array}{c}
 \\[-8mm]
| \\
| \\
| \\
|
\end{array}
 \hspace{4mm}
 \begin{array}{c@{\hspace{-2mm}}}
 \\[-6mm]
 \begin{tikzpicture}[node distance=1.5cm,scale=1]
        \node (square-w) [draw,minimum width=0.8cm,minimum height=0.8cm] {\large $\hh_4$};
        \node [state] (w-c) [right of = square-w, draw=none] {};
        \node [state] (w-a) [below of = square-w, draw=none] {};
        \node [state] (w-b) [above right of = square-w, draw=none] {};
        \draw [-stealth] (w-c) --  node {$\msg[c]$} (square-w);
        \draw [-stealth,line width=0.5mm] (square-w) --  node [right] {$\msg[a]$} (w-a);
        \draw [stealth-] (square-w) --  node {$\msg[b]$} (w-b);
 \end{tikzpicture}
 \hspace{-6mm}
 \begin{array}{l}
 \\[6mm]
 \text{\large\ \ $S_4$}
 \end{array} 
 \end{array}
 \\[-4mm]
 \end{array}
 \end{equation}

Clearly, even considering this abstract representation, the way the partial gateways can interact
is not uniquely determined by the participants we choose for the composition and by the 
interface actions we select.
In particular, for the participants and interface actions highlighted in (\ref{eq:four-ips}) above,
 one could decide that message $\msg[a]$ received by $\hh_1$ has 
 to be forwarded
 to $\hh_4$; the $\msg[a]$ received by $\hh_3$
 to $\hh_2$; the $\msg[b]$ received by $\hh_2$ to $\hh_3$.
 Another possible choice 
could be similar to the previous one but for the forwarding
of the messages $\msg[a]$: the one received by $\HH_1$ could be forwarded now to $\hh_2$
whereas the one received by $\hh_3$ could be forwarded to $\hh_4$. 
So, it is possible to have the following two connection policies, both coherent
with  (\ref{fig:four-ips}).

 \begin{equation}
    \raisebox{15mm}{$\cs_\mathrm{A}$}
    \begin{array}{c@{\qquad\qquad\qquad\qquad\qquad}c}
\dbox{
 \begin{tikzpicture}[node distance=1.5cm,scale=1]
        \node (square-h) [draw,minimum width=0.8cm,minimum height=0.8cm] {\large $\hh_1$};
        \node (square-k) [draw,minimum width=0.8cm,minimum height=0.8cm, right of = square-h] {\large $\hh_2$};
        \node (square-v)  [draw,minimum width=0.8cm,minimum height=0.8cm, below of = square-h, yshift=-4mm] {\large $\hh_3$};
        \node (square-w)  [draw,minimum width=0.8cm,minimum height=0.8cm, below of = square-k, yshift=-4mm] {\large $\hh_4$};
        %\draw[-stealth]  (square-w) to[out=-135,in=-45]  node {$\msg[b]$} (square-v);
        %
        %
        \draw [-stealth] (0.4,0)  -- node {$\msg[a]$}  (1.5,-1.5 ); % carryng a from h to w    
        % \draw [stealth-] (0,-0.4)  -- node {$\msg[c]$} (1.1,-1.9); % carryng c from w to h 
        \draw [-stealth] (0.4,-1.9)  --  node {$\msg[a]$} (1.5,-0.4 ); % carrying a from v to k
        \draw [stealth-] (0,-1.5)  --  node {$\msg[b]$} (1.1,0); % carrying  b from k to v
 \end{tikzpicture}
 }
&
\dbox{
 \begin{tikzpicture}[node distance=1.5cm,scale=1]
        \node (square-h) [draw,minimum width=0.8cm,minimum height=0.8cm] {\large $\HH_1$};
        \node (square-k) [draw,minimum width=0.8cm,minimum height=0.8cm, right of = square-h] {{\large $\hh_2$}};
        \node (square-v)  [draw,minimum width=0.8cm,minimum height=0.8cm, below of = square-h, yshift=-4mm] {{\large $\hh_3$}};
        \node (square-w)  [draw,minimum width=0.8cm,minimum height=0.8cm, below of = square-k, yshift=-4mm] {{\large $\hh_4$}};
        %
        %\draw [-stealth] (square-w) to[out=-135,in=-45]  node {$\msg[b]$} (square-v);
        %
        %
        \draw [-stealth] (0.4,-1.9) to[out=45,in=90] node {$\msg[a]$} (1.5,-1.5 ); % carryng a from v to w    
         %\draw [stealth-] (0,-0.4)  -- node {$\msg[c]$} (1.1,-1.9) ; % carryng c from w to h 
        \draw [-stealth]  (0.4,0)  to[out=-45,in=-90]  node {$\msg[a]$} (1.5,-0.4 ); % carrying a from h to k
        %\draw [-stealth]  (square-v)  to[out=45,in=-90]  node {$\msg[a]$} (square-k); % carrying a from h to k
        \draw [stealth-] (0,-1.5)  --  node {$\msg[b]$} (1.1,0); % carrying  b from k to v
 \end{tikzpicture}
 }
 \raisebox{14.5mm}{$\,\,\,\,\cs_\mathrm{B}$}
 \end{array}
  \end{equation}
 
 \noindent
  Similarly to the binary case, the PaI multicomposition via partial gateways of the four
 systems consists in fact in replacing the choosen participants  $\hh_1$, $\hh_2$, $\hh_3$ and $\hh_4$, -- one per system -- by gateways whose task stays the same for what concerns the non interface-actions,
whereas messages of interfaces actions are simply forwarded.
 The  architectures  of the resulting composed systems, according to the connection policies   
diagrammatically represented  above are described in (\ref{eq:multiconnection}),  where the names $\hh_1, \hh_2, \hh_3$
and $\hh_4$ are now partial gateways.
\vspace{-4mm} 
\begin{equation}
 \label{eq:multiconnection}
    \begin{array}{c@{\quad\qquad}c}
 \begin{tikzpicture}[node distance=1.5cm,scale=1]
        \node (square-h) [draw,minimum width=0.8cm,minimum height=0.8cm] {\large $\hh_1$};
        \node [state] (h-a) [above of = square-h, draw=none] {};
        \node [state] (h-c) [left of = square-h, draw=none] {};
        \draw [-stealth] (h-a) --  node {$\msg[a]$} (square-h);
        \draw [-stealth] (square-h) --  node {$\msg[c]$} (h-c);
        \node (square-k) [draw,minimum width=0.8cm,minimum height=0.8cm, right of = square-h] {\large $\hh_2$};
        \node [state] (k-b) [above of = square-k, draw=none] {};
        \node [state] (k-a) [right of = square-k, draw=none] {};
        \draw [-stealth] (k-b) --  node {$\msg[b]$} (square-k);
        \draw [-stealth] (square-k) --  node {$\msg[a]$} (k-a);
        \node (square-v)  [draw,minimum width=0.8cm,minimum height=0.8cm, below of = square-h, yshift=-4mm] {\large $\hh_3$};
        \node [state] (v-b) [left of = square-v, draw=none] {};
        \node [state] (v-a) [below of = square-v, draw=none] {};
        \draw [-stealth] (v-a) --  node {$\msg[a]$} (square-v);
        \draw [-stealth] (square-v) --  node {$\msg[b]$} (v-b);
        \node (square-w)  [draw,minimum width=0.8cm,minimum height=0.8cm, below of = square-k, yshift=-4mm] {\large $\hh_4$};
        \node [state] (w-c) [right of = square-w, draw=none] {};
        \node [state] (w-a) [below of = square-w, draw=none] {};
        \node [state] (w-b) [above right of = square-w, draw=none] {};
        \draw [-stealth] (w-c) --  node {$\msg[c]$} (square-w);
        \draw [-stealth] (square-w) --  node {$\msg[a]$} (w-a);
        \draw [stealth-] (square-w) --  node {$\msg[b]$} (w-b);
        %
      %  \draw (square-h)  to[out=-90,in=90]   node {} (square-w);
       % \draw (square-k) to[out=-90,in=90]  node {} (square-v);
      %  \draw (square-h) to[out=0,in=180]  node {} (square-w);
      %  \draw [-stealth] (square-w) to[out=-135,in=-45]  node {$\msg[b]$} (square-v);
      %  \draw (square-k) to[out=-135,in=45]  node {} (square-v);
        %
        \draw (0,0.4)[dotted,thick]  --  (0.4,0); % carrying a inside h  
        %\draw (-0.4,0)[dotted,thick]  --  (0,-0.4); % carrying c inside h
        %
        \draw (1.5,0.4)[dotted,thick]  --  (1.1,0); % carrying b inside k
        \draw (1.5,-0.4)[dotted,thick]  --  (1.9,0); % carrying a inside k
        \draw (0.4,-1.9)[dotted,thick]  --  (0,-2.3); % carrying a inside v
        \draw (0,-1.5)[dotted,thick]  --  (-0.4,-1.9); % carrying k's b inside v
        %\draw (0.4,-2.3)[dotted,thick]  --  (-0.4,-1.9); % carrying b inside v
        %
        \draw (1.5,-1.5)[dotted,thick]  --  (1.5,-2.3); % carrying a inside w
        %\draw (1.1,-1.9)[dotted,thick]  --  (1.9,-1.9); % carrying c inside w
        %\draw (1.1,-2.3)[dotted,thick]  to[out=30,in=260]  (1.9,-1.5); % carrying b inside w
        %
        %
        \draw[-stealth]   (0.4,0)  -- node  {$\msg[a]$}   (1.5,-1.5 ); % carryng a from h to w    
         %\draw  [stealth-]  (0,-0.4)  --  node  {$\msg[c]$}  (1.1,-1.9); % carryng c from w to h 
        \draw [-stealth]  (0.4,-1.9)  --   node  {$\msg[a]$} (1.5,-0.4 ); % carrying a from v to k
        \draw [stealth-]  (0,-1.5)  --  node {$\msg[b]$} (1.1,0); % carrying  b from k to v
 \end{tikzpicture}
& 
\begin{tikzpicture}[node distance=1.5cm,scale=1]
        \node (square-h) [draw,minimum width=0.8cm,minimum height=0.8cm] {\large $\HH$};
        \node [state] (h-a) [above of = square-h, draw=none] {};
        \node [state] (h-c) [left of = square-h, draw=none] {};
        \draw [-stealth] (h-a) --  node {$\msg[a]$} (square-h);
        \draw [-stealth] (square-h) --  node {$\msg[c]$} (h-c);
        \node (square-k) [draw,minimum width=0.8cm,minimum height=0.8cm, right of = square-h] {\large $\KK$};
        \node [state] (k-b) [above of = square-k, draw=none] {};
        \node [state] (k-a) [right of = square-k, draw=none] {};
        \draw [-stealth] (k-b) --  node {$\msg[b]$} (square-k);
        \draw [-stealth] (square-k) --  node {$\msg[a]$} (k-a);
        \node (square-v)  [draw,minimum width=0.8cm,minimum height=0.8cm, below of = square-h, yshift=-4mm] {\large $\hh_3$};
        \node [state] (v-b) [left of = square-v, draw=none] {};
        \node [state] (v-a) [below of = square-v, draw=none] {};
        \draw [-stealth] (v-a) --  node {$\msg[a]$} (square-v);
        \draw [-stealth] (square-v) --  node {$\msg[b]$} (v-b);
        \node (square-w)  [draw,minimum width=0.8cm,minimum height=0.8cm, below of = square-k, yshift=-4mm] {\large $\hh_4$};
        \node [state] (w-c) [right of = square-w, draw=none] {};
        \node [state] (w-a) [below of = square-w, draw=none] {};
        \node [state] (w-b) [above right of = square-w, draw=none] {};
        \draw [-stealth] (w-c) --  node {$\msg[c]$} (square-w);
        \draw [-stealth] (square-w) --  node {$\msg[a]$} (w-a);
        \draw [stealth-] (square-w) --  node {$\msg[b]$} (w-b);
        %
        %\draw [-stealth] (square-w) to[out=-135,in=-45]  node {$\msg[b]$} (square-v);
        %
        \draw (0,0.4)[dotted,thick]  --  (0.4,0); % carrying a inside h  
        %\draw (-0.4,0)[dotted,thick]  --  (0,-0.4); % carrying c inside h
        %
        \draw (1.5,0.4)[dotted,thick]  --  (1.1,0); % carrying b inside k
        \draw (1.5,-0.4)[dotted,thick]  --  (1.9,0); % carrying a inside k
        \draw (0.4,-1.9)[dotted,thick]  --  (0,-2.3); % carrying a inside v
        \draw (0,-1.5)[dotted,thick]  --  (-0.4,-1.9); % carrying k's b inside v
        %\draw (0.4,-2.3)[dotted,thick]  --  (-0.4,-1.9); % carrying b inside v
        %
        \draw (1.5,-1.5)[dotted,thick]  --  (1.5,-2.3); % carrying a inside w
        %\draw (1.1,-1.9)[dotted,thick]  --  (1.9,-1.9); % carrying c inside w
        %\draw (1.1,-2.3)[dotted,thick]  to[out=30,in=260]  (1.9,-1.5); % carrying b inside w
        %
        %
        \draw [-stealth]  (0.4,-1.9) to[out=45,in=90]  node [pos=0.6] {$\msg[a]$} (1.5,-1.5 ); % carryng a from v to w    
         %\draw [stealth-]  (0,-0.4)  --  node  {$\msg[c]$} (1.1,-1.9); % carryng c from w to h 
        \draw [-stealth]  (0.4,0)  to[out=-45,in=-90]  node [pos=0.6]  {$\msg[a]$} (1.5,-0.4 ); % carrying a from h to k
        \draw [stealth-]  (0,-1.5)  -- node  {$\msg[b]$}  (1.1,0); % carrying  b from k to v
 \end{tikzpicture}\\[-4mm]
 \text{\small (Using $\cs_\mathrm{A}$)}
 &
 \text{\small (Using $\cs_\mathrm{B}$)}
 \end{array}
  \end{equation}
 
%  }
%  \vspace{-1mm}
%  \caption{Static description of two PaI multicompositions via partial gateways}
%  \label{fig:multiconnection}
% \end{figure}
}

One of the paper's main findings is that PaI composition via partial gateways preserves a number of communication properties if the connection policy used for the composition satisfies the same properties and some specific requirements due to partiality. For all the properties, but orphan-message-freedom, the no-mixed-state condition
is required for interfaces.
This result is obtained as a corollary of a preservation result for a restricted and simple form of binary PaI composition, which we call {\em partial-fusion composition}.
All forms of PaI composition for asynchronous CFSMs that have been investigated so far, as well as the one via partial gateways that we introduce in this paper, can be conceivably obtained via a number of binary partial-fusion compositions. This composition can therefore be considered at the very core of the PaI approach to system composition.

\paragraph{Partial-Fusion Composition}

This particular form of composition is binary.
To get an intuition, let us consider the following simple example, in which
both $S_1$ and $S_2$ have a participant named $\hh$ that we select as their respective participant
for the composition.\\[-15mm]
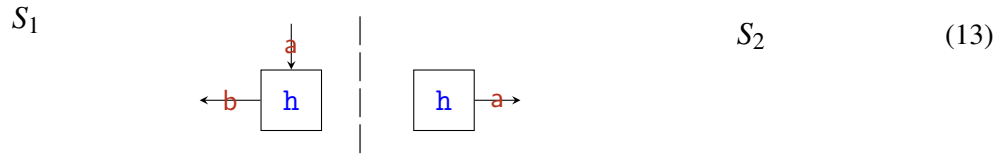
\begin{equation}
\label{eq:exps}
\raisebox{2mm}{\text{\large $S_1$}\,\,}
%    \dbox{
\hspace{10mm} \begin{tikzpicture}[node distance=1.5cm,scale=1]
        \node (square-h) [draw,minimum width=0.8cm,minimum height=0.8cm] {\large $\hh$};
        \node [state] (h-a) [above of = square-h, draw=none] {};
        \node [state] (h-c) [left of = square-h, draw=none, xshift=-2mm] {};
        \draw [-stealth] (h-a) --  node {$\msg[a]$} (square-h);
        \draw [stealth-] (h-c) --  node {$\msg[b]$} (square-h);
 \end{tikzpicture}
%            }
\hspace{2mm}
 \begin{array}{c}
 \\[8mm]
| \\
| \\
|\\
|\\
\end{array}
\hspace{4mm}
%     \dbox{
 \begin{tikzpicture}[node distance=1.5cm,scale=1]
        \node (square-k) [draw,minimum width=0.8cm,minimum height=0.8cm] {\large $\hh$};
        \node [state] (k-b) [above of = square-k, draw=none] {};
        \node [state] (k-a) [right of = square-k, draw=none] {};
        %\draw [stealth-] (k-b) --  node[right] {$\msg[inf]$} (square-k);
        \draw [-stealth] (square-k) --  node {$\msg[a]$} (k-a);
       % \draw  [-stealth] (0.4,-0.2)   --  node [below] {$\msg[par]$} (1.2,-0.2);
 \end{tikzpicture} \hspace{18mm}
 %            }
 \raisebox{0mm}{\text{\large $\,\,S_2$}}
 \end{equation}
The composition method requires that some actions are chosen as interface actions in only one
of the two participants so that, if possible, we get two participants that complement each other perfectly. For instance, by choosing\\[-2mm]\\[-2mm]
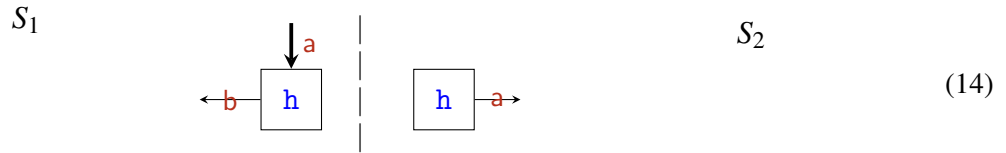
\begin{equation}
\label{eq:expsia}
\begin{array}{c}
\\[-18mm]
\raisebox{2mm}{\text{\large $S_1$}\,\,}
%    \dbox{
\hspace{10mm} \begin{tikzpicture}[node distance=1.5cm,scale=1]
        \node (square-h) [draw,minimum width=0.8cm,minimum height=0.8cm] {\large $\hh$};
        \node [state] (h-a) [above of = square-h, draw=none] {};
        \node [state] (h-c) [left of = square-h, draw=none, xshift=-2mm] {};
        \draw [-stealth,line width=0.5mm] (h-a) --  node[right] {$\msg[a]$} (square-h);
        \draw [stealth-] (h-c) --  node {$\msg[b]$} (square-h);
 \end{tikzpicture}
%            }
\hspace{2mm}
 \begin{array}{c}
 \\[8mm]
| \\
| \\
|\\
|\\
\end{array}
\hspace{4mm}
%     \dbox{
 \begin{tikzpicture}[node distance=1.5cm,scale=1]
        \node (square-k) [draw,minimum width=0.8cm,minimum height=0.8cm] {\large $\hh$};
        \node [state] (k-b) [above of = square-k, draw=none] {};
        \node [state] (k-a) [right of = square-k, draw=none] {};
        %\draw [stealth-] (k-b) --  node[right] {$\msg[inf]$} (square-k);
        \draw [-stealth] (square-k) --  node {$\msg[a]$} (k-a);
       % \draw  [-stealth] (0.4,-0.2)   --  node [below] {$\msg[par]$} (1.2,-0.2);
 \end{tikzpicture} \hspace{18mm}
 %            }
 \raisebox{0mm}{\text{\large $\,\,S_2$}}
 \end{array}
 \end{equation}
 \noindent
and  ``restricting'' the behaviour of  $\hh$ in $S_1$ (the  participant with interface actions)
to its interface actions only, we obtain
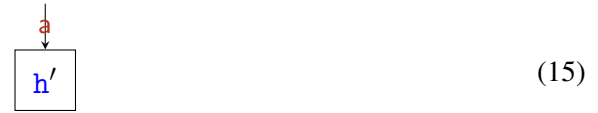
\begin{equation}
\begin{array}{c}
\\[-18mm]
\begin{tikzpicture}[node distance=1.5cm,scale=1]
        \node (square-h) [draw,minimum width=0.8cm,minimum height=0.8cm] {\large $\hh'$};
        \node [state] (h-a) [above of = square-h, draw=none] {};
        \node [state] (h-c) [left of = square-h, draw=none, xshift=-2mm] {};
        \draw [-stealth] (h-a) --  node {$\msg[a]$} (square-h);
 \end{tikzpicture}
  \end{array}
\end{equation}
Now  $\hh'$ above and $\hh$ of $S_2$ in (\ref{eq:expsia}) are perfectly complementary
(roughly, one has an input where the other has an output, and vice versa).
 Such a complementarity relation enables the $\hh$ in $S_1$ and  the
 $\hh$ in $S_2$ represented in (\ref{eq:exps}) to be partially {\em fused}.
The fusion makes the two $\hh$s a single, partially forwarding gateway. Messages pertaining to interface actions are forwarded, while the partial gateway continues to behave as the  $\hh$ of $S_1$ with regard to non-interface actions.\\[-4mm]
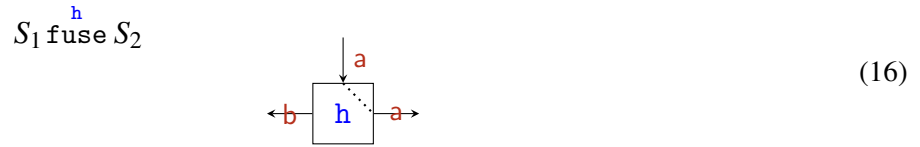
\begin{equation}
\label{fig:bincomp}
\begin{array}{l}
\text{\large $S_1$}\!\stackrel{\hh}{ \mathtt{fuse}}\text{\large $S_2$}\\[12mm]
\end{array}
\hspace{12mm}
% \dbox{ \hspace{28mm}
 \begin{tikzpicture}[node distance=1.5cm,scale=1]
        \node (square-h) [draw,minimum width=0.8cm,minimum height=0.8cm] {\large $\hh$};
         \node[draw=none,fill=none] (h-a) [above = 6mm  of square-h]{};
         \node[draw=none,fill=none] (a-h) [left = 6mm  of square-h]{};
          \node[draw=none,fill=none] (a-hr) [right = 6mm  of square-h]{};
        \draw [-stealth] (h-a) --  node [right] {$\msg[a]$} (square-h);
        \draw [-stealth] (square-h) --  node {$\msg[a]$} (a-hr);
        \draw (0,0.4)[dotted,thick]  --  (0.4,0); 
        \draw [-stealth] (square-h) --  node {$\msg[b]$} (a-h); 
 \end{tikzpicture}
 \hspace{22mm}
%       }
\end{equation}
Notably, the relationship between fusible participants corresponds to that between interface participants and their respective participants in a connection policy within the PaI composition via partial gateways.
\medskip

We now informally show how PaI composition via partial gateways can be achieved through 
a number of partial-fusion compositions. This is the primary objective of partial-fusion.
From this perspective, the requirement to consider participants with the same name in the two systems is not as restrictive as it seems at first glance. 

Consider the interface participants for $S_1$ and $S_2$ in (\ref{eq:sbspar}) and the following selection of interface actions.\\[0mm]
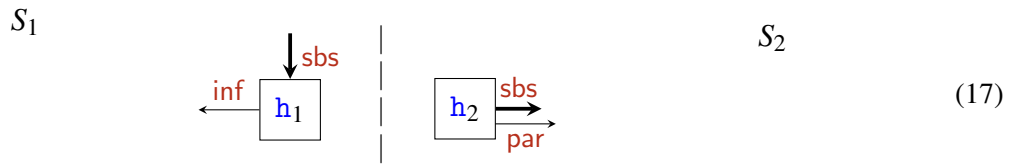
\begin{equation}
\label{eq:simpleexia}
\begin{array}{c}
\\[-18mm]
\raisebox{3mm}{\text{\large $S_1$}\,\,}
%    \dbox{
\hspace{10mm} \begin{tikzpicture}[node distance=1.5cm,scale=1]
        \node (square-h) [draw,minimum width=0.8cm,minimum height=0.8cm] {\large $\hh_1$};
        \node [state] (h-a) [above of = square-h, draw=none] {};
        \node [state] (h-c) [left of = square-h, draw=none, xshift=-2mm] {};
        \draw [-stealth,line width=0.5mm] (h-a) --  node[right] {$\msg[sbs]$} (square-h);
        \draw [stealth-] (h-c) --  node[above] {$\msg[inf]$} (square-h);
 \end{tikzpicture}
%            }
\hspace{2mm}
 \begin{array}{c}
 \\[8mm]
| \\
| \\
|\\
|\\
\end{array}
\hspace{4mm}
%     \dbox{
 \raisebox{-1mm}{\begin{tikzpicture}[node distance=1.5cm,scale=1]
        \node (square-k) [draw,minimum width=0.8cm,minimum height=0.8cm] {\large $\hh_2$};
        \node [state] (k-b) [above of = square-k, draw=none] {};
        \node [state] (k-a) [right of = square-k, draw=none] {};
        %\draw [stealth-] (k-b) --  node[right] {$\msg[inf]$} (square-k);
        \draw [-stealth,line width=0.5mm] (square-k) --  node[above] {$\msg[sbs]$} (k-a);
        \draw  [-stealth] (0.4,-0.2)   --  node [below] {$\msg[par]$} (1.2,-0.2);
 \end{tikzpicture}} \hspace{18mm}
 %            }
 \raisebox{1mm}{\text{\large $\,\,S_2$}}
 \end{array}
 \end{equation}

Since we are considering a binary case, the connection policy is uniquely determined
by the selected interface actions.

%\begin{figure}[h]
%    \centering{\small
%    $
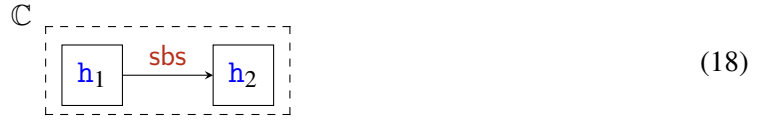
\begin{equation}
\label{fig:pcp}
\begin{array}{l}
\text{$\cs$}\\[12mm]
\end{array}
 \dbox{
 \begin{tikzpicture}[node distance=1.5cm,scale=1]
        \node (square-h) [draw,minimum width=0.8cm,minimum height=0.8cm] {\large $\hh_1$};
        \node (square-k) [draw,minimum width=0.8cm,minimum height=0.8cm, right of = square-h, xshift=5mm] {\large $\hh_2$};
        \draw [-stealth] (0.4,0)  --  node [above] {$\msg[sbs]$} (1.6,0); % carrying sbs from h1 to h2 h2
 \end{tikzpicture}
 }
 \end{equation}
% $
% \caption{\label{fig:pcp} A connection policy for composing the systems in \cref{fig:twoipnterfaces}}
% }
%\end{figure}

Now we consider systems $S_1$ and $\cs$ (recall that here $\cs$ represents a system of the same sort as $S_1$ and $S_2$).

% \begin{figure}[h]
%    \centering{\small
%    $
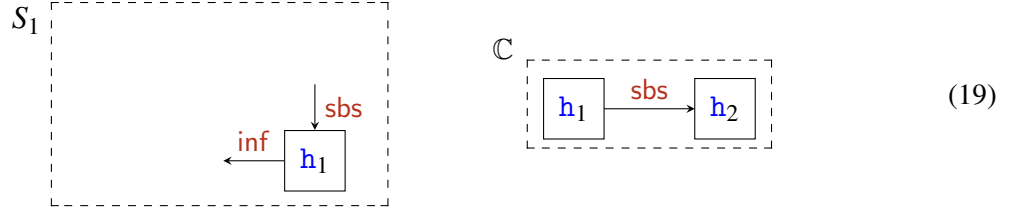
\begin{equation}
\label{fig:twoipnterfaces}
\raisebox{10mm}{\text{\large $S_1$}\,\,}
    \dbox{
\hspace{10mm} \begin{tikzpicture}[node distance=1.5cm,scale=1]
        \node (square-h) [draw,minimum width=0.8cm,minimum height=0.8cm] {\large $\hh_1$};
        \node [state] (h-a) [above of = square-h, draw=none] {};
        \node [state] (h-c) [left of = square-h, draw=none, xshift=-2mm] {};
        \draw [-stealth] (h-a) --  node[right] {$\msg[sbs]$} (square-h);
        \draw [stealth-] (h-c) --  node[above] {$\msg[inf]$} (square-h);
 \end{tikzpicture}
            }
\hspace{12mm}
 \begin{array}{l}
\text{$\cs$}\\[12mm]
\end{array}
 \dbox{
 \begin{tikzpicture}[node distance=1.5cm,scale=1]
        \node (square-h) [draw,minimum width=0.8cm,minimum height=0.8cm] {\large $\hh_1$};
        \node (square-k) [draw,minimum width=0.8cm,minimum height=0.8cm, right of = square-h, xshift=5mm] {\large $\hh_2$};
        \draw [-stealth] (0.4,0)  --  node [above] {$\msg[sbs]$} (1.6,0); % carrying sbs from h1 to h2 h2
 \end{tikzpicture}
 }
 \end{equation}
% $\vspace{-2mm}
% \caption{\label{fig:twoipnterfaces} Two interfaces participants belonging to, respectively, systems $S_1$ and $S_2$.}
% }
%\end{figure}

The above systems are eligible for fusion composition, which returns the following system.

%\begin{figure}[h]
%    \centering{\small
%    $
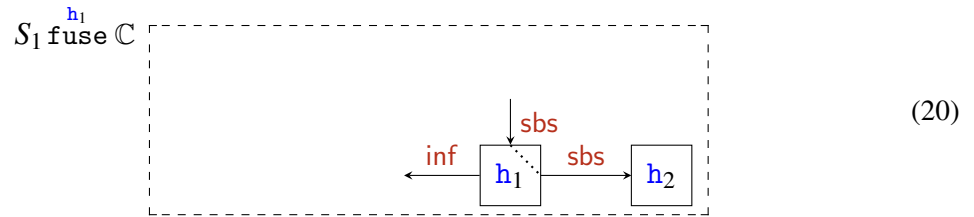
\begin{equation}
\label{fig:binpcomp}
\begin{array}{l}
\text{\large $S_1$}\!\stackrel{\hh_1}{ \mathtt{fuse}}\cs\\[22mm]
\end{array}
 \dbox{ \hspace{28mm}
 \begin{tikzpicture}[node distance=1.5cm,scale=1]
        \node (square-h) [draw,minimum width=0.8cm,minimum height=0.8cm] {\large $\hh_1$};
        \draw [-stealth] (h-a) --  node [right] {$\msg[sbs]$} (square-h);
        \node (square-k) [draw,minimum width=0.8cm,minimum height=0.8cm, right of = square-h, xshift=5mm] {\large $\hh_2$};
         \node[draw=none,fill=none] (phantom) [above = 12mm  of square-h]{};
         \node[draw=none,fill=none] (phantom2) [left = 10mm  of square-h]{};
        \draw [stealth-] (phantom2) --  node [above]{$\msg[inf]$} (square-h);
        \draw (0,0.4)[dotted,thick]  --  (0.4,0); % carrying sbs inside h 
        \draw [-stealth] (0.4,0)  -- node [above]{$\msg[sbs]$}  (1.6,0); % carrying sbs from h1 to h2
 \end{tikzpicture}
        }
\end{equation}
% $
% \caption{\label{fig:binpcomp} Pai binary composition via partial gateways}
% }
%\end{figure}

We now consider the following two systems.\\[-3mm]
%\begin{figure}[h]
%    \centering{\small
%    $
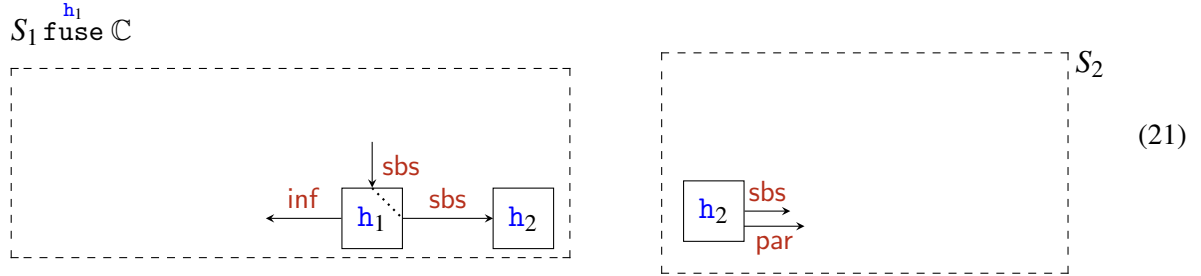
\begin{equation}
\label{fig:binpcomp}
\begin{array}{l}
\text{\large $S_1$}\!\stackrel{\hh_1}{ \mathtt{fuse}}\cs
\\
 \dbox{ \hspace{28mm}
 \begin{tikzpicture}[node distance=1.5cm,scale=1]
        \node (square-h) [draw,minimum width=0.8cm,minimum height=0.8cm] {\large $\hh_1$};
        \draw [-stealth] (h-a) --  node [right] {$\msg[sbs]$} (square-h);
        \node (square-k) [draw,minimum width=0.8cm,minimum height=0.8cm, right of = square-h, xshift=5mm] {\large $\hh_2$};
         \node[draw=none,fill=none] (phantom) [above = 12mm  of square-h]{};
         \node[draw=none,fill=none] (phantom2) [left = 10mm  of square-h]{};
        \draw [stealth-] (phantom2) --  node [above]{$\msg[inf]$} (square-h);
        \draw (0,0.4)[dotted,thick]  --  (0.4,0); % carrying sbs inside h 
        \draw [-stealth] (0.4,0)  -- node [above]{$\msg[sbs]$}  (1.6,0); % carrying sbs from h1 to h2
 \end{tikzpicture}
        }
        \hspace{12mm}
     \dbox{
 \begin{tikzpicture}[node distance=1.5cm,scale=1]
        \node (square-k) [draw,minimum width=0.8cm,minimum height=0.8cm] {\large $\hh_2$};
        \node [state] (k-b) [above of = square-k, draw=none] {};
        \node [state] (k-a) [right of = square-k, draw=none] {};
        %\draw [stealth-] (k-b) --  node[right] {$\msg[inf]$} (square-k);
        \draw [-stealth] (square-k) --  node[above] {$\msg[sbs]$} (k-a);
        \draw  [-stealth] (0.4,-0.2)   --  node [below] {$\msg[par]$} (1.2,-0.2);
 \end{tikzpicture} \hspace{24mm}
             }
 \raisebox{12mm}{\text{\large $\,\,S_2$}}
 \end{array}
 \end{equation}
% $
% \caption{\label{fig:binpcomp} Two systems amenable for fusion composition via $\hh_2$}
% }
%\end{figure}

By applying fusion composition again, we get the following system.\\[-3mm]
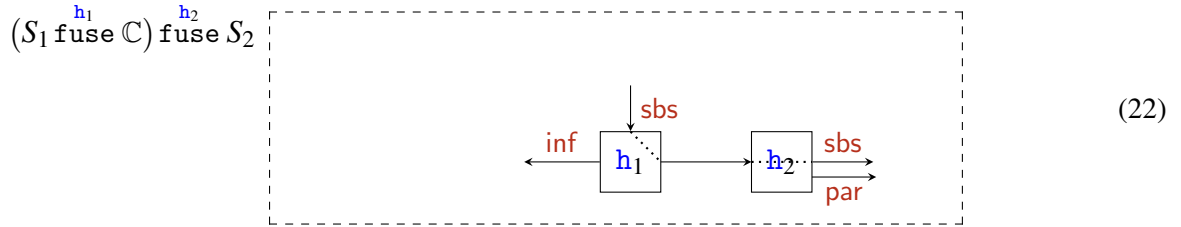
\begin{equation}
\label{fig:bincomp}
\hspace{-22mm}
\begin{array}{l}\big(\text{\large $S_1$}\!\stackrel{\hh_1}{ \mathtt{fuse}}\cs\big)\!\stackrel{\hh_2}{ \mathtt{fuse}}\text{\large $S_2$}\\[22mm]
\end{array}
 \dbox{ \hspace{28mm}
 \begin{tikzpicture}[node distance=1.5cm,scale=1]
        \node (square-h) [draw,minimum width=0.8cm,minimum height=0.8cm] {\large $\hh_1$};
        \draw [-stealth] (h-a) --  node [right] {$\msg[sbs]$} (square-h);
        \node (square-k) [draw,minimum width=0.8cm,minimum height=0.8cm, right of = square-h, xshift=5mm] {\large $\hh_2$};
        \node [state] (k-a) [right of = square-k, draw=none,xshift=2mm] {};
         \node[draw=none,fill=none] (phantom) [above = 12mm  of square-h]{};
         \node[draw=none,fill=none] (phantom2) [left = 10mm  of square-h]{};
        \draw [-stealth] (square-k) --  node [above]{$\msg[sbs]$} (k-a);
        \draw [stealth-] (phantom2) --  node [above]{$\msg[inf]$} (square-h);
        \draw (0,0.4)[dotted,thick]  --  (0.4,0); % carrying sbs inside h 
        \draw [-stealth] (0.4,0)  --  (1.6,0); % carrying sbs from h1 to h2
        \draw  [-stealth] (2.4,-0.2)   --  node [below] {$\msg[par]$} (3.25,-0.2); % carrying par outside h2
        \draw (1.6,0)[dotted,thick]  --  (2.4,0); % carrying sbs inside h2
 \end{tikzpicture}
       }
\end{equation}
 We have  that 
 $
 \big(\text{\large $S_1$}\!\stackrel{\hh_1}{ \mathtt{fuse}}\cs\big)\!\stackrel{\hh_2}{ \mathtt{fuse}}\text{\large $S_2$}\
  \equiv\
   \text{\large $S_1$\!}\!\stackrel{\hh_1{\leftrightarrow}\hh_2}{ \phantom{\mathtt{p\texttt{-}gw}}}\text{\large \!$S_2$}
 $,
 where $\text{\large $S_1$\!}\!\stackrel{\hh_1{\leftrightarrow}\hh_2}{ \phantom{\mathtt{p\texttt{-}gw}}}\text{\large \!$S_2$}$ is the system obtained %from $S_1$ and $S_2$ 
 by PaI composition via partial gateways using the connection policy $\cs$ as depicted in (\ref{eq:pgcompsbspar}).
% This procedures applies both to multicomposition and orchestrated multicomposition as 
% defined in \cite{BH24,BH26}. 

%!TEX root = Main-CFSM-partial-fusion.tex
\section{Systems of Communicating Finite State Machines}
\label{sect:cfsm}

Communicating Finite State Machines (CFSM)   are 
 a widely investigated automata-based
formalism for the description and analysis of distributed systems, originally proposed in \cite{BZ83}.

\begin{definition}[FSA]
\label{def:fsa}A {\em Finite State Automaton} is a quadruple $(Q,q_0,\mathcal{L},\delta)$ where
$Q$ is a finite set of states, $q_0\in Q$ is the initial state, $\mathcal{L}$ a set of labels and
$\delta\subseteq Q\times\mathcal{L}\times Q$ is a set of transitions.\\
An {\em $\varepsilon$-FSA} is an FSA such that $\varepsilon\in\mathcal{L}$.
\end{definition}

CFSMs are particular finite state automata representing processes which communicate by asynchronous exchanges of messages via FIFO channels. 
We now recall (partly following \cite{CF05,DY12,TY15}) the definitions of CFSM and systems of CFSMs.
We assume a countably infinite set  
$\roles_\mathfrak{U}$ of participant names (ranged over by $\ttp,\ttq,\ttr,\HH,\KK,\ttv,\ttw\ldots$) and a countably infinite alphabet $\mathbb{A}_\mathfrak{U}$ 
of messages (ranged over by $\msg[a]$, $\msg[b]$, $\msg[c]$, $\msg[m],\ldots$). We also assume an infinite set of indexes, ranged over by $i$, $j$, $\ldots$ and use $I$, $J$, $\ldots$ to range over finite sets of indexes.\\

\begin{definition}[Communicating Finite State Machine]\label{def:cfsm}%\hfill\\
Let $\roles$  and $\mathbb{A}$ be finite subsets of $\roles_\mathfrak{U}$ and $\mathbb{A}_\mathfrak{U}$, respectively.
\begin{enumerate}[i)] 
\item
The set $C_\roles$ of {\em channels} over $\roles$ is defined by\ \
$C_\roles=\Set{\ttp\ttq \mid \ttp,\ttq\in \roles, \ttp\neq\ttq}$
\item
The set $\mathit{Act}_{\roles,\mathbb{A}}$ of {\em actions}  over $\roles$ and $\mathbb{A}$ is defined by\ 
$\textit{Act}_{\roles,\mathbb{A}} = C_\roles\times\Set{!,?}\times\mathbb{A}$

Given an action $\elle$ we define $\subj{\elle}$ ({\em the subject of $\elle$}) by  \  \
$\subj{\elle}=\ttp \text{ if }  \elle=\ttp\ttq!\msg[m] \text{ or }\elle=\ttq\ttp?\msg[m]$.
\item
\label{def:cfsm-iii}
A {\em communicating finite state machine $M$ over} $\roles$ \emph{and} $\mathbb{A}$
is an FSA where  $\mathcal{L}= \textit{Act}_{\roles,\mathbb{A}}$ and 
%finite transition system given by a tuple\\
%\centerline{ $M=(Q,q_0,\mathbb{A},\delta)$ }
%where $Q$ is a finite set of states, $q_0\in Q$ is the initial state, and
%$\delta\subseteq Q\times\textit{Act}_{\roles,\mathbb{A}}\times Q$ is a set of transitions
such that all the actions have the same subject, to which we refer as the {\em name} of $M$.
\end{enumerate}
\end{definition}
\noindent
We shall write $M_{\ttp}$ to denote a CFSM with name $\ttp$. 
Where no ambiguity arises we shall refer to a CFSM by its name.
%\brc Let us check whether we ever use $C$ and $\mathit{Act}$.
%But even then, I would prefer to see $C_\roles$ and $\mathit{Act}_{\roles,\mathbb{A}}$ which is minimal longer.\erc
 We assume $\elle,\elle',\ldots$ to range over actions %$\textit{Act}$
%$\varphi,\varphi',\ldots$ to range over $\textit{Act}^*$ (the set of finite words over $\textit{Act}$), 
and $w,w',\ldots$ to range over $\mathbb{A}_\mathfrak{U}^*$ (the set of finite words over $\mathbb{A}_\mathfrak{U}$).
The symbol $\varepsilon\,(\notin \mathbb{A}_\mathfrak{U})$ denotes the empty word and 
$\mid w\mid$ the length of a word $w$. %\in \mathbb{A}^*$.
%$\mid v\mid$ the lenght of a word $v\in \textit{Act}^*\cup\mathbb{A}^*$.
%%>>>>>>>> DEFINITIONS NOT USED in the present paper
%Given a word $v$ with prefix $v'$, i.e. such that $v=v'\cdot v''$ for a certain $v''$, we define $v\setminus v' =v''$.
%Moreover, given a  word $v$ with $\msg[a]$ as last  element, i.e. $v=v'\cdot \msg[a]$ for a certain (possibly empty) $v'$, we define
%$\mathsf{init}(v) = v'$ and  $\mathsf{last}(v) = \msg[a]$. 
%Moreover, we shall denote by $\widetilde{v}$ the reverse of the word $v$. \\
The transitions of a CFSM are labelled by actions; a label $\tts\ttr!\msg[a]$ represents
the asynchronous sending of message $\msg[a]$ from machine $\tts$ to $\ttr$ through channel $\tts\ttr$ and, dually,
$\tts\ttr?\msg[a]$ represents the reception (consumption) of $\msg[a]$ by $\ttr$ from channel
$\tts\ttr$. 
% Given a CFSM $M=(Q,q_0,\mathbb{A},\delta)$,
%we also define \\
%\centerline{$\inn{M}=\Set{\msg[a] \mid (\_,\_\,\_?\msg[a],\_)\in \delta }$
%\quad \text{ and }\quad $\outt{M}=\Set{\msg[a] \mid (\_,\_\,\_!\msg[a],\_)\in \delta }$.}
% If $M$ is a CFSM with name $\ttp$, we also write $\inn{\ttp}$ for $\inn{M}$ and
% $\outt{\ttp}$ for $\outt{M}$.
%  Note that, in concrete examples, the name of a CFSM together with its input and output messages can be graphically depicted as in the introduction. 
%We write $\lang{M}\subseteq\textit{Act}^*$ for
%the language over $\textit{Act}$ accepted by the automaton corresponding
%to machine $M$, where each state of $M$ is an accepting state. 
A state
$q\in Q$ with no outgoing transition is {\em final}; 
$q$ is a {\em sending} (resp. {\em receiving}) state if it is not final and
all outgoing transitions are labelled with sending (resp. receiving) actions;
$q$  is a {\em mixed} state if there are at least two outgoing transitions such that one is labelled with a sending action and the other one is labelled with a receiving action.

%
%  ?!-DETERMINISM DEFINITION <<<<<<<<<<<<<<<<<<<<<<<<
% 
%\vspace{2mm}
%A CFSM $M = (Q,q_0,\mathbb{A},\delta)$ is:
%\begin{enumerate}[a)]
%\item
% {\em deterministic} if for all transitions:\quad % states $q\in Q$ and all actions $\elle$: 
%$(q,\elle, q'), (q,\elle,q'')\in \delta$ imply $q'=q''$;
%\item
%{\em ?-deterministic} (resp. {\em !-deterministic}) if for all transitions:\\ % all states  $q\in Q$ and all actions:\\
%$\qquad$ $(q,\ttr\tts?\msg[a], q'), (q,\ttp\ttq?\msg[a],q'')\in \delta$ (resp. $(q,\ttr\tts!\msg[a], q'), (q,\ttp\ttq!\msg[a],q'')\in \delta$) imply $q'=q''$;\footnote{Note that, by Definition \ref{def:cfsm}(\ref{def:cfsm-iii}), we have
%necessarily that $\tts=\ttq$ in the clause for ?-determinism and $\ttr=\ttp$ in the one for
%!-determinism.}
%\item
%{\em ?!-deterministic} if it is both ?-deterministic and !-deterministic.
%\end{enumerate}
%
%The notion of ?!-deterministic machine is more demanding than in usual CFSM settings. It will be needed in order to guarantee preservation of communication properties when systems are connected. 
%Note that a ?!-deterministic CFSM is also deterministic, but the converse does not hold
%(since the channel names are abstracted away in the definition of ?!-determinism). \\

\smallskip
A {\em communicating system} % called ``protocol'' in \cite{BZ83}, 
is a finite set of 
CFSMs referred to as {\em participants} of the communicating system.
% We use the term ``participant'' both for a CFSM and its name.
% over some vocabulary of messages such that senders and receivers are identified by the 
%names of CFSMs. 
\Commented{In~\cite{CF05,DY12,TY15} the names of the CFSMs in a system are called {\em roles}. In the present paper we call them {\em participants}.}

\begin{definition}[Communicating system]%\hfill\\
Let $\roles$  and $\mathbb{A}$ be as in Def.~\ref{def:cfsm}.
%\begin{enumerate}[i)]
%\item
A {\em communicating system (CS)
over} $\roles$ \emph{and} $\mathbb{A}$ is a  set  %tuple 
$S= (M_\ttp)_{\ttp\in\roles}$
%\centerline{$S= (M_\ttp)_{\ttp\in\roles}$}
where
%\\
%-  $\roles\subseteq_{\text{fin}}\roles_\mathfrak{U}$   is the set of {\em roles} (participants) of $S$, and\\
for each $\ttp\in \roles$,
$M_\ttp=(Q_\ttp,q_{0\ttp},\mathit{Act}_{\roles,\mathbb{A}},\delta_\ttp)$ is a CFSM  over $\roles$ and $\mathbb{A}$.
%\end{enumerate}
\end{definition}

The dynamics of a system are
formalised as a transition relation on configurations, where a configuration is a
pair of tuples: a tuple of states of the machines in the system and a tuple of buffers representing the content of the channels. A buffer is described as an element of $\mathbb{A}_\mathfrak{U}^*$. 

\begin{definition}[Configuration]%\hfill\\
Let $S$ be a  communicating system over $\roles$ and $\mathbb{A}$.
A {\em configuration} of $S$ is a pair $s = (\vec{q},\vec{w})$
%\centerline{$s = (\vec{q},\vec{w})$}
where\\
\centerline{$\vec{q}= (q_\ttp)_{\ttp\in\roles}$ with $q_\ttp \in Q_\ttp$,
\qquad and \qquad  $\vec{w}  = (w_{\ttp\ttq})_{\ttp\ttq\in C}$ with $w_{\ttp\ttq}\in\mathbb{A^*}$.}
%\begin{itemize}
%\item[-]  $\vec{q}= (q_\ttp)_{\ttp\in\roles}$ with $q_\ttp \in Q_\ttp$,
%\item[-]  $\vec{w}  = (w_{\ttp\ttq})_{\ttp\ttq\in C}$ with $w_{\ttp\ttq}\in\mathbb{A^*}$.
%\end{itemize}

The component $\vec{q}$ is the {\em control state\/} of the system and $q_\ttp \in Q_\ttp$ is the 
{\em local state\/} of machine $M_\ttp$. 
The component $\vec{w}$ represents the state of the channels of the system and $w_{\ttp\ttq} \in \mathbb{A}^*$ is the state of the channel $\ttp\ttq$, i.e. the unread messages sent from $\ttp$ to $\ttq$. The initial configuration of $S$ is $s_0=  (\vec{q_0},\vec{\varepsilon})$
with $\vec{{q_0}} = (q_{0_\ttp})_{\ttp\in\roles}$.
\end{definition}

\noindent
In the following we will often denote a communicating system $(M_{\ttp})_{\ttp\in \Set{\ttr_i}_{i\in I}}$ by $(M_{\ttr_i})_{i\in I}$.

Notice that the above definition of CFSM is generic with respect to the underlying sets
$\roles$ and $\mathbb{A}$.
%Since we not deal with a single system of CFSMs but with a number of  systems of CFSMs that can be {\em composed},
We shall often use $C$ and $\mathit{Act}$ for specific $C_{\roles}$ and $\mathit{Act}_{\roles,\mathbb{A}}$ when no ambiguity can arise.

\begin{definition}[Transitions and reachable configurations]
\label{def:opsem}
Let $S$ be a communicating system over $\roles$ and $\mathbb{A}$, and let $s= (\vec{q},\vec{w})$ and $s'= (\vec{q'},\vec{w'})$ 
be two configurations of $S$. %\\
Configuration $s'$ {\em is reachable from} $s$
{\em by firing  a transition} with action $\elle$, written $s\lts{\elle}s'$, if there is $\msg[a]\in\mathbb{A}$
such that one of the following conditions holds:
%\begin{center}
%\begin{tabular}{ r l r l}
%$1.$ & $\elle = \tts\ttr!\msg[a]$ and $(q_\tts,\elle,q'_\tts)\in\delta_\tts$ and &  
%$2.$ & $\elle = \tts\ttr?\msg[a]$ and $(q_\ttr,\elle,q'_\ttr)\in\delta_\ttr$ and\\
%        & $a)$ for all $\ttp\neq\tts: ~ q'_\ttp =  q_\ttp$  and &
%        & $a)$ for all $\ttp\neq\ttr: ~ q'_\ttp =  q_\ttp$  and \\
%        & $b)$ $w'_{\tts\ttr} =  w_{\tts\ttr}\cdot \msg[a]$ and for all $\ttp\ttq\neq\tts\ttr: ~ w'_{\ttp\ttq} =  w_{\ttp\ttq}$; &
%        & $b)$  $w_{\tts\ttr} =  \msg[a]\cdot w'_{\tts\ttr}$ and for all $\ttp\ttq\neq\tts\ttr: ~w'_{\ttp\ttq} =  w_{\ttp\ttq}$.
%\end{tabular}
%\end{center}
\begin{enumerate}
\item
$\elle = \tts\ttr!\msg[a]$ and $(q_\tts,\elle,q'_\tts)\in\delta_\tts$ and\\
$a)$
for all $\ttp\neq\tts: ~ q'_\ttp =  q_\ttp$  \quad and \quad 
$b)$
$w'_{\tts\ttr} =  w_{\tts\ttr}\cdot \msg[a]$ and for all $\ttp\ttq\neq\tts\ttr: ~ w'_{\ttp\ttq} =  w_{\ttp\ttq}$;
%\begin{enumerate}
%\item
%$\elle = \tts\ttr!\msg[a]$ and $(q_\tts,\elle,q'_\tts)\in\delta_\tts$ and
%\qquad \begin{enumerate}[a)]
%\item
%for all $\ttp\neq\tts: ~ q'_\ttp =  q_\ttp$  and
%\item
%$w'_{\tts\ttr} =  w_{\tts\ttr}\cdot \msg[a]$ and for all $\ttp\ttq\neq\tts\ttr: ~ w'_{\ttp\ttq} =  w_{\ttp\ttq}$;
%\end{enumerate}
\item 
$\elle = \tts\ttr?\msg[a]$ and $(q_\ttr,\elle,q'_\ttr)\in\delta_\ttr$ and\\
$a)$
for all $\ttp\neq\ttr: ~ q'_\ttp =  q_\ttp$  \quad and \quad
$b)$
$w_{\tts\ttr} =  \msg[a]\cdot w'_{\tts\ttr}$ and for all $\ttp\ttq\neq\tts\ttr: ~w'_{\ttp\ttq} =  w_{\ttp\ttq}$.
%\begin{enumerate}[a)]
%\item
%for all $\ttp\neq\ttr: ~ q'_\ttp =  q_\ttp$  and
%\item
%$w_{\tts\ttr} =  \msg[a]\cdot w'_{\tts\ttr}$ and for all $\ttp\ttq\neq\tts\ttr: ~w'_{\ttp\ttq} =  w_{\ttp\ttq}$.
%\end{enumerate}
\end{enumerate}
We write $s\lts{}s'$ if there exists $\elle$ such that  $s\lts{\elle}s'$
 and we write $s\notlts{}\hspace{2mm}$ if no $s'$ and no $\elle$ exist with
$s\lts{\elle}s'$.
As usual, we denote the reflexive and transitive 
closure of $\lts{}$ by $\to^*$.
The set of {\em reachable configurations} of S is $\RS(S) = \Set{s \mid s_0 \to^* s}.$
\end{definition}
\noindent
According to the above definition, communication happens asynchronously via buffered channels following the FIFO principle.

\smallskip
%The overall behaviour of a system can be described (at least) by the traces of configurations that are reachable from a distinguished initial one. Configurations may exhibit some pathological properties, like various forms of {\em deadlock} or {\em progress violation}, channels containing messages that will never be consumed ({\em orphan messages}) or 
%participants expecting messages which are different from those 
%present in their input channels
%({\em unspecified receptions}). 
The aim of analysing communication systems is to verify the properties that ensure certain pathological configurations, such as deadlock or reception errors, cannot be reached. 
\Commented{
Although the desirable system properties are undecidable in general~\cite{BZ83}, sufficient conditions are known that are effectively checkable
relying, for instance, on half-duplex communication~\cite{CF05}, on the form of network topologies~\cite{DBLP:conf/concur/ClementeHS14}, or on synchronous compatibility checking~\cite{HB18}.} % END Commented
We formalise now a number of relevant communication properties for systems of CFSMs
that we deal with in the present paper.  

\begin{definition}[Communication properties]%\hfill\\
\label{def:safeness}
Let $S$ be a communicating system, and let $s= (\vec{q},\vec{w})$ be a configuration of $S$.
\begin{enumerate}[i)]
\item
\label{def:safeness-i}
$s$ is a {\em deadlock configuration} of $S$ if \hspace{2mm}
$\vec{w}=\vec{\varepsilon}\quad\text{and}\quad \forall \ttp\in\roles.~q_\ttp \text{ is a receiving state}$.\\
I.e. all buffers are empty, but all machines are waiting for a message.\\
We say that $S$ is {\em deadlock-free} whenever, for any $s\in \RS(S)$, $s$ is not a  deadlock configuration.

%\item
%\label{def:safeness-i}
%$s$ is a {\em deadlock configuration} if $s\, \not\!\!\lts{}$ and either
%\begin{enumerate}[a)]
%\item $\exists \ttr\in\roles$ such that $q_\ttr \lts{\ttr\tts?a} q'_\ttr$ , or
%\item
%\label{def:safeness-wnotem}
%$\vec{w}\neq\vec{\varepsilon}$ 
%\end{enumerate}
%i.e. $s$ is stuck  because all machines which are not in a final state are in a receiving state waiting for messages that cannot be read from the buffer; moreover if  $\vec{q}$ is final all buffers are empty.\\
%We say that $S$ is {\em deadlock-free} whenever, for any $s\in \RS(S)$, $s$ is not a  deadlock configuration.

\item
$s$ is an {\em  orphan-message  configuration} of $S$ if \hspace{2mm}
$\forall \ttp\in\roles. ~ q_\ttp \text{ is final} \quad\text{and}\quad  \vec{w}\neq \vec{\varepsilon}$.\\
I.e. each machine is in a final state, but there is still  at least one non-empty buffer.
We say that $S$ is {\em orphan-message free} whenever, for any $s\in \RS(S)$, $s$ is not an orphan-message configuration.

\item
\label{def:safeness-ur}
$s$ is an {\em unspecified reception configuration} of $S$  if ~$\exists \ttr \in\roles$ such that  
\begin{enumerate}[a)]
\item
%$\exists \ttr \in\roles. ~ 
$q_\ttr \text{ is a receiving state}$; and
\item
$\forall\tts\in\roles.[~(q_\ttr,\tts\ttr?\msg[a],q'_\ttr)\in\delta_\ttr  \implies
(|w_{\tts\ttr}| > 0~~\wedge~~ w_{\tts\ttr}\not\in  \msg[a]\cdot\mathbb{A}^*)~ ]$.
\end{enumerate}
I.e. there is a receiving  state $q_\ttr$ 
which is prevented from
receiving any message from any of its buffers.
(In other words, in each channel $\tts\ttr$ from which participant $\ttr$ could consume, there
is a message which cannot be received by $\ttr$ in state $q_\ttr$.)
We say that $S$ is {\em reception-error free} whenever, for any $s\in \RS(S)$, $s$ is not an unspecified reception configuration.
\item
\label{def:progress-i}
$S$ satisfies the {\em progress property} if for all $s= (\vec{q},\vec{w}) \in \RS(S)$, either there exists $s'$ such that $s\lts{} s'$
or $~\forall \ttp\in\roles. ~ q_\ttp \text{ is final}$. 
\Commented{%<<<<<<<<<<<<<<<<<<<<<<<<<< BEGIN-COMMENTED
\item
\label{def:lock-freedom}
$s$ is a $\ttp$-{\em lock configuration} of $S$ if $\ttp\in\roles$ and such that
\begin{enumerate}[a)]
\item
$q_{\ttp}$ is a receiving state; and
\item 
 $\ttp$ does not appear as subject in any label of any transition sequence from $s$.
\end{enumerate}
I.e. $\ttp$ remains stuck in all possible transition sequences from $s$.
We say that $S$ is {\em lock-free} whenever, for each $\ttp\in\roles$ and each $s\in \RS(S)$, $s$ is not a $\ttp$-lock configuration.
}%<<<<<<<<<<<<<<<<<<<<<<<<<<<<<<<<< END-COMMENTED
\end{enumerate}
\end{definition}
Note that the progress property (\cref{def:progress-i}) implies deadlock freedom, but the converse is not true. For example, consider a non-final stuck configuration with a non-empty buffer.
\Commented{%<<<<<<<<<<<<<<<<<<<
Furthermore, any unspecified reception configuration trivially constitutes a $\ttp$-lock for some $\ttp$, which implies that lock freedom entails reception-error freedom.
It is also easy to show that lock freedom implies both deadlock freedom and progress.
The remaining properties are mutually independent.
}%<<<<<<<<<<<<<<<<<<<<<<<<<<<
%\brc I believe that lock-freedom also implies progress.
%Then we have that lock-freedom implies everything but orphan-message freedom
%and the converse does also not hold. Also we know from our previous paper that
%all other properties are mutually independent which then should also hold here.
%But before we must say that  lock-freedom also implies progress if you agree.
%\erc
 Communication properties above are essentially as presented in~\cite{BZ83,CF05,DY12,TY15}. 

%The above definitions of communication properties (\ref{def:safeness-i})--(\ref{def:progress-i}) are the same as the properties considered in~\cite{DY12},
%though the above formulation of progress is slightly simpler but equivalent to the one in~\cite{DY12}.
%The notions of orphan message and unspecified reception are also the same as in~\cite{TY15}.
%The same notions of deadlock and unspecified reception are given in~\cite{CF05} and inspired by~\cite{BZ83}. The deadlock notions in~\cite{BZ83} and~\cite{TY15} coincide with~\cite{CF05} and~\cite{DY12} if the local CFSMs have no final states. Otherwise, deadlock in~\cite{TY15} is weaker than deadlock above.
%A still weaker notion of deadlock configuration, and hence a stronger notion of deadlock-freedom, has been suggested in~\cite{TG18}. 
%This deadlock notion has been formally related to the above 
%communication properties in~\cite{BdLH19}.

%\brc
%A further comment: If we can save enough space, I would so much
%prefer to move the definition of projection and~\cref{lem:nohatrestrict}
%to the main part of the paper as a hint for the most important proposition in our proofs.
%\erc

%To distinguish it from the notion above, we call it \emph{strong deadlock-freedom}, as done in \cite{BdLH19}.

\section{PaI composition via partial gateways}
\label{sec:cpg}

In this section we formally define the PaI composition of CFSM systems via partial gateways.
As mentioned before,
for the sake of simplicity and space motivations, we only consider the binary case. 
However, the definitions we provide naturally scale up to the composition of an arbitrary number of CFSM systems.
To introduce and discuss the main issues, we use two communicating systems that implement the communications of the following systems as a running example.\\[-4mm]
\begin{quote}
Systems, $S_1$ and $S_2$ describe two simple home automation protocols.
Participant $\hh_1$ in  $S_1$ is an autonomous robot that performs various housekeeping tasks once activated. 
It also functions as a humidifier. 
In order to perform this task, the robot remote controller $\pr$ must send either the humidity parameter or the message $\msg[sbs]$ (set-by-sensor), indicating that the parameter is set by the robot's own sensor rule rather than the controller.
System $S_2$, on the other hand, is a humidification system. 
Through the remote controller $\hh_2$, one can instruct the controller $\tts$ (through message $\msg[hum]$) to compute the humidity parameter or to set it according to the humidifier's sensor (through message $\msg[sbs]$). 
If the parameter is computed by the controller, the actual humidity level ($\msg[deg]$) is sent back to the controller for display after a certain amount of time. Alternatively, one can send a parameter manually set by the user ($\msg[mus]$). \\[-4mm]
\end{quote}
For enhanced readability, all parts relating to our running example will be delimited by 
``$\blacktriangleright$'' and ``$\blacktriangleleft$''.\\
\brunningex 
Focusing only on communications and abstracting from the actual values carried by the messages, the behaviour of the systems $S_1$ 
and $S_2$ previously described can be represented by communicating systems, as follows.\\[-5mm]
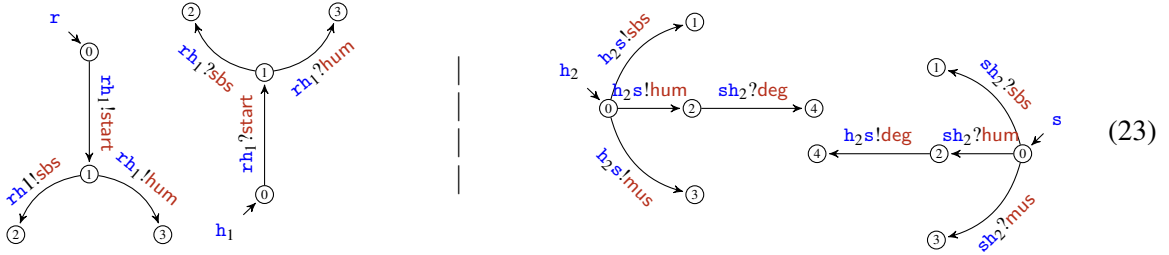
\begin{equation}
\label{eq:runex}
\begin{array}{c@{\qquad}c@{\hspace{1cm}}c@{\hspace{-4mm}}c}
    \begin{array}{c@{\hspace{-4mm}}c}
      \begin{tikzpicture}[mycfsm]
   \node[state]           (0)                        {$0$};
   \node[draw=none,fill=none] (start) [above left = 0.3cm  of 0]{$\ttr$};
   \node[state]            (1) [below of=0] {$1$};
   \node[state]            (2) [below left of=1, yshift=4mm,xshift=2mm] {$2$};
   \node[state]            (3) [below right of=1, yshift=4mm,xshift=-2mm] {$3$};
   \path  (start) edge node {} (0)
            (0)  edge    node [above] {$\ttr\hh_1!\msg[start]$} (1) 
            (1)  edge[bend right]    node [above] {$\ttr\hh1!\msg[sbs]$} (2)
            (1)  edge[bend left]    node [above] {$\ttr\hh_1!\msg[hum]$} (3) 
            ;
       \end{tikzpicture}
&
      \begin{tikzpicture}[mycfsm]
   \node[state]           (0)                        {$0$};
   \node[draw=none,fill=none] (start) [below left = 0.3cm  of 0]{$\hh_1$};
   \node[state]            (1) [above of=0] {$1$};
   \node[state]            (2) [above left of=1, yshift=-4mm,xshift=2mm] {$2$};
   \node[state]            (3) [above right of=1, yshift=-4mm,xshift=-2mm] {$3$};
   \path  (start) edge node {} (0)
            (0)  edge                    node [above] {$\ttr\hh_1?\msg[start]$} (1) 
            (1)  edge[bend left]    node [below] {$\ttr\hh_1?\msg[sbs]$} (2)
            (1)  edge[bend right]    node [below] {$\ttr\hh_1?\msg[hum]$} (3) 
            ;
       \end{tikzpicture}
    \end{array}
       &
       \begin{array}{c}
       |\\
       |\\
       |\\
       |
       \end{array}
       &
      \raisebox{3mm}{\begin{tikzpicture}[mycfsm]
  \node[state]           (0)              {$0$};
   \node[draw=none,fill=none] (start) [above left = 0.3cm  of 0]{$\hh_2$};
  \node[state]            (1) [above right of=0] {$1$};
   \node[state]           (2) [right of=0,xshift=-6mm] {$2$};
   \node[state]           (3) [below right of=0] {$3$};
   \node[state]           (4) [right of=2] {$4$};
   \path  (start) edge node {} (0) 
            (0)  edge     [bend left]      node [above] {$\hh_2\tts!\msg[sbs]$} (1)
                   edge                          node [above]  {$\hh_2\tts!\msg[hum]$} (2)
                   edge    [bend right]     node [below]  {$\hh_2\tts!\msg[mus]$} (3)
            (2)  edge                           node [above]  {$\tts\hh_2?\msg[deg]$} (4)
                   ;
       \end{tikzpicture}
        }
&
      \raisebox{-3mm}{ \begin{tikzpicture}[mycfsm]
  \node[state]           (0)            {$0$};
   \node[draw=none,fill=none] (start) [above right = 0.3cm  of 0]{$\tts$};
  \node[state]            (1) [above left of=0] {$1$};
   \node[state]           (2) [left of=0,xshift=6mm] {$2$};
   \node[state]           (3) [below left of=0] {$3$};
   \node[state]           (4) [left of=2] {$4$};
   \path  (start) edge node {} (0) 
            (0)  edge     [bend right]      node [above] {$\tts\hh_2?\msg[sbs]$} (1)
                   edge                          node [above]  {$\tts\hh_2?\msg[hum]$} (2)
                   edge    [bend left]     node [below]  {$\tts\hh_2?\msg[mus]$} (3)
            (2)  edge                           node [above]  {$\hh_2\tts!\msg[deg]$} (4)
                   ;
       \end{tikzpicture}
       }
\end{array}
\vspace{-2mm}
\end{equation}
\\[-2mm]
\noindent
The standard PaI composition \cite{BH24} of $S_1$ and $S_2$  using, respectively,
$\hh_1$ and $\hh_2$ as interfaces, 
requires the interfaces to be replaced by forwarders, the gateways.
In the present case, this replacement would be completely
\Commented{
The (unique) connection policy below is non deadlock-free. (FALSE)
$$
\dbox{
     \begin{tikzpicture}[mycfsm]
   \node[state]           (0)                        {$0$};
   \node[draw=none,fill=none] (start) [below left = 0.3cm  of 0]{$\hh_1$};
   \node[state]            (1) [above of=0] {$1$};
   \node[state]            (2) [above left of=1, yshift=-4mm,xshift=2mm] {$2$};
   \node[state]            (3) [above right of=1, yshift=-4mm,xshift=-2mm] {$3$};
   \path  (start) edge node {} (0)
            (0)  edge                    node [above] {$\hh_1\hh_2!\msg[start]$} (1) 
            (1)  edge[bend left]    node [below] {$\hh_1\hh_2!\msg[sbs]$} (2)
            (1)  edge[bend right]    node [below] {$\hh_1\hh_2!\msg[hum]$} (3) 
            ;
       \end{tikzpicture}
       \qquad
     \begin{tikzpicture}[mycfsm]
  \node[state]           (0)              {$0$};
   \node[draw=none,fill=none] (start) [above left = 0.3cm  of 0]{$\hh_2$};
  \node[state]            (1) [above right of=0] {$1$};
   \node[state]           (2) [right of=0,xshift=-6mm] {$2$};
   \node[state]           (3) [below right of=0] {$3$};
   \node[state]           (4) [right of=2] {$4$};
   \path  (start) edge node {} (0) 
            (0)  edge     [bend left]      node [above] {$\hh_1\hh_2?\msg[sbs]$} (1)
                   edge                          node [above]  {$\hh_1\hh_2?\msg[hum]$} (2)
                   edge    [bend right]     node [below]  {$\hh_1\hh_2?\msg[mus]$} (3)
            (2)  edge                           node [above]  {$\hh_2\hh_1!\msg[deg]$} (4)
                   ;
       \end{tikzpicture}
}
$$
}
%However, using the standard PaI composition in the present case would not be very
unreasonable, since it would mean to replace the entire robot with a gateway.
A more reasonable approach would be to keep the robot as it is and use, for humidification, the more sophisticated humidification tool in $S_2$.
To achieve this, rather than considering entire participants as interfaces, we can consider only some specific transitions as description of the behaviour of external systems.
In the present example, for instance, we could still decide to compose $S_1$ and $S_2$ through 
$\hh_1$ and $\hh_2$.
However, when interpreting the diagram (\ref{eq:1}) below from the perspective of $S_1$ and $S_2$, only the bold edges should be considered as actions belonging to an outer system.\\[-4mm]  
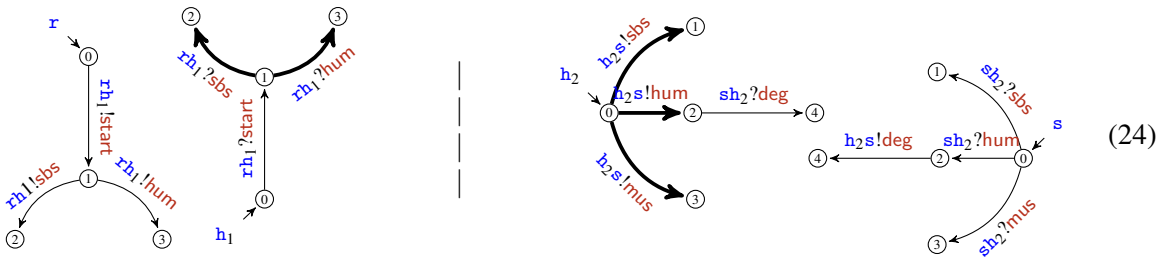
\begin{equation}
\label{eq:1}
\vspace{-0mm}
\begin{array}{c@{\qquad}c@{\hspace{1cm}}c@{\hspace{-4mm}}c}
    \begin{array}{c@{\hspace{-4mm}}c}
      \begin{tikzpicture}[mycfsm]
   \node[state]           (0)                        {$0$};
   \node[draw=none,fill=none] (start) [above left = 0.3cm  of 0]{$\ttr$};
   \node[state]            (1) [below of=0] {$1$};
   \node[state]            (2) [below left of=1, yshift=4mm,xshift=2mm] {$2$};
   \node[state]            (3) [below right of=1, yshift=4mm,xshift=-2mm] {$3$};
   \path  (start) edge node {} (0)
            (0)  edge    node [above] {$\ttr\hh_1!\msg[start]$} (1) 
            (1)  edge[bend right]    node [above] {$\ttr\hh1!\msg[sbs]$} (2)
            (1)  edge[bend left]    node [above] {$\ttr\hh_1!\msg[hum]$} (3) 
            ;
       \end{tikzpicture}
&
      \begin{tikzpicture}[mycfsm]
   \node[state]           (0)                        {$0$};
   \node[draw=none,fill=none] (start) [below left = 0.3cm  of 0]{$\hh_1$};
   \node[state]            (1) [above of=0] {$1$};
   \node[state]            (2) [above left of=1, yshift=-4mm,xshift=2mm] {$2$};
   \node[state]            (3) [above right of=1, yshift=-4mm,xshift=-2mm] {$3$};
   \path  (start) edge node {} (0)
            (0)  edge                    node [above] {$\ttr\hh_1?\msg[start]$} (1) 
            (1)  edge[bend left, line width=0.5mm]    node [below] {$\ttr\hh_1?\msg[sbs]$} (2)
            (1)  edge[bend right, line width=0.5mm]    node [below] {$\ttr\hh_1?\msg[hum]$} (3) 
            ;
       \end{tikzpicture}
    \end{array}
       &
       \begin{array}{c}
       |\\
       |\\
       |\\
       |
       \end{array}
       &
      \raisebox{3mm}{\begin{tikzpicture}[mycfsm]
  \node[state]           (0)              {$0$};
   \node[draw=none,fill=none] (start) [above left = 0.3cm  of 0]{$\hh_2$};
  \node[state]            (1) [above right of=0] {$1$};
   \node[state]           (2) [right of=0,xshift=-6mm] {$2$};
   \node[state]           (3) [below right of=0] {$3$};
   \node[state]           (4) [right of=2] {$4$};
   \path  (start) edge node {} (0) 
            (0)  edge     [bend left, line width=0.5mm]      node [above] {$\hh_2\tts!\msg[sbs]$} (1)
                   edge     [line width=0.5mm]                     node [above]  {$\hh_2\tts!\msg[hum]$} (2)
                   edge    [bend right, line width=0.5mm]     node [below]  {$\hh_2\tts!\msg[mus]$} (3)
            (2)  edge                           node [above]  {$\tts\hh_2?\msg[deg]$} (4)
                   ;
       \end{tikzpicture}
        }
&
      \raisebox{-3mm}{ \begin{tikzpicture}[mycfsm]
  \node[state]           (0)            {$0$};
   \node[draw=none,fill=none] (start) [above right = 0.3cm  of 0]{$\tts$};
  \node[state]            (1) [above left of=0] {$1$};
   \node[state]           (2) [left of=0,xshift=6mm] {$2$};
   \node[state]           (3) [below left of=0] {$3$};
   \node[state]           (4) [left of=2] {$4$};
   \path  (start) edge node {} (0) 
            (0)  edge     [bend right]      node [above] {$\tts\hh_2?\msg[sbs]$} (1)
                   edge                          node [above]  {$\tts\hh_2?\msg[hum]$} (2)
                   edge    [bend left]     node [below]  {$\tts\hh_2?\msg[mus]$} (3)
            (2)  edge                           node [above]  {$\hh_2\tts!\msg[deg]$} (4)
                   ;
       \end{tikzpicture}
       }
\end{array}
\end{equation}
Intuitively, by considering the bold edges above as interface transitions expresses the intention of replacing the robot's humidification task with the function of $S_2$. 
To achieve this, only messages relating to these transitions need to be forwarded.
A possible connection policy should be then represented as a CFSM system that only deals with the interactions between the 'interface parts' of $\hh_1$ and $\hh_2$, as shown below.
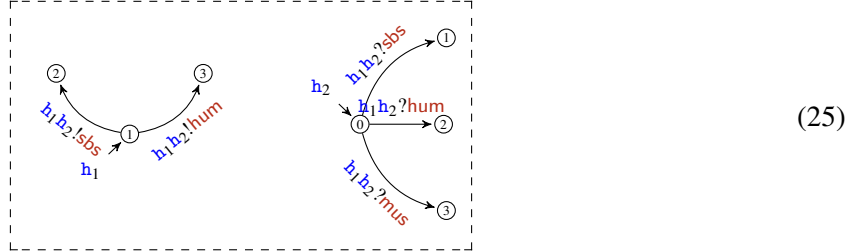
\begin{equation}
\label{eq:cp1}
\dbox{
     \begin{tikzpicture}[mycfsm]
   \node[state]            (1) [above of=0] {$1$};
   \node[draw=none,fill=none] (start) [below left = 0.3cm  of 1]{$\hh_1$};
   \node[state]            (2) [above left of=1, yshift=-4mm,xshift=2mm] {$2$};
   \node[state]            (3) [above right of=1, yshift=-4mm,xshift=-2mm] {$3$};
   \path  (start) edge node {} (1)
            (1)  edge[bend left]    node [below] {$\hh_1\hh_2!\msg[sbs]$} (2)
            (1)  edge[bend right]    node [below] {$\hh_1\hh_2!\msg[hum]$} (3) 
            ;
       \end{tikzpicture}
       \qquad
     \begin{tikzpicture}[mycfsm]
  \node[state]           (0)              {$0$};
   \node[draw=none,fill=none] (start) [above left = 0.3cm  of 0]{$\hh_2$};
  \node[state]            (1) [above right of=0] {$1$};
   \node[state]           (2) [right of=0,xshift=-6mm] {$2$};
   \node[state]           (3) [below right of=0] {$3$};
   \path  (start) edge node {} (0) 
            (0)  edge     [bend left]      node [above] {$\hh_1\hh_2?\msg[sbs]$} (1)
                   edge                          node [above]  {$\hh_1\hh_2?\msg[hum]$} (2)
                   edge    [bend right]     node [below]  {$\hh_1\hh_2?\msg[mus]$} (3)
                   ;
       \end{tikzpicture}
}
\end{equation}
Such a communicating system  enjoys all the communication properties of \cref{def:safeness},
even if the $\hh_1$ and $\hh_2$ are not perfectly complementary, in particular even if 
state $3$ of $\hh_2$ cannot belong to any reachable configuration.
\erunningex\\
In our envisaged approach, the composition would be achieved by replacing {\em participants with interface transitions} with {\em partial gateways} that forward only the interface parts' messages.
The non-interface parts would continue to describe what the participants used for the connection are still in charge of.
Each partial gateway is therefore constructed in a similar way to the standard composition via gateways, using the CFSMs of the interface participants and those forming the connection policy.\\
\brunningex
In our example, out of the interface transition
\raisebox{2mm}
{\begin{tikzpicture}[mycfsm]
      \node[state] (zero) [yshift=-4mm] {$0$};
      \node[state] (one) [right of=zero, xshift=-2mm]   {$1$};
      \path
      (zero) edge[bend left=15, line width=0.5mm] node[above] {$\aout[h_2][s][][sbs]$} (one)
      ;
 \end{tikzpicture}
 } 
 of $\hh_2$ in (\ref{eq:1}) and of the transition
 \raisebox{2mm}
{\begin{tikzpicture}[mycfsm]
      \node[state] (zero) [yshift=-4mm] {$0$};
      \node[state] (one) [right of=zero, xshift=-2mm]   {$1$};
      \path
      (zero) edge[bend left=15] node[above] {$\ain[\hh_1][\hh_2][][sbs]$} (one)
      ;
 \end{tikzpicture}
 }  
 of $\hh_2$ in the connection policy (\ref{eq:cp1}) , we introduce 
  \raisebox{2mm}
{\begin{tikzpicture}[mycfsm]
      \node[state] (zero) [yshift=-4mm] {$0$};
      \node[state] (one) [right of=zero, xshift=-2mm, yshift=2mm]   {$\widehat 1$};
       \node[state] (two) [right of=one, xshift=-2mm, yshift=-2mm,]   {$1$};
      \path
      (zero) edge[bend left=10] node[above] {$\ain[h_1][h_2][][sbs]$} (one)
      (one) edge[bend left=10] node[above] {$\aout[\hh_2][s][][sbs]$} (two)
      ;
 \end{tikzpicture}
 }
in the resulting partial gateway, where $\widehat 1$ is specifically introduced 
by the partial gateway construction.
Note that, to enforce {\em conservativity} \cite{BH24} -- that is, to minimise the impact of the composition on the systems involved -- partial gateways are given the same names as the corresponding participants with interface transitions.
The above discussion applies, dually, for the interface transitions of $\hh_1$ labelled with output actions.
Non-interface transitions are instead left unchanged by the partial gateway construction.
So, the resulting composed system in our running example is as follows.\\[-2mm]
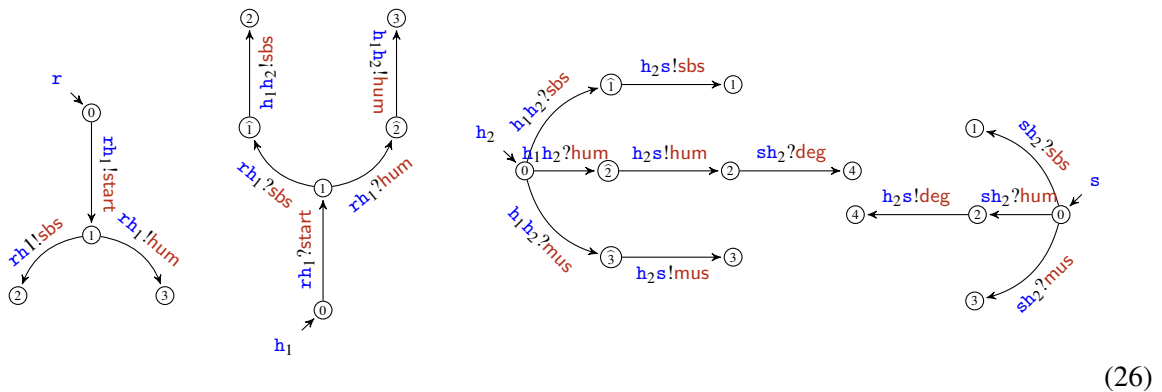
\begin{equation}
\label{eq:fincomp}
\begin{array}{cc@{\hspace{-4mm}}c}
    \begin{array}{cc}
      \begin{tikzpicture}[mycfsm]
   \node[state]           (0)                        {$0$};
   \node[draw=none,fill=none] (start) [above left = 0.3cm  of 0]{$\ttr$};
   \node[state]            (1) [below of=0] {$1$};
   \node[state]            (2) [below left of=1, yshift=4mm,xshift=2mm] {$2$};
   \node[state]            (3) [below right of=1, yshift=4mm,xshift=-2mm] {$3$};
   \path  (start) edge node {} (0)
            (0)  edge    node [above] {$\ttr\hh_1!\msg[start]$} (1) 
            (1)  edge[bend right]    node [above] {$\ttr\hh1!\msg[sbs]$} (2)
            (1)  edge[bend left]    node [above] {$\ttr\hh_1!\msg[hum]$} (3) 
            ;
       \end{tikzpicture}
&
      \begin{tikzpicture}[mycfsm]
   \node[state]           (0)                        {$0$};
   \node[draw=none,fill=none] (start) [below left = 0.3cm  of 0]{$\hh_1$};
   \node[state]            (1) [above of=0] {$1$};
   \node[state]            (1hat) [above left of=1, yshift=-4mm,xshift=2mm] {$\widehat 1$};
   \node[state]            (2) [above of=1hat, yshift=-2mm] {$2$};
   \node[state]            (2hat) [above right of=1, yshift=-4mm,xshift=-2mm] {$\widehat 2$};
   \node[state]            (3) [above of=2hat, yshift=-2mm] {$3$};
   \path  (start) edge node {} (0)
            (0)  edge                    node [above] {$\ttr\hh_1?\msg[start]$} (1) 
            (1)  edge[bend left]    node [below] {$\ttr\hh_1?\msg[sbs]$} (1hat)
             (1hat)  edge   node [below] {$\hh_1\hh_2!\msg[sbs]$} (2)
            (1)  edge[bend right]    node [below] {$\ttr\hh_1?\msg[hum]$} (2hat) 
             (2hat)  edge   node [below] {$\hh_1\hh_2!\msg[hum]$} (3) 
            ;
       \end{tikzpicture}
    \end{array}
  &
      \raisebox{3mm}{\begin{tikzpicture}[mycfsm]
  \node[state]           (0)              {$0$};
   \node[draw=none,fill=none] (start) [above left = 0.3cm  of 0]{$\hh_2$};
  \node[state]            (1hat) [above right of=0] {$\widehat 1$};
    \node[state]            (1) [right of=1hat] {$1$};
   \node[state]           (2hat) [right of=0,xshift=-6mm] {$\widehat 2$};
    \node[state]           (2) [right of=2hat] {$2$};
   \node[state]           (3hat) [below right of=0] {$\widehat 3$};
   \node[state]           (3) [ right of=3hat] {$3$};
   \node[state]           (4) [right of=2] {$4$};
   \path  (start) edge node {} (0) 
            (0)  edge     [bend left]      node [above] {$\hh_1\hh_2?\msg[sbs]$} (1hat)
                   edge                          node [above]  {$\hh_1\hh_2?\msg[hum]$} (2hat)
                   edge    [bend right]     node [below]  {$\hh_1\hh_2?\msg[mus]$} (3hat)
            (3hat)  edge                      node [below]  {$\hh_2\tts!\msg[mus]$} (3)
            (1hat)  edge                      node [above]  {$\hh_2\tts!\msg[sbs]$} (1)
            (2hat)  edge                      node [above]  {$\hh_2\tts!\msg[hum]$} (2)
            (2)  edge                           node [above]  {$\tts\hh_2?\msg[deg]$} (4)
                   ;
       \end{tikzpicture}
        }
&
      \raisebox{-3mm}{ \begin{tikzpicture}[mycfsm]
  \node[state]           (0)            {$0$};
   \node[draw=none,fill=none] (start) [above right = 0.3cm  of 0]{$\tts$};
  \node[state]            (1) [above left of=0] {$1$};
   \node[state]           (2) [left of=0,xshift=6mm] {$2$};
   \node[state]           (3) [below left of=0] {$3$};
   \node[state]           (4) [left of=2] {$4$};
   \path  (start) edge node {} (0) 
            (0)  edge     [bend right]      node [above] {$\tts\hh_2?\msg[sbs]$} (1)
                   edge                          node [above]  {$\tts\hh_2?\msg[hum]$} (2)
                   edge    [bend left]     node [below]  {$\tts\hh_2?\msg[mus]$} (3)
            (2)  edge                           node [above]  {$\hh_2\tts!\msg[deg]$} (4)
                   ;
       \end{tikzpicture}
       }
\end{array}
\end{equation}
Our results guarantee that the above PaI composition via partial gateways
satisfies the same communication properties (among those of \cref{def:safeness}) 
satisfied by $S_1$ in $S_2$ in (\ref{eq:runex}) and the connection policy in (\ref{eq:cp1}). 
\erunningex

In general, once two interface participants have been chosen, some care must be taken when deciding which of their transitions to consider as interface transitions.
Not all transitions are suitable for this purpose.  That is why in \cref{def:IDM} below we define the set
of the {\em interface decorations} of a CFSM, where an interface decoration corresponds to a possible suitable  selection of  interface transitions preventing issues like those we discuss below. \\
\brunningex
Consider again $\hh_1$ and 
$\hh_2$ of our running example, in particular  the following interface transitions.
 \vspace{-2mm}
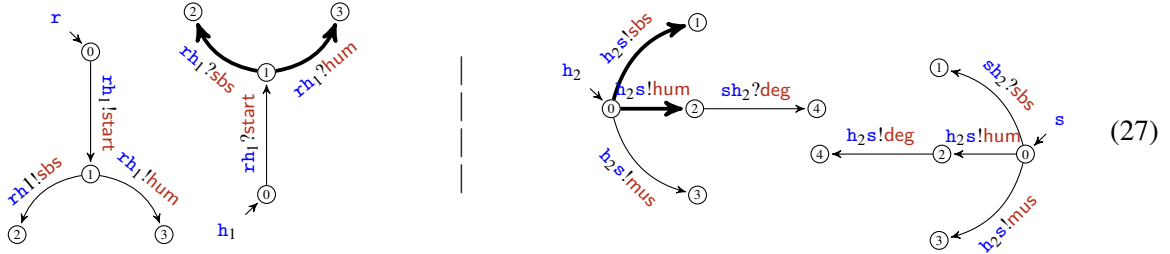
 \begin{equation}
 \label{eq:2}
\begin{array}{c@{\qquad}c@{\hspace{1cm}}c@{\hspace{-4mm}}c}
    \begin{array}{c@{\hspace{-4mm}}c}
      \begin{tikzpicture}[mycfsm]
   \node[state]           (0)                        {$0$};
   \node[draw=none,fill=none] (start) [above left = 0.3cm  of 0]{$\ttr$};
   \node[state]            (1) [below of=0] {$1$};
   \node[state]            (2) [below left of=1, yshift=4mm,xshift=2mm] {$2$};
   \node[state]            (3) [below right of=1, yshift=4mm,xshift=-2mm] {$3$};
   \path  (start) edge node {} (0)
            (0)  edge    node [above] {$\ttr\hh_1!\msg[start]$} (1) 
            (1)  edge[bend right]    node [above] {$\ttr\hh1!\msg[sbs]$} (2)
            (1)  edge[bend left]    node [above] {$\ttr\hh_1!\msg[hum]$} (3) 
            ;
       \end{tikzpicture}
&
      \begin{tikzpicture}[mycfsm]
   \node[state]           (0)                        {$0$};
   \node[draw=none,fill=none] (start) [below left = 0.3cm  of 0]{$\hh_1$};
   \node[state]            (1) [above of=0] {$1$};
   \node[state]            (2) [above left of=1, yshift=-4mm,xshift=2mm] {$2$};
   \node[state]            (3) [above right of=1, yshift=-4mm,xshift=-2mm] {$3$};
   \path  (start) edge node {} (0)
            (0)  edge                    node [above] {$\ttr\hh_1?\msg[start]$} (1) 
            (1)  edge[bend left, line width=0.5mm]    node [below] {$\ttr\hh_1?\msg[sbs]$} (2)
            (1)  edge[bend right, line width=0.5mm]    node [below] {$\ttr\hh_1?\msg[hum]$} (3) 
            ;
       \end{tikzpicture}
    \end{array}
       &
       \begin{array}{c}
       |\\
       |\\
       |\\
       |
       \end{array}
       &
      \raisebox{3mm}{\begin{tikzpicture}[mycfsm]
  \node[state]           (0)              {$0$};
   \node[draw=none,fill=none] (start) [above left = 0.3cm  of 0]{$\hh_2$};
  \node[state]            (1) [above right of=0] {$1$};
   \node[state]           (2) [right of=0,xshift=-6mm] {$2$};
   \node[state]           (3) [below right of=0] {$3$};
   \node[state]           (4) [right of=2] {$4$};
   \path  (start) edge node {} (0) 
            (0)  edge     [bend left, line width=0.5mm]      node [above] {$\hh_2\tts!\msg[sbs]$} (1)
                   edge     [line width=0.5mm]                     node [above]  {$\hh_2\tts!\msg[hum]$} (2)
                   edge    [bend right]     node [below]  {$\hh_2\tts!\msg[mus]$} (3)
            (2)  edge                           node [above]  {$\tts\hh_2?\msg[deg]$} (4)
                   ;
       \end{tikzpicture}
        }
&
      \raisebox{-3mm}{ \begin{tikzpicture}[mycfsm]
  \node[state]           (0)            {$0$};
   \node[draw=none,fill=none] (start) [above right = 0.3cm  of 0]{$\tts$};
  \node[state]            (1) [above left of=0] {$1$};
   \node[state]           (2) [left of=0,xshift=6mm] {$2$};
   \node[state]           (3) [below left of=0] {$3$};
   \node[state]           (4) [left of=2] {$4$};
   \path  (start) edge node {} (0) 
            (0)  edge     [bend right]      node [above] {$\tts\hh_2?\msg[sbs]$} (1)
                   edge                          node [above]  {$\hh_2\tts!\msg[hum]$} (2)
                   edge    [bend left]     node [below]  {$\hh_2\tts!\msg[mus]$} (3)
            (2)  edge                           node [above]  {$\hh_2\tts!\msg[deg]$} (4)
                   ;
       \end{tikzpicture}
       }
\end{array}
\end{equation}
The corresponding connection policy would now be 
$$
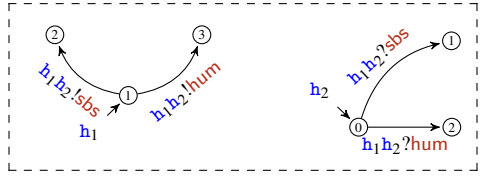
\dbox{
     \begin{tikzpicture}[mycfsm]
   \node[state]            (1) [above of=0] {$1$};
   \node[draw=none,fill=none] (start) [below left = 0.3cm  of 1]{$\hh_1$};
   \node[state]            (2) [above left of=1, yshift=-4mm,xshift=2mm] {$2$};
   \node[state]            (3) [above right of=1, yshift=-4mm,xshift=-2mm] {$3$};
   \path  (start) edge node {} (1)
            (1)  edge[bend left]    node [below] {$\hh_1\hh_2!\msg[sbs]$} (2)
            (1)  edge[bend right]    node [below] {$\hh_1\hh_2!\msg[hum]$} (3) 
            ;
       \end{tikzpicture}
       \qquad
     \begin{tikzpicture}[mycfsm]
  \node[state]           (0)              {$0$};
   \node[draw=none,fill=none] (start) [above left = 0.3cm  of 0]{$\hh_2$};
  \node[state]            (1) [above right of=0, xshift=1mm] {$1$};
   \node[state]           (2) [right of=0, xshift=-4.5mm] {$2$};
   \path  (start) edge node {} (0) 
            (0)  edge     [bend left]      node [above] {$\hh_1\hh_2?\msg[sbs]$} (1)
                   edge                          node [below]  {$\hh_1\hh_2?\msg[hum]$} (2)
                   ;
       \end{tikzpicture}
}
$$
This communicating system, as well as $S_1$ and $S_2$, is orphan-message free.
 The composition obtained by building partial gateways out of such connection policy is then
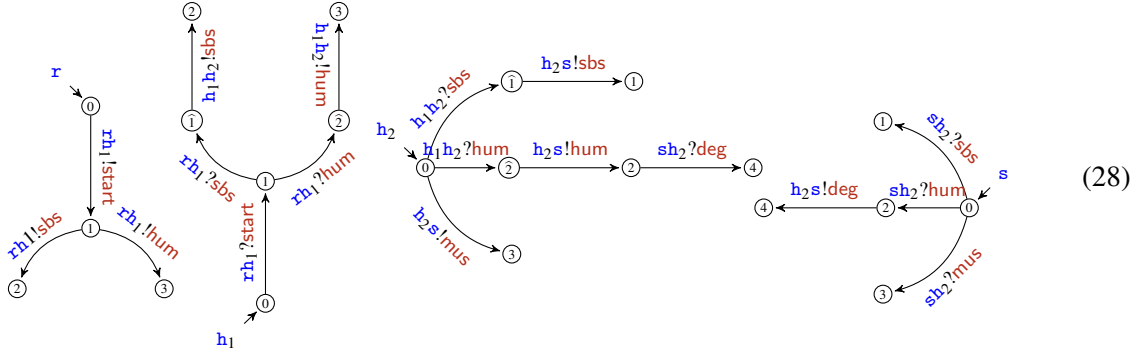
\begin{equation}
\label{eq:comprunex}
\begin{array}{c@{\hspace{-2mm}}c@{\hspace{-3mm}}c}
    \begin{array}{c@{\hspace{-4mm}}c}
      \begin{tikzpicture}[mycfsm]
   \node[state]           (0)                        {$0$};
   \node[draw=none,fill=none] (start) [above left = 0.3cm  of 0]{$\ttr$};
   \node[state]            (1) [below of=0] {$1$};
   \node[state]            (2) [below left of=1, yshift=4mm,xshift=2mm] {$2$};
   \node[state]            (3) [below right of=1, yshift=4mm,xshift=-2mm] {$3$};
   \path  (start) edge node {} (0)
            (0)  edge    node [above] {$\ttr\hh_1!\msg[start]$} (1) 
            (1)  edge[bend right]    node [above] {$\ttr\hh1!\msg[sbs]$} (2)
            (1)  edge[bend left]    node [above] {$\ttr\hh_1!\msg[hum]$} (3) 
            ;
       \end{tikzpicture}
&
      \begin{tikzpicture}[mycfsm]
   \node[state]           (0)                        {$0$};
   \node[draw=none,fill=none] (start) [below left = 0.3cm  of 0]{$\hh_1$};
   \node[state]            (1) [above of=0] {$1$};
   \node[state]            (1hat) [above left of=1, yshift=-4mm,xshift=2mm] {$\widehat 1$};
   \node[state]            (2) [above of=1hat, yshift=-2mm] {$2$};
   \node[state]            (2hat) [above right of=1, yshift=-4mm,xshift=-2mm] {$\widehat 2$};
   \node[state]            (3) [above of=2hat, yshift=-2mm] {$3$};
   \path  (start) edge node {} (0)
            (0)  edge                    node [above] {$\ttr\hh_1?\msg[start]$} (1) 
            (1)  edge[bend left]    node [below] {$\ttr\hh_1?\msg[sbs]$} (1hat)
             (1hat)  edge   node [below] {$\hh_1\hh_2!\msg[sbs]$} (2)
            (1)  edge[bend right]    node [below] {$\ttr\hh_1?\msg[hum]$} (2hat) 
             (2hat)  edge   node [below] {$\hh_1\hh_2!\msg[hum]$} (3) 
            ;
       \end{tikzpicture}
    \end{array}
  &
      \raisebox{3mm}{\begin{tikzpicture}[mycfsm]
  \node[state]           (0)              {$0$};
   \node[draw=none,fill=none] (start) [above left = 0.3cm  of 0]{$\hh_2$};
  \node[state]            (1hat) [above right of=0] {$\widehat 1$};
    \node[state]            (1) [right of=1hat] {$1$};
   \node[state]           (2hat) [right of=0,xshift=-6mm] {$\widehat 2$};
    \node[state]           (2) [right of=2hat] {$2$};
   \node[state]           (3) [below right of=0] {$3$};
   \node[state]           (4) [right of=2] {$4$};
   \path  (start) edge node {} (0) 
            (0)  edge     [bend left]      node [above] {$\hh_1\hh_2?\msg[sbs]$} (1hat)
                   edge                          node [above]  {$\hh_1\hh_2?\msg[hum]$} (2hat)
                   edge    [bend right]     node [below]  {$\hh_2\tts!\msg[mus]$} (3)
            (1hat)  edge                      node [above]  {$\hh_2\tts!\msg[sbs]$} (1)
            (2hat)  edge                      node [above]  {$\hh_2\tts!\msg[hum]$} (2)
            (2)  edge                           node [above]  {$\tts\hh_2?\msg[deg]$} (4)
                   ;
       \end{tikzpicture}
        }
&
      \raisebox{-3mm}{ \begin{tikzpicture}[mycfsm]
  \node[state]           (0)            {$0$};
   \node[draw=none,fill=none] (start) [above right = 0.3cm  of 0]{$\tts$};
  \node[state]            (1) [above left of=0] {$1$};
   \node[state]           (2) [left of=0,xshift=6mm] {$2$};
   \node[state]           (3) [below left of=0] {$3$};
   \node[state]           (4) [left of=2] {$4$};
   \path  (start) edge node {} (0) 
            (0)  edge     [bend right]      node [above] {$\tts\hh_2?\msg[sbs]$} (1)
                   edge                          node [above]  {$\tts\hh_2?\msg[hum]$} (2)
                   edge    [bend left]     node [below]  {$\tts\hh_2?\msg[mus]$} (3)
            (2)  edge                           node [above]  {$\hh_2\tts!\msg[deg]$} (4)
                   ;
       \end{tikzpicture}
       }
\end{array}
\end{equation}
This communicating system, however, is not orphan-message free.
In fact the following configuration, made by final states only, is reachable:
$s=((3_\ttr,3_{\hh_1},3_{\hh_2},3_\tts),\vec{w})$,
where $\vec{w}\neq\vec{\varepsilon}$, in particular $w_{\hh_1\hh_2}=\langle\msg[hum]\rangle$.
Hence $s$ is an orphan-message configuration.
\erunningex

The problem in the above discussion arises from the fact that the resulting partial gateway $\hh_2$ is actually a CFSM with mixed states. 
However, issues can arise also  in the absence of mixed states.
%We now present an example that demonstrates how unrestricted selection of interface transitions can hinder progress preservation through composition, even in the partial gateways.
Consider the two following systems $S_1$ and $S_2$ we wish to compose through,
$\hh_1$ and $\hh_2$. Both the systems satisfy the progress property.\\[-3mm]
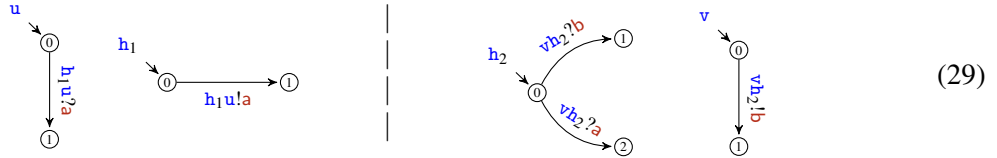
\begin{equation}
\label{eq:pre3}
\begin{array}{c@{\qquad}c@{\hspace{1cm}}c@{\qquad}c}
    \begin{array}{cc}
      \begin{tikzpicture}[mycfsm]
  \node[state]           (0)                        {$0$};
   \node[draw=none,fill=none] (start) [above left = 0.3cm  of 0]{$\ttu$};
   \node[state]            (1) [below of=0, yshift=4mm] {$1$};

   \path  (start) edge node {} (0)
            (0)  edge    node [above] {$\hh_1\ttu?\msg[a]$} (1) ;
       \end{tikzpicture}
&
       \begin{tikzpicture}[mycfsm]
  \node[state]           (0)                        {$0$};
   \node[draw=none,fill=none] (start) [above left = 0.3cm  of 0]{$\hh_1$};
  \node[state]            (1) [right of=0] {$1$};
  %\node[state]           (2) [above right of=0] {$2$};

   \path  (start) edge node {} (0) 
            (0)  edge   node [below] {$\hh_1\ttu!\msg[a]$} (1);
       \end{tikzpicture}
    \end{array}
       &
       \begin{array}{c}
       |\\
       |\\
       |\\
       |
       \end{array}
       &
       \begin{tikzpicture}[mycfsm]
  \node[state]           (0)                        {$0$};
   \node[draw=none,fill=none] (start) [above left = 0.3cm  of 0]{$\hh_2$};
  \node[state]            (1) [above right of=0,yshift=-5mm] {$1$};
  \node[state]           (2) [below right of=0,yshift=5mm] {$2$};

   \path  (start) edge node {} (0) 
            (0)  edge     [bend left]      node [above] {$\ttv\hh_2?\msg[b]$} (1)
            (0)   edge    [bend right]            node [above]  {$\ttv\hh_2?\msg[a]$} (2);
       \end{tikzpicture}
&
      \begin{tikzpicture}[mycfsm]
  \node[state]           (0)                        {$0$};
   \node[draw=none,fill=none] (start) [above left = 0.3cm  of 0]{$\ttv$};
   \node[state]            (1) [below of=0, yshift=4mm] {$1$};

   \path  (start) edge node {} (0)
            (0)  edge    node [above] {$\ttv\hh_2!\msg[b]$} (1) ;
       \end{tikzpicture}
\end{array}
\end{equation}
\\[-3mm]
We then make the following choice of interface transitions for the interface participants.\\[-3mm]
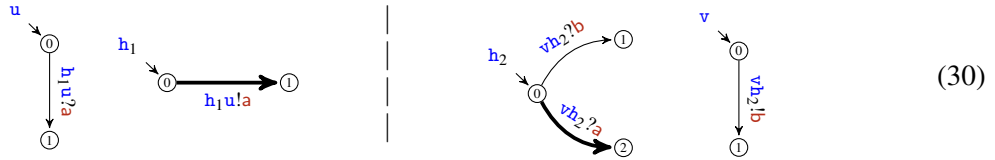
\begin{equation}
\label{eq:3}
\begin{array}{c@{\qquad}c@{\hspace{1cm}}c@{\qquad}c}
    \begin{array}{cc}
      \begin{tikzpicture}[mycfsm]
  \node[state]           (0)                        {$0$};
   \node[draw=none,fill=none] (start) [above left = 0.3cm  of 0]{$\ttu$};
   \node[state]            (1) [below of=0, yshift=4mm] {$1$};

   \path  (start) edge node {} (0)
            (0)  edge    node [above] {$\hh_1\ttu?\msg[a]$} (1) ;
       \end{tikzpicture}
&
       \begin{tikzpicture}[mycfsm]
  \node[state]           (0)                        {$0$};
   \node[draw=none,fill=none] (start) [above left = 0.3cm  of 0]{$\hh_1$};
  \node[state]            (1) [right of=0] {$1$};
  %\node[state]           (2) [above right of=0] {$2$};

   \path  (start) edge node {} (0) 
            (0)  edge [line width=0.5mm]     node [below] {$\hh_1\ttu!\msg[a]$} (1);
       \end{tikzpicture}
    \end{array}
       &
       \begin{array}{c}
       |\\
       |\\
       |\\
       |
       \end{array}
       &
       \begin{tikzpicture}[mycfsm]
  \node[state]           (0)                        {$0$};
   \node[draw=none,fill=none] (start) [above left = 0.3cm  of 0]{$\hh_2$};
  \node[state]            (1) [above right of=0,yshift=-5mm] {$1$};
  \node[state]           (2) [below right of=0,yshift=5mm] {$2$};

   \path  (start) edge node {} (0) 
            (0)  edge     [bend left]      node [above] {$\ttv\hh_2?\msg[b]$} (1)
            (0)   edge    [bend right, line width=0.5mm]            node [above]  {$\ttv\hh_2?\msg[a]$} (2);
       \end{tikzpicture}
&
      \begin{tikzpicture}[mycfsm]
  \node[state]           (0)                        {$0$};
   \node[draw=none,fill=none] (start) [above left = 0.3cm  of 0]{$\ttv$};
   \node[state]            (1) [below of=0, yshift=4mm] {$1$};

   \path  (start) edge node {} (0)
            (0)  edge    node [above] {$\ttv\hh_2!\msg[b]$} (1) ;
       \end{tikzpicture}
\end{array}
\end{equation}

The  following (unique) connection policy satisfies the progress property.\\

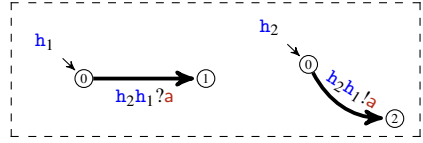
\centerline{ $
\dbox{
       \begin{tikzpicture}[mycfsm]
  \node[state]           (0)                        {$0$};
   \node[draw=none,fill=none] (start) [above left = 0.3cm  of 0]{$\hh_1$};
  \node[state]            (1) [right of=0] {$1$};
   \path  (start) edge node {} (0) 
            (0)  edge [line width=0.5mm]     node [below] {$\hh_2\hh_1?\msg[a]$} (1);
       \end{tikzpicture}
\quad
       \begin{tikzpicture}[mycfsm]
  \node[state]           (0)                        {$0$};
   \node[draw=none,fill=none] (start) [above left = 0.3cm  of 0]{$\hh_2$};
  \node[state]           (2) [below right of=0,yshift=5mm] {$2$};

   \path  (start) edge node {} (0)
            (0)   edge    [bend right, line width=0.5mm]            node [above]  {$\hh_2\hh_1!\msg[a]$} (2);
       \end{tikzpicture}
       }
$}
\vspace{2mm}

\noindent
However,
the PaI composition via partial gateway below does not satisfies the progress property.\\[-4mm]
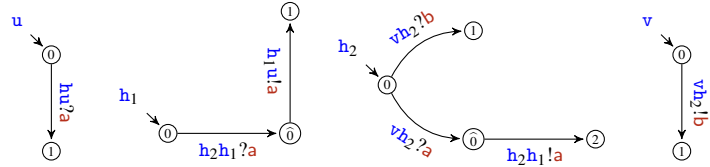
$$
      \begin{tikzpicture}[mycfsm]
  \node[state]           (0)                        {$0$};
   \node[draw=none,fill=none] (start) [above left = 0.3cm  of 0]{$\ttu$};
   \node[state]            (1) [below of=0, yshift=4mm] {$1$};
   \path  (start) edge node {} (0)
            (0)  edge    node [above] {$\hh\ttu?\msg[a]$} (1) ;
       \end{tikzpicture}
\quad
        \begin{tikzpicture}[mycfsm]
  \node[state]           (0)                        {$0$};
   \node[draw=none,fill=none] (start) [above left = 0.3cm  of 0]{$\hh_1$};
  \node[state]            (0hat) [right of=0] {$\widehat 0$};
  \node[state]           (1) [above of=0hat] {$1$};
   \path  (start) edge node {} (0) 
            (0)  edge     node [below] {$\hh_2\hh_1?\msg[a]$} (0hat)
            (0hat)  edge     node [below] {$\hh_1\ttu!\msg[a]$} (1)
            ;
       \end{tikzpicture}
\quad
            \begin{tikzpicture}[mycfsm]
  \node[state]           (0)                        {$0$};
  \node[state]           (hat0)          [below right of=0, yshift=5mm]              {$\widehat{0}$};
   \node[draw=none,fill=none] (start) [above left = 0.3cm  of 0]{$\HH_2$};
  \node[state]            (2) [right of=hat0] {$2$};
  \node[state]           (1) [above right of=0, yshift=-5mm] {$1$};
  %\node[state]           (2) [right of=hat0'] {$2$};

   \path  (start) edge node {} (0) 
            (0)         edge   [bend right]        node [below] {${\ttv\hh_2}?{\msg[a]}$} (hat0)
                         edge   [bend left]      node [above]  {${\ttv\hh_2}?{\msg[b]}$} (1)
             (hat0)  edge        node [below] {${\HH_2\hh_1}!{\msg[a]}$} (2);       
             \end{tikzpicture}
\quad
      \begin{tikzpicture}[mycfsm]
  \node[state]           (0)                        {$0$};
   \node[draw=none,fill=none] (start) [above left = 0.3cm  of 0]{$\ttv$};
   \node[state]            (1) [below of=0, yshift=4mm] {$1$};

   \path  (start) edge node {} (0)
            (0)  edge    node [above] {$\ttv\hh_2!\msg[b]$} (1) ;
       \end{tikzpicture}
$$
\\[-0mm]
In fact, some states in $s=(0_\ttu,0_{\hh_1},1_{\hh_2},1_\ttv)$ are not final, $s$ is reachable and $s\notlts{}$. 

In both of the previous examples, there is a state that contains both normal outgoing transitions and outgoing interface transitions.
However, the following example shows that progress preservation can also be disrupted by states with no outgoing interface transitions.\\[-4mm]
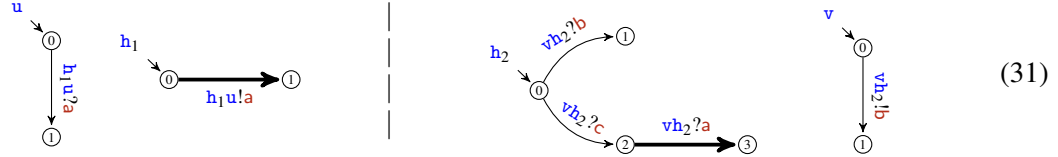
\begin{equation}
\label{eq:4}
\begin{array}{c@{\qquad}c@{\hspace{1cm}}c@{\qquad}c}
    \begin{array}{cc}
      \begin{tikzpicture}[mycfsm]
  \node[state]           (0)                        {$0$};
   \node[draw=none,fill=none] (start) [above left = 0.3cm  of 0]{$\ttu$};
   \node[state]            (1) [below of=0, yshift=4mm] {$1$};

   \path  (start) edge node {} (0)
            (0)  edge    node [above] {$\hh_1\ttu?\msg[a]$} (1) ;
       \end{tikzpicture}
&
       \begin{tikzpicture}[mycfsm]
  \node[state]           (0)                        {$0$};
   \node[draw=none,fill=none] (start) [above left = 0.3cm  of 0]{$\hh_1$};
  \node[state]            (1) [right of=0] {$1$};
  %\node[state]           (2) [above right of=0] {$2$};

   \path  (start) edge node {} (0) 
            (0)  edge  [line width=0.5mm]     node [below] {$\hh_1\ttu!\msg[a]$} (1);
       \end{tikzpicture}
    \end{array}
       &
       \begin{array}{c}
       |\\
       |\\
       |\\
       |
       \end{array}
       &
       \begin{tikzpicture}[mycfsm]
  \node[state]           (0)                        {$0$};
   \node[draw=none,fill=none] (start) [above left = 0.3cm  of 0]{$\hh_2$};
  \node[state]            (1) [above right of=0,yshift=-5mm] {$1$};
  \node[state]           (2) [below right of=0,yshift=5mm] {$2$};
  \node[state]           (3) [right of=2] {$3$};
   \path  (start) edge node {} (0) 
            (0)  edge     [bend left]      node [above] {$\ttv\hh_2?\msg[b]$} (1)
            (0)   edge    [bend right]            node [above]  {$\ttv\hh_2?\msg[c]$} (2)
            (2)   edge    [line width=0.5mm]            node [above]  {$\ttv\hh_2?\msg[a]$} (3);
       \end{tikzpicture}
&
      \begin{tikzpicture}[mycfsm]
  \node[state]           (0)                        {$0$};
   \node[draw=none,fill=none] (start) [above left = 0.3cm  of 0]{$\ttv$};
   \node[state]            (1) [below of=0, yshift=4mm] {$1$};

   \path  (start) edge node {} (0)
            (0)  edge    node [above] {$\ttv\hh_2!\msg[b]$} (1) ;
       \end{tikzpicture}
\end{array}
\end{equation}
The (unique) connection policy satisfies progress.
The composition below, however, does not.
$$
      
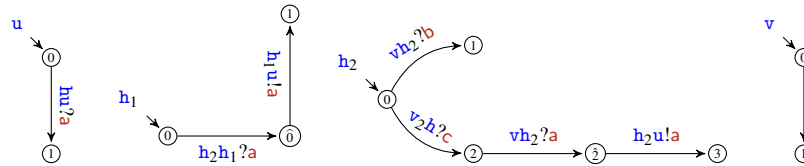
\begin{tikzpicture}[mycfsm]
  \node[state]           (0)                        {$0$};
   \node[draw=none,fill=none] (start) [above left = 0.3cm  of 0]{$\ttu$};
   \node[state]            (1) [below of=0, yshift=4mm] {$1$};

   \path  (start) edge node {} (0)
            (0)  edge    node [above] {$\hh\ttu?\msg[a]$} (1) ;
       \end{tikzpicture}
\quad
        \begin{tikzpicture}[mycfsm]
  \node[state]           (0)                        {$0$};
   \node[draw=none,fill=none] (start) [above left = 0.3cm  of 0]{$\hh_1$};
  \node[state]            (0hat) [right of=0] {$\widehat 0$};
  \node[state]           (1) [above of=0hat] {$1$};
   \path  (start) edge node {} (0) 
            (0)  edge     node [below] {$\hh_2\hh_1?\msg[a]$} (0hat)
            (0hat)  edge     node [below] {$\hh_1\ttu!\msg[a]$} (1)
            ;
       \end{tikzpicture}
\quad
       \begin{tikzpicture}[mycfsm]
  \node[state]           (0)                        {$0$};
   \node[draw=none,fill=none] (start) [above left = 0.3cm  of 0]{$\hh_2$};
  \node[state]            (1) [above right of=0,yshift=-5mm] {$1$};
  \node[state]           (2) [below right of=0,yshift=5mm] {$2$};
  \node[state]           (2hat) [right of=2] {$\hat{2}$};
  \node[state]           (3) [right of=2hat] {$3$};
   \path  (start) edge node {} (0) 
            (0)  edge     [bend left]      node [above] {$\ttv\hh_2?\msg[b]$} (1)
            (0)   edge    [bend right]            node [above]  {$\ttv_2\hh?\msg[c]$} (2)
            (2)   edge           node [above]  {$\ttv\hh_2?\msg[a]$} (2hat)
            (2hat)   edge      node [above]  {$\hh_2\ttu!\msg[a]$} (3)
            ;
       \end{tikzpicture}
\quad
      \begin{tikzpicture}[mycfsm]
  \node[state]           (0)                        {$0$};
   \node[draw=none,fill=none] (start) [above left = 0.3cm  of 0]{$\ttv$};
   \node[state]            (1) [below of=0, yshift=4mm] {$1$};

   \path  (start) edge node {} (0)
            (0)  edge    node [above] {$\ttv\hh_2!\msg[b]$} (1) ;
       \end{tikzpicture}
$$
In fact, some states in the configuration $s=(0_\ttu,0_{\hh_1},1_{\hh_2},1_\ttv)$ are not final, $s$ is reachable and $s\notlts{}$. 
Intuitively, the main problem with some of the previous examples is that the choice of transition in the composition does not depend on a single participant anymore, involving actually both
$\hh_1$ and $\hh_2$.
In the last example, problems also arise when a choice that depends on $\hh_2$ conflicts with what the other partial gateways does.
We therefore need to avoid such issues, possibly by syntactic means.
Our proposal requires outgoing non-interface transitions must always be single, i.e.
if a state has an outgoing non-interface transition, this must be its only outgoing transition.
%\footnote{We claim this condition to be unnecessary for deadlock-freedom preservation.}.
We formalise such restriction in the following definition of CFSM with interface transitions, where  the symbol $\intf$ is intended to identify
interface transitions and $\nintf$  is intended to identify non-interface transitions.

\begin{definition}[CFSM with interface transitions]\label{def:cfsmie}
\label{def:cfsmintftrans}
A {\em CFSM with interface transitions} ($CFSM^{\mathsf{it}}$ for short) is a tuple $M=(Q,q_0,\textit{Act},\bm{\delta})$ 
where $Q$, $q_0$ and $\textit{Act}$ are as in the definition of CFSM, whereas\\
\centerline{
$\bm{\delta} \subseteq Q\times\textit{Act}\times Q \times \Set{\intf,\nintf}$}
\begin{tabular}{lc@{\hspace{4pt}}l}
and such that & - & $(q_1,\elle,q_2,x),(q_1,\elle,q_2,y)\in\bm{\delta} \implies x=y$.\\
                     & - & $(q_1,\elle,q_2,\nintf), (q_1,\elle',q'_2,z)\in\bm{\delta} \implies (\elle=\elle' \text{ and } q_2=q'_2)$\\
                     &    & \hspace{51mm}  (and hence $z=\nintf$ by the previous item).
\end{tabular}\\
An element of $\bm{\delta}$ of the form $(\_,\_,\_,\intf)$ is called {\em interface transition}.
\end{definition}
\noindent 
As done in the previous examples, we use the notation $q\lts{l}q'$ for $(q,l,q',\nintf)$ and
 $q
 \raisebox{2.7mm}
{\begin{tikzpicture}[mycfsm]
      \node[state, draw=none] (zero) [yshift=-4mm, xshift=5mm] {$~$};
      \node[state, draw=none] (one) [right of=zero, xshift=-10mm]   {$~$};
      \draw (zero) edge[-to,line width=0.5mm] node[above]{$l$} (one)
      ;
 \end{tikzpicture}
 } 
\!\! q'$ for $(q,l,q',\intf)$.

It is possible to check that $\hh_2$ in (\ref{eq:1}) above is a CFSM with interface transitions,
whereas participants $\hh_2$ in (\ref{eq:2}), (\ref{eq:3}) and (\ref{eq:4}) are not.
It is worth also noting that the usual PaI composition via gateway is a particular case 
of composition via partial gateways where all transitions are interface transitions.
Absence of non interface transition makes conditions in \cref{def:cfsmintftrans} above
vacuously satisfied.

We now describe the various stages of the PaI composition via partial gateways that have
been roughly presented so far, and provide formal definitions of the involved notions.
We continue to use our running example to illustrate the various stages. 

\paragraph{Indentifying interface participants and selecting interface transitions.}
PaI composition via partial gateways consists, first of all, in identifying one participant
per system and then properly selecting some transitions in order to get CFSMs with
interface transitions. Given a CFSM $M$, an interface decoration is one of the possible
CFSM with interface transitions that we can get out of $M$.

\begin{definition}[Interface decorations]\label{def:IDM}
Let $M=(Q,q_0,\textit{Act},\delta)$ be a CFSM. We define the {\em interface decoration set} of $M$
as the following set of CFSM$^{\mathsf{it}}$\!s:\\
\centerline{$\IDS(M) = \Set{(Q,q_0,\textit{Act},\bm{\delta}) \mid (Q,q_0,\textit{Act},\bm{\delta}) \text{ is a CFSM$^\mathsf{it}$ with } \proj{\bm{\delta}}{Q\times\textit{Act}\times Q} =\delta}$}
where $\proj{\bm{\delta}}{Q\times\textit{Act}\times Q} = \Set{(q,l,q')\mid \exists x\  s.t.  (q,l,q',x)\in\bm{\delta} }$.\\ 
We refer to an element of $\IDS(M)$ as {\em an interface decoration of} $M$.
\end{definition}
\brunningex By the above definition, $\hh_1$ and $\hh_2$ in (\ref{eq:1}) are interface decorations for $\hh_1$ and $\hh_2$ in (\ref{eq:runex}).
In the following we focus for simplicity just on the interface of $S_2$.\erunningex
\\[-8mm]
\paragraph{Extract the intended interface behaviours from the chosen decorations.}
Given a CFSM with interface transitions -- for example $\hh_2$ in (\ref{eq:1}) -- these transitions identify the behaviour of the external system that we intend to connect to via the composition.
Such a behaviour is represented by focusing on interface
transitions only, disregarding the non-interface ones, which we label by $\varepsilon$.\\
\brunningex From $\hh_2$ in (\ref{eq:1}) we hence intend to get,\\[-4mm]
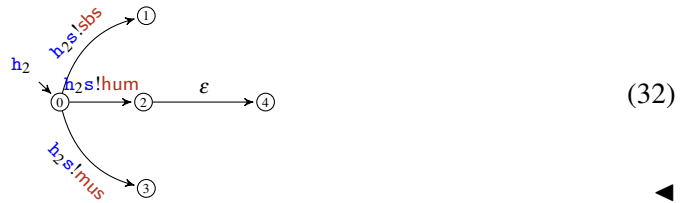
\begin{equation}
\label{eq:epsrunex}
\hspace{18mm}
\begin{tikzpicture}[mycfsm]
  \node[state]           (0)              {$0$};
   \node[draw=none,fill=none] (start) [above left = 0.3cm  of 0]{$\hh_2$};
  \node[state]            (1) [above right of=0] {$1$};
   \node[state]           (2) [right of=0,xshift=-6mm] {$2$};
   \node[state]           (3) [below right of=0] {$3$};
   \node[state]           (4) [right of=2] {$4$};
   \path  (start) edge node {} (0) 
            (0)  edge     [bend left]      node [above] {$\hh_2\tts!\msg[sbs]$} (1)
                   edge                          node [above]  {$\hh_2\tts!\msg[hum]$} (2)
                   edge    [bend right]     node [below]  {$\hh_2\tts!\msg[mus]$} (3)
            (2)  edge                           node [above]  {$\varepsilon$} (4)
                   ;
       \end{tikzpicture}
       \vspace{-8mm}
\end{equation}
\begin{flushright}\erunningex\end{flushright}
%where the label $\varepsilon$ in interpreted as the silent action $\tau$ commonly used in concurrency theory [[[*not sure that such an interpretation is helpful*]]]. 

\noindent
We use the $\varepsilon$ symbol since we manipulate it as the empty string in automata theory.\\
To formalise the process through which we get (\ref{eq:epsrunex}) from the $\hh_2$ of (\ref{eq:1}), we use the following function.
\begin{definition}[$\varepsilon(M)$]
\label{def:epsfun}
\item
Let $M=(Q,q_0,\textit{Act},\bm{\delta})$ be a CFSM$^{\mathsf{it}}$. We define
$\varepsilon(M)$ as  the $\varepsilon$-FSA $(Q,q_0,\textit{Act}\cup\Set{\varepsilon},\delta')$   where\\
\centerline{
$\delta' = \Set{(q,\varepsilon,q') \mid (q,\elle,q',\nintf) \in \bm{\delta}}\cup \Set{(q,l,q') \mid (q,\elle,q',\intf) \in \bm{\delta}}$  }
\end{definition}

\noindent
From an $\varepsilon$-FSA like (\ref{eq:epsrunex}) we need now to get a proper CFSM  which behaves the same\footnote{The usual algorithm for $\varepsilon$-FSA are not useful for our purposes.}.\\
%corresponding to the behaviour of the interface participant restricted to what specified by the 
%interface transitions.
%We obtain such restricted behaviour out of the $\varepsilon$-FSA returned by the application of
%the function $\varepsilon(\_)$ on a CFSM with interface transitions.
\brunningex
In our running example such a behaviour should hence correspond to the following CFSM.\\[-4mm]
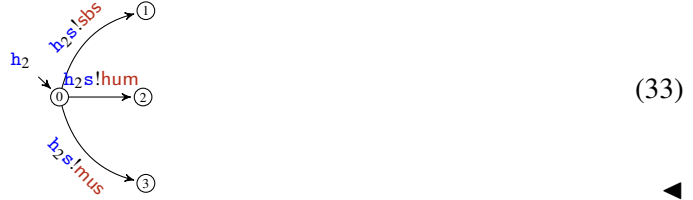
\begin{equation}
\label{eq:cfsmnoeps}
\vspace{-4mm}
\begin{tikzpicture}[mycfsm]
  \node[state]           (0)              {$0$};
   \node[draw=none,fill=none] (start) [above left = 0.3cm  of 0]{$\hh_2$};
  \node[state]            (1) [above right of=0] {$1$};
   \node[state]           (2) [right of=0,xshift=-6mm] {$2$};
   \node[state]           (3) [below right of=0] {$3$};
   %\node[state]           (4) [right of=2] {$4$};
   %
   \path  (start) edge node {} (0) 
            (0)  edge     [bend left]      node [above] {$\hh_2\tts!\msg[sbs]$} (1)
                   edge                          node [above]  {$\hh_2\tts!\msg[hum]$} (2)
                   edge    [bend right]     node [below]  {$\hh_2\tts!\msg[mus]$} (3)
            %(2)  edge                           node [above]  {$\varepsilon$} (4)
                   ;
\end{tikzpicture}
       \vspace{-0mm}
\end{equation}
\\[-12mm]
\begin{flushright}\erunningex\end{flushright}

According to the conditions in \cref{def:cfsmintftrans}, outgoing transitions in a CFSM$^{\mathsf{it}}$
are single if they are non-interface transitions. Therefore, the $\varepsilon$ transitions
in the CFSM obtained by the application of $\varepsilon(\_)$ are also single.
With this restriction, the interface behaviour can be obtained 
 by simply making all the $\varepsilon$-transition ``collapse''.
This transformation is a much simpler algorithm than the standard algorithms for 
eliminating $\varepsilon$-transitions in $\varepsilon$-FSA, due to our strong
restriction on $\varepsilon$-transitions.
In general, however, the CFSM that describes the behaviour of the $\varepsilon$-FSA might even
differ from that obtained using the aforementioned algorithm, provided it is possible
to establish a precise correspondence between the non-$\varepsilon$ transitions in the 
$\varepsilon$-FSA and the CFSM transitions.
We therefore introduce the following definition which focuses not on a particular algorithm
for eliminating $\varepsilon$-transitions, but rather on the properties that the result must satisfy.
%This definition will also be more useful when we deal with partial-fusion later on.

\begin{definition}[$\noeps$-versions]
\label{def:noepsver}
Let $M=(Q,q_0,\textit{Act}\cup\Set{\varepsilon},\delta)$ be a $\varepsilon$-FSA
and $M'=(Q',q'_0,\textit{Act},\delta)$ be a CFSM.
We say that $M'$ is a {\em $\noeps$-version of $M$ via $f$} ($\noeps_f$-version of $M$ for short)
whenever
\begin{enumerate}[a)]
\item
$M'$ is deterministic;
\item 
$f:Q\to Q'$ is onto and such that $f(q_0) = f(q'_0)$;
\item
$q_1\lts{\epsilon}q_2 \quad\text{implies}\quad f(q_1)=f(q_2)$; 
\item
$q_1\lts{\elle}q_2 \quad\text{implies}\quad f(q_1)\lts{\elle}f(q_2)$; 
\item
\label{def:noepsver-e}
$ f(q_1)\lts{\elle}f(q_2)\quad\text{implies}\quad q_1\LTS{\elle}q_2$ 
(i.e. $q_1\lts{\varepsilon}^{\hspace{-2mm}*}q'_1\lts{\elle}q'_2\lts{\varepsilon}^{\hspace{-2mm}*}q_2$ for some $q'_1$ and $q'_2$).
\end{enumerate}
\end{definition}
\Commented{
\bfc $f$ is a sort of weak bisimulation? In the last two items $q_1\lts{\elle}q_2$ should be $q_1\LTS{\elle}q_2$ \efc } % END Commented

\brunningex
The CFSM in (\ref{eq:cfsmnoeps} is a $\noeps_f$-version of the $\varepsilon$-FSA in
(\ref{eq:epsrunex}) where \\[-4mm]
\begin{equation}
\label{eq:f}
f:\Set{0,1,2,3,4}\to\Set{0,1,2,3} \text{ with } f(4)=2 \text{ and } f(q)=q \text{ for } q\in\Set{0,1,2,3}.
\vspace{-3mm}
\end{equation}
\erunningex

\Commented{%<<<<<<<<<<<<< BEGIN COMMENTED
\begin{remark}
\label{rem:neccond}
{\em
Note that it might be impossible to find an $f$ that satisfies the conditions in \cref{def:noepsver} without requiring  \ 
$(q_1,\elle,q_2,\nintf), (q_1,\elle',q'_2,z)\in\bm{\delta} \implies (\elle=\elle' \text{ and } q_2=q'_2)$ \
in \cref{def:cfsmie}. \finex
}
\end{remark}
} %<<<<<<<<<<<<<<<<<<<<< END COMMENTED

\paragraph{Describing the forwarding behaviours i.e. the connection policy.}
As previously mentioned, a connection policy is a communicating system that describes how in the intended composition, partial gateways forward the messages involved in
interface transitions. 
We can obtain an element for the desired connection policy from a $\noeps$-version of an $\varepsilon$-FSA by $(a)$ ``dualising'' the actions labelling the transitions and $(b)$ replacing the senders in inputs and the receivers in outputs with participant names of the connection policy. Note that, unlike in the composition of multiple systems, such a replacement is uniquely determined in the binary composition case. 
In the following definition of $\roles$-duality, $\roles$ is the set of the names of the other participants in the intended connection policy.
%\begin{definition}[$\I(M)$]\label{def:IM}%\hfill\\
%Let $M=(Q,q_0,\textit{Act},\bm{\delta})$ be a CFSM with interface edges
%and let $M'=(Q',q'_0,\textit{Act},\delta')$ be a standard CFSM. We say that
%$M'$ is an {\em interface for $M$ via $(f,g)$}, $\I^M_{\!\!(f,g)}(M')$, whenever
%\begin{itemize}
%\item[-]
%$f:Q\to Q'$  is onto and such that, for all $q\in Q$, $\langin(q)=\langin(f(q))$;
%\item[-]
%$g:\delta'\to\Set{e\in\delta\mid e \text{ is an interface edge}}$ is such that 
%$\I(M)$ as the CFSM obtained out of $\bm{\varepsilon}(M)$ using the standard procedure
%to get a FSA without $\varepsilon$-transitions out of a $\varepsilon$-FSA \cite[someStandardReference]. 
%Roughly:\\ 
%- one first calculate the $\varepsilon$-closure for each state, which is the set of all states reachable from a given state using only $\varepsilon$-transitions;\\ 
%- then, for each element of $\textit{Act}$, define new transitions for each state by considering the $\varepsilon$-closures of the states reachable via the original transition function.
%\end{itemize}
%\end{definition}
\begin{definition}[$\roles$-duality]\label{def:PD}%\hfill\\
\begin{enumerate}[i)]
%\item
%Let $M=(Q,q_0,\textit{Act},\bm{\delta})$ be a CFSM with interface edges. We define
%$\varepsilon(M)$ as  the $\varepsilon$-FSA $(Q,q_0,\textit{Act}\cup\Set{\varepsilon},\delta')$   where\\
%\centerline{
%$\delta' = \Set{(q,l,q') \mid (q,\varepsilon,q',\nintf) \in \bm{\delta}}\cup \Set{(q,\varepsilon,q') \mid (q,\elle,q',\intf) \in \bm{\delta}}$  }
\item
Let $l,l'\in\textit{Act}$ and  let $\roles\neq\emptyset$ be a set of participants.
We say that $l'$ is {\em a $\roles$-dual of $l$} whenever 
\begin{itemize}
\item[-]
$l = \ttr\ttq?\msg[m] \implies l'= \ttq\tts!\msg[m] \text{ with } \tts\in\roles$; %, $\tts\neq\ttq$;
\item[-]
$l = \ttq\ttr!\msg[m] \implies l'= \tts\ttq?\msg[m] \text{ with } \tts\in\roles$; %, $\tts\neq\ttq$.
\end{itemize}
\item
Let $\delta,\delta'\in Q\times\textit{Act}\times Q$ and  let $\roles$ be a set of participants.
We say that $\delta'$ is {\em a $\roles$-dual of $\delta$} whenever $\delta'$ is a minimal relation over
 $Q\times\textit{Act}\times Q$ such that\\
\centerline{
$q\lts{l}q'\in\delta\ $ implies $\ q\lts{l'}q'\in\delta'$, where $l'$ is a $\roles$-dual of $l$.
}
\item
Let $M=(Q,q_0,\textit{Act},\delta)$ and $M'=(Q,q_0,\textit{Act},\delta')$ be two CFSMs and  let $\roles$ be a set of participants.
We say that $M'$ is {\em a $\roles$-dual of $M$} whenever $\delta'$ is a $\roles$-dual of $\delta$.
\end{enumerate}
\end{definition}

\brunningex
We have that $\hh_2$ in (\ref{eq:cp1}) is a $\Set{\hh_1}$-dual of (\ref{eq:cfsmnoeps}) (actually the
only $\Set{\hh_1}$-dual, since $\Set{\hh_1}$ is a singleton by the fact that in our running example we consider binary composition).
\erunningex

In order to formalise the notion of connection policy, we now define a binary relation on CFSMs.
Two CFSMs are related if the first CFSM is an element of a possible connection policy, and the second CFSM is a participant whose certain transitions are intended to serve as interface transitions. This relation depends on the concepts of interface decoration,
$\varepsilon(\_)$, $\noeps$-version and $\roles$-duality.
 \begin{definition}[$\roles$-embeddability]
\label{def:Pcomplementarity}
Let $M^1_\hh = (Q, q_0, \textit{Act}, \delta^1)$ and 
$M^2_\hh = (Q, q_0, \textit{Act}, \delta^2)$  
be two CFSMs with the same name $\hh$, and let $\roles$ be a set of participants.
Moreover, let $\bm{\delta} \subseteq Q\times\textit{Act}\times Q \times \Set{\intf,\nintf}$.\\
We say that
$M^1_\hh$ is {\em $\roles$-embeddable into $M^2_\hh$ via $\bm{\delta}$ and $f$}, 
written $\emb{f}{\roles}{\bm{\delta}}{M^1_\hh}{M^2_\hh}$, whenever \\
\centerline{$M^1_\hh \text{ is a $\roles$-dual of a $\noeps_{\!f}$-version of } \varepsilon(M'_\hh)$}
where $M'_\hh= (Q, q_0, \textit{Act}, \bm{\delta})\in\IDS(M^2_\hh)$.
We refer to $\varepsilon(M'_\hh)$ as the {\em $\epsilon$-counterpart of $M^1_\hh$}.
We write simply $M^1_\hh \embd M^2_\hh$ whenever there exist
$\roles$, $\bm\delta$ and $f$ such that $\emb{f}{\roles}{\bm{\delta}}{M^1_\hh}{M^2_\hh}$
and in this case we say that  $M^1_\hh$ {\em is embeddable into} $M^2_\hh$.
\end{definition}

\brunningex
 In our running example we have that\\[-2mm]
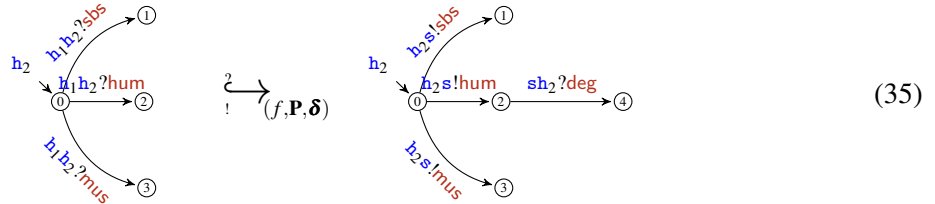
\begin{equation}
\label{eq:emb}
\begin{tikzpicture}[mycfsm]
  \node[state]           (0)              {$0$};
   \node[draw=none,fill=none] (start) [above left = 0.3cm  of 0]{$\hh_2$};
  \node[state]            (1) [above right of=0] {$1$};
   \node[state]           (2) [right of=0,xshift=-6mm] {$2$};
   \node[state]           (3) [below right of=0] {$3$};
   %\node[state]           (4) [right of=2] {$4$};
   %
   \path  (start) edge node {} (0) 
            (0)  edge     [bend left]      node [above] {$\hh_1\hh_2?\msg[sbs]$} (1)
                   edge                          node [above]  {$\hh_1\hh_2?\msg[hum]$} (2)
                   edge    [bend right]     node [below]  {$\hh_1\hh_2?\msg[mus]$} (3)
            %(2)  edge                           node [above]  {$\varepsilon$} (4)
                   ;
       \end{tikzpicture}
 \qquad
\text{\LARGE $\embd$}_{\hspace{-2mm}(f,\roles,\bm{\delta})}
\quad
\begin{tikzpicture}[mycfsm]
  \node[state]           (0)              {$0$};
   \node[draw=none,fill=none] (start) [above left = 0.3cm  of 0]{$\hh_2$};
  \node[state]            (1) [above right of=0] {$1$};
   \node[state]           (2) [right of=0,xshift=-6mm] {$2$};
   \node[state]           (3) [below right of=0] {$3$};
   \node[state]           (4) [right of=2] {$4$};
   \path  (start) edge node {} (0) 
            (0)  edge     [bend left]      node [above] {$\hh_2\tts!\msg[sbs]$} (1)
                   edge                          node [above]  {$\hh_2\tts!\msg[hum]$} (2)
                   edge    [bend right]     node [below]  {$\hh_2\tts!\msg[mus]$} (3)
            (2)  edge                           node [above]  {$\tts\hh_2?\msg[deg]$} (4)
                   ;
       \end{tikzpicture}
\end{equation}
where $f$ is as in (\ref{eq:f}), $\roles = \Set{\hh_1}$ and $\bm{\delta}$ is as for $\hh_2$ 
in (\ref{eq:1}).
\erunningex

We can now formally define the notion of connection policy, intended as the description of the 
way partial gateways has to communicate in a PaI composition. 

 \begin{definition}[Connection policy]\label{def:cp}
 Let $S_i=(M^i_{\ttx})_{\ttx\in\roles_i}$ be two communicating systems with interface participants $\hh_i$ ($i=1,2$).
 A {\em connection policy} for the set of interface participants $H=\Set{\hh_1,\hh_2}$ is a communicating system\ \ $\cs = (M^\cs_\ttu)_{\ttu\in H}$\ 
 such that,
 for each $i\in\Set{1,2}$, for some $f_i$ and $\bm{\delta_i}$: \  $\emb{f_i}{H\setminus\Set{\hh_i}}{\bm{\delta_i}}{M^\cs_{\hh_i}}{M^i_{\hh_i}}$.\\[-6mm]
\end{definition}
\brunningex
The communicating system of (\ref{eq:cp1}) is hence a connection policy for our running example.
\erunningex

\paragraph{Build the partial gateways.}
%Binary PaI composition via partial gateways is obtained by replacing the interface participants with partial gateways. 
In the PaI composition via partial gateways, such gateways are constructed using the interface participants and their corresponding participants in a connection policy (those possessing the same name).

\begin{definition}[Partial Gateway]
\label{def:gatewaycs} 
\label{def:gatewaymc}
\label{def:parfus}
Let $M^1_{\hh}= (Q_1, q^1_0,\textit{Act},\delta^1)$ and 
$M^2_{\hh} = (Q_2, {q^2_0},\textit{Act},\delta^2)$ such that \linebreak
$\emb{f}{\roles}{\bm{\delta}}{M^1_\hh}{M^2_\hh}$.
The {\em partial gateway} $M^1_{\hh}{\gts} M^2_{\hh}$ obtained from  $M^1_\HH$ and $M^2_\hh$  is the CFSM with name $\hh$ defined by  \\
\centerline{$
M^1_{\hh}{\gts} M^2_{\hh} = (Q^2\cup\widehat Q, q^2_0, \textit{Act},\widehat{\delta})
$}
\begin{tabular}{l@{\hspace{4pt}}c@{\hspace{2mm}}l}
where &  $\bullet$  & $\widehat{Q} =\bigcup_{q\in Q}\Set{q^{(q, l,q')} \mid (q, l,q',\intf)\in\bm{\delta}}$; \\[1mm]
          &  $\bullet$  & $\widehat\delta = \Set{(q,{\ttr}\HH?\msg[a],\widehat q), (\widehat q,\HH\tts!\msg[a],q') \mid  (q,\HH\tts!\msg[a],q',\intf)\in\bm{\delta}, (f(q),\ttr\hh?\msg[a],\ {f(q')})\in\delta^1,\ \widehat q=q^{(q,\hh\tts!\msg[a],q')}}\, \cup$ \\
                &    & ${\hspace{20pt}}\Set{(q,\tts\HH?\msg[a],\widehat q), (\widehat q,\HH {\ttr}!\msg[a],q') \mid  (q,\tts\HH?\msg[a],q',\intf)\in\bm{\delta},\ (f(q),\hh\ttr!\msg[a],{f(q')})\in\delta^1,\ \widehat q=q^{(q,\tts\HH?\msg[a],q')}}\, \cup$ \\
                 &    & ${\hspace{20pt}}\Set{(q,\elle, q') \mid (q,\elle, q',\nintf) \in\bm{\delta}}.$
 \end{tabular} 
 
\smallskip
\noindent
We refer to $\widehat\delta$ as $\widehat\delta_{\HH}$ whenever $\HH$ is not clear from the
context; similarly for $\widehat Q$.
\end{definition}
 
\brunningex
Using the two participants $\hh_2$ %in (\ref{eq:runex}) and $\hh_2$ 
in (\ref{eq:emb})  
we can build the partial gateway intended to be substituted for $\hh_2$ in the composition.
Namely $\hh_2$ in (\ref{eq:fincomp}). %(\ref{eq:comprunex}).
\erunningex

A simple condition has to be satisfied in order two communicating systems can be elegible for PaI composition via partial gateways.

 \begin{definition}[Composability]
 \label{def:composability}
Let $S_1$ and $S_2$  be two communicating systems such that $S_i=(M^i_{\ttx})_{\ttx\in\roles_i}$ ($i=1,2$). 
%Moreover, let $\hh_1$ and $\hh_2$ be two participants of, respectively,  $S_1$ and $S_2$
%and appointed as interfaces.
We say that $S_1$ and $S_2$ are {\em composable} whenever $\roles_1\cap\roles_2=\emptyset$.
\end{definition}

\begin{definition}[PaI composition of communicating systems via partial gateways]
\label{def:comppgw} 
Let $S_1$ and $S_2$ be two composable communicating systems
such that $S_i=(M^i_{\ttx})_{\ttx\in\roles_i}$ ($i\in\Set{1,2}$)
and let  $\cs=(M^\cs_{\ttu})_{\ttu\in H}$  
 be a connection policy for the interface participants $H=\Set{\hh_1,\hh_2}$. 
The {\em PaI composition via partial gateways of 
 $S_1$ and $S_2$ with respect to $\cs$} is the communicating system \\
\centerline{$\PC(\Set{S_i}_{i\in \Set{1,2}}, \cs) =  (M'_\ttp)_{\ttp\in\roles_1\cup\roles_2}$}
where\\
${\qquad\qquad}M'_\ttp = \left\{ \begin{array}{ll}
                          M^i_\ttp 
                                    &  \text{ if }\ \ttp\not\in H \text{ and } \ttp\in\roles_i
                          \\[2mm]
                          M^\cs_{\hh_i}{\gts\,}M^i_{\hh_i} & \text{ if } \ttp=\hh_i \text{ and $i\in \Set{1,2}$}
                           \end{array}
                 \right.$
\end{definition}

%We define $\noeps(M)$ as the CFSM obtained out of $M$ 
%using the following version of the standard procedure to get a FSA without $\varepsilon$-transitions
%out of a $\varepsilon$-FSA \cite{sipser96}, where final states are not taken into account and
%the set of states of $\I(M)$ and $\noeps(M)$ stay the same . 
%Roughly:\\ 
%- Add an arc from p to q labeled a iff there is an arc labeled a in N from some state in eps-CLOSE(p) to q.;\\
%- Delete all arcs labeled with epsilon.
%%- one first calculate the $\varepsilon$-closure for each state, which is the set of all states reachable from a given state using only $\varepsilon$-transitions;\\ 
%%- then, for each element of $\textit{Act}$, define new transitions for each state by considering the $\varepsilon$-closures of the states reachable via the original transition function.
%\end{enumerate}

%Notice that the states of $\I(M)$ and $M$ are the same.

\medskip
We claim that all the previous definitions can be extended in a natural way for the composition of
an arbitrary number of systems where in each of them an interface participant is
selected.

For most properties, preservation by composition requires the no-mixed-state condition (see the notations after \cref{def:cfsm}) to be met.
Indeed, the presence of mixed states among the interface participants would otherwise result in some counterexamples that can be obtained by adapting those in \cite{BH24, BH26}.

\begin{theorem}[Safety of PaI composition via partial gateways]
\label{th:paisafenesse}
Let $S_1$ and $S_2$ be two composable communicating systems with, respectively,
interface participants $\hh_1$ and $\hh_2$ with no mixed states; 
 and let $\cs$ be a connection policy for $H=\Set{\hh_1, \hh_2}$.

Let $\mathcal{P}$ be
either the property of {\em deadlock-freedom} or {\em reception-error-freedom} 
or {\em progress}
(as defined in \cref{def:safeness}).
If $\mathcal{P}$ holds for both $S_1$ and $S_2$, as well as for $\cs$, 
then $\mathcal{P}$ holds for $\PC(\Set{S_i}_{i\in \Set{1,2}}, \cs)$.
Moreover, the above holds also if the no-mixed-state condition is removed and
$\mathcal{P}$ is {\em orphan-message-freedom}.
\end{theorem}

The proof of the above theorem descends from the fact that any system obtained by 
PaI composition via partial gateways can be actually got by two applications of
(binary) partial-fusion composition (\cref{prop:paifromfusion})  and that this particular composition is property preserving (\cref{th:pfusionsafenesse}).
In the following section we define partial-fusion composition and provide the mentioned results.

\section{Partial-fusion Composition}
\label{sec:pfc}

We consider now a particular composition of two communicating systems where a participant of a system is embeddable in a participant of the other system. Their {\em partial-fusion} composition
is obtained by replacing the two participants with a single participant obtained through the 
gateway construction of \cref{def:gatewaycs} that in the present context we call {\em partial fusion}.

\Commented{\begin{definition}[Partial Fusion]
\label{def:parfus}
Let $M^1_\hh = (Q^1, q^1_0, \textit{Act}, \delta^1)$ and $M^2_\hh = (Q^2, q^2_0, \textit{Act}, \delta^2)$  be two CFSMs with the same name $\hh$ such that 
$\emb{f}{\bm{\delta}}{\roles\setminus\Set{\hh}}{M^1_{\hh}}{M^2_\hh}$.
We define the {\em partial fusion of $M^1_{\hh}$ and $M^2_{\hh}$ (via $f$, $\bm{\delta}$ and $\roles$}) as
$$\fusion(M^1_{\hh},M^2_{\hh}) = M^1_{\hh}{\gts}M^2_{\hh}$$
%$$\fusion_{\!\!\bm{\delta}}^{\roles}(M^1_{\hh},M^2_{\hh}) = (Q^2\cup\widehat{Q},q_0,\textit{Act},\widehat{\delta})$$
%\begin{tabular}{l@{\hspace{1mm}}c@{\hspace{2mm}}l}
%where &  $\bullet$  & $\widehat{Q} =\Set{q^{(q, l,q')} \mid (q, l,q',\intf)\in\bm{\delta}}$; \\[1mm]
%          &  $\bullet$  & $\widehat\delta = \Set{(q, l,q') \mid (q, l,q',\nintf)\in\bm{\delta}}\,\cup$\\ 
%           &    & ${\hspace{20pt}}\Set{(q,\ttr\HH?\msg[a],\widehat q), (\widehat q,\HH\tts!\msg[a],q') \mid  (q,\HH\tts!\msg[a],q',\intf)\in\bm{\delta}, (q,\ttr\hh?\msg[a],\ q')\in\delta^1,\ \widehat q=q^{(q,\HH\tts!\msg[a],q')}}\, \cup$ \\
%                &    & ${\hspace{20pt}}\Set{(q,\tts\HH?\msg[a],\widehat q), (\widehat q,\HH'\ttr!\msg[a],q') \mid  (q,\tts\HH?\msg[a],q',\intf)\in\bm{\delta},\ (q,\hh\ttr!\msg[a],q')\in\delta^1,\ \widehat q=q^{(q,\tts\HH?\msg[a],q')}}.$
% \end{tabular} 
\end{definition}
}

\begin{definition}[Composition by Partial Fusion]
\label{def:cpf}
Let $S_1=(M^1_\ttx)_{\ttx\in\roles_1}$ and $S_2=(M^2_\ttx)_{\ttx\in\roles_2}$ be two communicating systems such that $\roles_1\cap\roles_2=\Set{\hh}$
and $\emb{f}{\bm{\delta}}{\roles_1\setminus\Set{\hh}}{M^1_{\hh}}{M^2_{\hh}}$.
We define the {\em composition of $S_1$ and $S_2$ via partial fusion of $\hh$} by
$$\fusioncomp_{\!\hh}(S_1,S_2) = (\widetilde{M}_\ttx)_{\ttx\in\roles_1\cup\roles_2}$$ 
\begin{tabular}{lc@{\hspace{2mm}}l@{\hspace{4mm}}l}
where &  $\bullet$  & $\widetilde M_\ttx = M^1_\ttx$  & $\text{if}\quad \ttx\in\roles_1\setminus\Set{\hh} $; \\[1mm]
          &   $\bullet$  & $\widetilde M_\ttx = M^2_\ttx$ &  $\text{if}\quad \ttx\in\roles_2\setminus\Set{\hh}$; \\[1mm]
                    &   $\bullet$  & $\widetilde M_{\hh} = %\fusion_{\!\!\bm{\delta}}^{{\roles_1\setminus\Set{\hh}}}
                   M^1_{\hh}{\gts}M^2_{\hh}$ % (M^1_{\hh},M^2_{\hh})$.
 \end{tabular} 
 
 \noindent
 We refer to $\widetilde M_{\hh}$ as the {\em connector} of $S_1$ and $S_2$ in the composition.
We say the connector
 $\widetilde M_{\hh}$ is obtained by {\em partial fusion} of $M^1_{\hh}$ and $M^2_{\hh}$.

\end{definition}

\begin{example}
{\em 
By taking as $S_1$ the system in (\ref{eq:cp1}) and as $S_2$ the system on the right in (\ref{eq:runex}) we can check, as done in (\ref{eq:emb}), that $M^1_{\hh_2}$ is embeddable into
$M^2_{\hh_2}$. Hence, the composition of $S_1$ and $S_2$ via  partial fusion of $\hh_2$ returns the system\\
\centerline{$
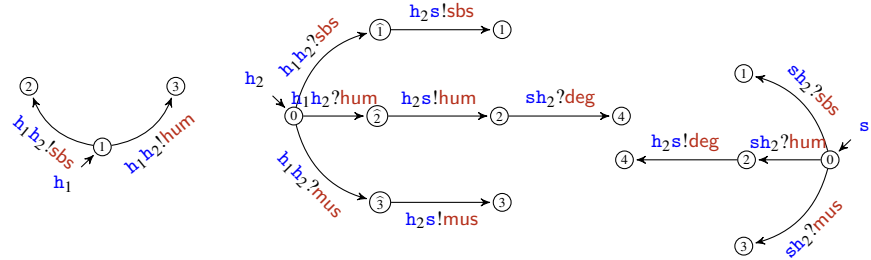
\begin{array}{cc@{\hspace{-4mm}}c}
\begin{tikzpicture}[mycfsm]
   \node[state]            (1) [above of=0] {$1$};
   \node[draw=none,fill=none] (start) [below left = 0.3cm  of 1]{$\hh_1$};
   \node[state]            (2) [above left of=1, yshift=-4mm,xshift=2mm] {$2$};
   \node[state]            (3) [above right of=1, yshift=-4mm,xshift=-2mm] {$3$};
   \path  (start) edge node {} (1)
            (1)  edge[bend left]    node [below] {$\hh_1\hh_2!\msg[sbs]$} (2)
            (1)  edge[bend right]    node [below] {$\hh_1\hh_2!\msg[hum]$} (3) 
            ;
       \end{tikzpicture}    
  &
      \raisebox{3mm}{\begin{tikzpicture}[mycfsm]
  \node[state]           (0)              {$0$};
   \node[draw=none,fill=none] (start) [above left = 0.3cm  of 0]{$\hh_2$};
  \node[state]            (1hat) [above right of=0] {$\widehat 1$};
    \node[state]            (1) [right of=1hat] {$1$};
   \node[state]           (2hat) [right of=0,xshift=-6mm] {$\widehat 2$};
    \node[state]           (2) [right of=2hat] {$2$};
   \node[state]           (3hat) [below right of=0] {$\widehat 3$};
   \node[state]           (3) [ right of=3hat] {$3$};
   \node[state]           (4) [right of=2] {$4$};
   \path  (start) edge node {} (0) 
            (0)  edge     [bend left]      node [above] {$\hh_1\hh_2?\msg[sbs]$} (1hat)
                   edge                          node [above]  {$\hh_1\hh_2?\msg[hum]$} (2hat)
                   edge    [bend right]     node [below]  {$\hh_1\hh_2?\msg[mus]$} (3hat)
            (3hat)  edge                      node [below]  {$\hh_2\tts!\msg[mus]$} (3)
            (1hat)  edge                      node [above]  {$\hh_2\tts!\msg[sbs]$} (1)
            (2hat)  edge                      node [above]  {$\hh_2\tts!\msg[hum]$} (2)
            (2)  edge                           node [above]  {$\tts\hh_2?\msg[deg]$} (4)
                   ;
       \end{tikzpicture}
        }
&
      \raisebox{-3mm}{ \begin{tikzpicture}[mycfsm]
  \node[state]           (0)            {$0$};
   \node[draw=none,fill=none] (start) [above right = 0.3cm  of 0]{$\tts$};
  \node[state]            (1) [above left of=0] {$1$};
   \node[state]           (2) [left of=0,xshift=6mm] {$2$};
   \node[state]           (3) [below left of=0] {$3$};
   \node[state]           (4) [left of=2] {$4$};
   \path  (start) edge node {} (0) 
            (0)  edge     [bend right]      node [above] {$\tts\hh_2?\msg[sbs]$} (1)
                   edge                          node [above]  {$\tts\hh_2?\msg[hum]$} (2)
                   edge    [bend left]     node [below]  {$\tts\hh_2?\msg[mus]$} (3)
            (2)  edge                           node [above]  {$\hh_2\tts!\msg[deg]$} (4)
                   ;
       \end{tikzpicture}
       }
\end{array}
$}
\finex}
\end{example}

\begin{theorem}[Safety of partial fusion composition]
\label{th:pfusionsafenesse}
Let $S_1=(M^1_\ttx)_{\ttx\in\roles_1}$ and $S_2=(M^2_\ttx)_{\ttx\in\roles_2}$ be two communicating systems such that $\roles_1\cap\roles_2=\Set{\hh}$ and where both $M^1_{\hh}$ and $M^2_{\hh}$ have no mixed state. Moreover, let  $\emb{f}{\bm{\delta}}{\roles_1\setminus\Set{\hh}}{M^1_{\hh}}{M^2_{\hh}}$.
Let now $\mathcal{P}$ be
either the property of {\em deadlock-freedom} or {\em reception-error-freedom} 
or {\em progress}
(as defined in \cref{def:safeness}).
If $\mathcal{P}$ holds for both $S_1$ and $S_2$, 
then $\mathcal{P}$ holds for $\fusioncomp_{\!\hh}(S_1,S_2)$.
Moreover, the above holds also if the no-mixed-state condition is removed and
$\mathcal{P}$ is {\em orphan-message-freedom}.
\end{theorem}

The theorem above can be proved by contradiction.
Specifically, one shows that if $\mathcal{P}$ fails to hold for $S$, then it must already fail for either $S_1$ or $S_2$.
Following the approach of \cite{BH24,BH26}, a central concept is that of {\em projection} of a reachable configuration of the composed system onto the corresponding configurations of the individual systems $S_1$ and $S_2$.
In fact, one can prove that the projections of reachable configurations that do not involve intermediate gateway states are themselves reachable configurations.
The proof of \cref{th:pfusionsafenesse} can be found in \cite{B26ext}.\\

\cref{th:paisafenesse} can be obtained as a corollary of \cref{th:pfusionsafenesse} above
by using the following proposition,
which states that any binary PaI-composition via partial gateways can actually be obtained by applying
 the partial-fusion composition twice. First between one of the systems to be composed and
the connection policy; then between  the other system and what has been obtained by the first fusion-composition. This result follows from the definitions of PaI composition via partial gateway and partial-fusion composition.

\begin{proposition}
\label{prop:paifromfusion}
Let $S_1$ and $S_2$ be two composable communicating systems
%such that $S_i=(M^i_{\ttx})_{\ttx\in\roles_i}$ ($i\in\Set{1,2}$)
and let  $\cs$ %$=(M^\cs_{\ttu})_{\ttu\in H}$  
 be a connection policy for their interface participants, respectively, $\hh_1$ and $\hh_2$.\\ %$H=\Set{\hh_1,\hh_2}$. \\
\centerline{$\PC(\Set{S_i}_{i\in \Set{1,2}}, \cs) = \fusioncomp_{\!\hh_1}(S_1,\fusioncomp_{\!\hh_2}(\cs,S_2)) $}
\end{proposition}

\section{Conclusions}
\label{sect:conclusions}
The {\em PaI composition via partial gateways} approach is a method of composing two systems. 
This approach transforms one participant from each system into a partial gateway. 
Any two participants can be chosen. The partial gateways are obtained from the chosen participants and a connection policy. 
One can choose one's preferred connection policy as long as certain requirements (such as embeddability) are met. 
The resulting composite system is guaranteed to satisfy many properties that are satisfied by all the individual systems and the connection policy. The way in which such properties are proven for the individual systems or the connection policy is beyond the scope of the present paper. The property preservation guaranteed by this method of system composition is shown to descend from a similar result for a simpler method of system composition: composition by fusion.

A thorough discussion on works related to system composition in general and PaI composition in particular,  can be
found in \cite[Sect.8]{BH26}.
Although numerous approaches to system composition in the literature can support some form of partial composition -- for example through selective interfaces \cite{H03}, partial specifications \cite{J20}, or modular synchronisation in process algebras like CCS or CSP -- the PaI approach via partial gateways looks conceptually distinct to the best of our knowledge.

We are planning to formally define partial versions of the PaI multicomposition and orchestrated multicomposition \cite{BH24,BH26} and to show that they can be obtained by appropriate numbers of application of the binary fusion-composition proposed in this paper. 
The property preservation of those composition approaches would then be entailed by the 
property preservation result of partial fusion composition.
It would also be interesting to investigate whether and how our approach fits into the
 {\em composition calculus} \cite{FR24}, a very abstract framework of modules and their composition.

Our results do not apply to the property of lock-freedom, i.e. the property guaranteeing 
the absence of reachable configurations where some participant remains stuck in all possible continuations (see \cite{BH24,BH26} for a formal definition).
Actually, counterexamples for lock-freedom preservation can be obtained by simply adapting \cite[Example 5.7]{BH24} for both PaI composition via partial gateways and partial fusion. 
We anticipate that the simple and straightforward nature of partial fusion will enable us to focus on the essential issues underlying the lack of lock-freedom preservation, and hence identify appropriate and manageable conditions ensuring it.

\paragraph{Acknowledgements}
We would like to express our gratitude to the anonymous referees for their careful reading of the paper, their helpful comments and suggestions, and their valuable interactions in the ICE discussion forum. 
We would also like to thank Emilio Tuosto for providing us with macros to draw automata.
Special thanks are also due to Mariangiola Dezani and Rolf Hennicker for their support and friendship.

%\paragraph*{\bf Acknowledgements}
% We warmly thank the ICE'24 reviewers for their careful reading, their thoughtful comments/suggestions and the helpful discussion in the forum. We also thank Emilio Tuosto for his nice tikz style for automata.
%
%
\setlength{\abovedisplayskip}{6pt}
\setlength{\belowdisplayskip}{\abovedisplayskip}

\bibliographystyle{eptcs}
\bibliography{session}

%%------APPENDIX--------
%\newpage
%\appendix
%%
%\input{safeness-of-binary-partial-fusion}
%\input{preservation-results}
%%---------------------------

%------------------------------------------------------------------------------
% Index
%\printindex

%------------------------------------------------------------------------------
\end{document}

% EOF